\documentclass[letterpaper,onecolumn,10pt]{article}
\PassOptionsToPackage{breaklinks, colorlinks}{hyperref}

\usepackage[10pt,nocopyright]{sigmin}
\usepackage{iris}

\makeatletter
\def\maketitle{\par
 \begingroup
   \def\thefootnote{\fnsymbol{footnote}}
   \def\@makefnmark{$^{\@thefnmark}$}
   \newpage\global\@topnum\z@
   \@maketitle
   \@thanks
 \endgroup
 \setcounter{footnote}{0}
 \let\maketitle\relax
 \let\@maketitle\relax
 \gdef\@thanks{}\gdef\@author{}\gdef\@title{}\gdef\@subtitle{}\let\thanks\relax}
\makeatother

\usepackage[letterpaper,margin=1in]{geometry}

\usepackage{libertine}
\usepackage{libertinust1math} 
\usepackage{inconsolata}

\usepackage{hyperref}
\usepackage{xcolor}
\definecolor{linkblue}{RGB}{0,40,200}
\hypersetup{%
  colorlinks=true,
  citecolor=linkblue,
  linkcolor=black,
  urlcolor=linkblue,
}

\input{glyphtounicode}
\usepackage[square,comma,numbers,sort&compress]{natbib}
\usepackage[T1]{fontenc}
\usepackage{xspace}
\usepackage{fancyvrb}
\usepackage{microtype}
\usepackage{graphicx}
\usepackage{amsmath,amssymb,amsthm,wasysym}
\usepackage{aliascnt}
\usepackage{booktabs}

\usepackage{tikz}
\usepackage{tikzpeople}
\usetikzlibrary{shapes,arrows,arrows.meta,positioning,automata,fit,calc,backgrounds,decorations.pathreplacing,calligraphy,patterns}

\usepackage{ifthen}

\usepackage{pgfplots}
\usepackage{pgfplotsthemetol}

\usepackage{proof}

\usepackage{underscore}
\usepackage{relsize}
\usepackage{bytefield}
\usepackage{enumitem}
\usepackage{subfigure}
\usepackage{rotating}
\usepackage{chngcntr}

\usepackage{stfloats}
\usepackage{pifont}
\usepackage{multirow}

\newcommand{\sys}{MachCSL\xspace}

\newif\ifdraft\drafttrue
\newif\ifnotes\notestrue
\ifdraft\else\notesfalse\fi

\definecolor{xxxcolor}{rgb}{0.8,0,0}
\definecolor{red}{RGB}{181, 23, 0}
\definecolor{blue}{RGB}{0, 118, 186}
\definecolor{gray}{RGB}{146, 146, 146}
\definecolor{green}{RGB}{26, 112, 30}
\definecolor{brown}{RGB}{186, 70, 0}
\makeatletter
\long\def\XXX{\@ifnextchar[{\@XXX}{\@XXX[]}}
\long\def\@XXX[#1]{\@ifnextchar[{\@@@XXX{#1}}{\@@XXX{#1}}}
\ifnotes
\long\def\@@XXX#1{{\color{xxxcolor} XXX #1}\xspace}
\long\def\@@@XXX#1[#2]{{\color{xxxcolor} XXX (#1) #2}\xspace}
\else
\long\def\@@XXX#1{\ignorespaces}
\long\def\@@@XXX#1[#2]{\ignorespaces}
\fi
\makeatother

\newcommand{\cc}[1]{\mbox{\smaller[0.5]\texttt{#1}}}

\fvset{fontsize=\small}
\fvset{xleftmargin=6pt}

\input{code/fmt}

\makeatletter
\newcommand{\oset}[3][0ex]{%
  \mathrel{\mathop{#3}\limits^{
    \vbox to#1{\kern-2\ex@
    \hbox{$\scriptstyle#2$}\vss}}}}
\makeatother

\makeatletter
\newcounter{@lena}
\newcommand{\lena}{\the\value{@lena}}

\makeatother

\theoremstyle{definition}

\newaliascnt{lemma}{thm}

\aliascntresetthe{lemma}

\def\Snospace~{\S{}}

\numberwithin{equation}{section}

\counterwithout{equation}{section}

  {\begin{list}{$\bullet$}%
     {\setlength{\parsep}{0pt}%
      \setlength{\topsep}{0pt}%
      \setlength{\itemsep}{0pt}}}%
  {\end{list}}

\bibpunct[: ]{[}{]}{,}{n}{XXX}{XXX}

\def\headline#1{\hbox to \hsize{\hrulefill\quad\lower0.5ex\hbox{#1}\quad\hrulefill}}

\renewcommand{\FancyVerbFormatLine}[1]{%
\ifnum\ifnum\value{FancyVerbLine}=101 1\else\ifnum\value{FancyVerbLine}=116 1\else0\fi\fi=1%
\raisebox{0.5em}{\headline{\textbf{\textrm{#1}}}}
\else#1\fi}

\RecustomVerbatimEnvironment
{Verbatim}{Verbatim}
{xleftmargin=2pt,baselinestretch=0.95}

\newcommand{\wpcycle}{\textlog{wp}\spac\textlog{CpuLoop}}
\newcommand{\wpcycleh}[1]{\textlog{wp}\spac(\textlog{CpuLoop}~{#1})}
\newcommand{\wpsail}[2]{\textlog{wp}\spac\textlog{Sail}({#1})\spac{\left\{#2\right\}}}
\newcommand{\ktext}{\textlog{kernel\_text}}
\newcommand{\instr}[2]{\textlog{instr}\spac{#1}~~{#2}}
\newcommand{\islock}[2]{\textlog{is\_lock}\spac{#1}\spac{#2}}
\newcommand{\regptsto}[2]{{#1}\spac\mapsto_\textlog{Reg}\spac{#2}}
\newcommand{\memptsto}[2]{{#1}\spac\mapsto_\textlog{Mem}\spac{#2}}
\newcommand{\tsoptsto}[2]{{#1}\spac\mapsto_\textlog{TSO}\spac{#2}}
\newcommand{\tsots}[1]{\textlog{timestamp\_is}\spac{#1}}
\newcommand{\ctxvalues}[3]{{#1}\spac\mapsto_\textlog{#2}\spac\{{#3}\}}
\newcommand{\vmemptsto}[2]{{#1}\spac\mapsto_\textlog{vMem}\spac{#2}}
\newcommand{\cons}{\mathbin{:\mkern-5mu:}}
\newcommand{\strptsto}[2]{{#1}\spac\mapsto_\textlog{vMemString}\spac{#2}}
\newcommand{\pc}[1]{\regptsto{\cc{pc}}{#1}}

\newenvironment{specblock}
  {%
   \[\begin{array}{@{}>{\displaystyle}l@{}>{\displaystyle}l@{}}}
  {\end{array}\]}
\newcommand{\specline}[1]{\multicolumn{2}{@{}l@{}}{\displaystyle #1}}

\newenvironment{specblockt}
  {%
   \noindent$\begin{array}[t]{@{}>{\displaystyle}l@{}>{\displaystyle}l@{}}}
  {\end{array}$}

\newcommand{\utext}[3]{{#2}\spac\mapsto_{\textlog{uText}\,#1}\spac{#3}}
\newcommand{\ubyte}[3]{{#2}\spac\mapsto_{\textlog{uData}\,#1}\spac{#3}}
\newcommand{\ubyteq}[3]{{#2}\spac\mapsto^{\textlog{ro}}_{\textlog{uData}\,#1}\spac{#3}}
\newcommand{\uword}[3]{{#2}\spac\mapsto_{\textlog{uWord}\,#1}\spac{#3}}
\newcommand{\ustack}[3]{\textlog{ustack}\spac{#1}\spac{#2}\spac{#3}}

\newenvironment{specblockc}
  {%
   \noindent$\begin{array}[t]{@{}>{\displaystyle}l@{}>{\displaystyle}l@{\qquad}l@{}}}
  {\end{array}$}
\newcommand{\speclinec}[1]{\multicolumn{3}{@{}l@{}}{\displaystyle #1}}

\xspaceaddexceptions{\%}

\newcommand{\nLogicFiles}{53\xspace}            
\newcommand{\nLogicLines}{42{,}948\xspace}
\newcommand{\nInstrFiles}{109\xspace}           
\newcommand{\nInstrLines}{104{,}989\xspace}
\newcommand{\nArchFiles}{23\xspace}             
\newcommand{\nArchLines}{17{,}411\xspace}
\newcommand{\nDevmodelFiles}{3\xspace}          
\newcommand{\nDevmodelLines}{4{,}535\xspace}
\newcommand{\nDevprotoFiles}{8\xspace}          
\newcommand{\nDevprotoLines}{18{,}718\xspace}
\newcommand{\nUsafeFiles}{33\xspace}            
\newcommand{\nUsafeLines}{41{,}018\xspace}
\newcommand{\nUlogicFiles}{59\xspace}           
\newcommand{\nUlogicLines}{54{,}448\xspace}
\newcommand{\nKinvFiles}{166\xspace}            
\newcommand{\nKinvLines}{145{,}938\xspace}
\newcommand{\nSpecFiles}{239\xspace}            
\newcommand{\nSpecLines}{85{,}030\xspace}
\newcommand{\nProofFiles}{322\xspace}           
\newcommand{\nProofLines}{445{,}970\xspace}
\newcommand{\nLinkFiles}{209\xspace}            
\newcommand{\nLinkLines}{3{,}288\xspace}
\newcommand{\nBootFiles}{37\xspace}             
\newcommand{\nBootLines}{25{,}723\xspace}
\newcommand{\nUprogFiles}{97\xspace}            
\newcommand{\nUprogLines}{135{,}960\xspace}
\newcommand{\nAppFiles}{166\xspace}             
\newcommand{\nAppLines}{127{,}841\xspace}
\newcommand{\nAuxFiles}{11\xspace}              
\newcommand{\nAuxLines}{5{,}728\xspace}
\newcommand{\nGenericLines}{265{,}349\xspace}   
\newcommand{\nKernelSideLines}{724{,}667\xspace} 
\newcommand{\nUserLines}{263{,}801\xspace}      
\newcommand{\nModelFiles}{2\xspace}             
\newcommand{\nModelLines}{59{,}887\xspace}
\newcommand{\nModelWrittenFiles}{2\xspace}
\newcommand{\nModelWrittenLines}{146\xspace}    
\newcommand{\nKernelDumpFiles}{5\xspace}
\newcommand{\nKernelDumpLines}{60{,}654\xspace} 
\newcommand{\nUserDumpFiles}{28\xspace}
\newcommand{\nUserDumpLines}{46{,}502\xspace}
\newcommand{\nDecodeFiles}{224\xspace}          
\newcommand{\nDecodeLines}{113{,}557\xspace}
\newcommand{\nTotalLines}{1{,}597{,}440\xspace} 
\newcommand{\nTotalMLines}{1.6\xspace}
\newcommand{\nGeneratedLines}{331{,}166\xspace}
\newcommand{\nWrittenLines}{1{,}266{,}274\xspace}

\newcommand{\nToolsLines}{10{,}358\xspace}      

\newcommand{\nNotesFiles}{238\xspace}
\newcommand{\nNotesLines}{165{,}505\xspace}     

\newcommand{\nXvKernelLines}{6{,}657\xspace}    

\newcommand{\nKernelInstrs}{8{,}725\xspace}

\newcommand{\nInvariants}{30\xspace}

\newcommand{\nTcbPowerLines}{6{,}660\xspace}    
\newcommand{\nTcbPowerDefs}{795\xspace}

\newcommand{\nTcbFsLines}{4{,}035\xspace}       
\newcommand{\nTcbFsDefs}{499\xspace}

\newcommand{\nTcbUnionLines}{4{,}627\xspace}    
\newcommand{\nTcbUnionDefs}{630\xspace}

\newcommand{\nVtestCases}{77\xspace}

\newcommand{\nVtestImageLines}{12{,}065\xspace} 
\newcommand{\nVtestRocqFiles}{13\xspace}
\newcommand{\nVtestRocqLines}{6{,}583\xspace}
\newcommand{\nVtestCaptureFiles}{442\xspace}
\newcommand{\nVtestCaptureLines}{50{,}566\xspace} 
\newcommand{\nVtestFindings}{35\xspace}         

\newcommand{\nProofBuildMin}{56\xspace}         
\newcommand{\nModelBuildMin}{2\xspace}          
\newcommand{\nAuditMin}{17\xspace}              
\newcommand{\nVtestMin}{6\xspace}               
\newcommand{\nBuildAuditVtestMin}{81\xspace}    

\newcommand{\nDevDays}{102\xspace}

\newcommand{\nSessions}{492\xspace}
\newcommand{\nWorkDirs}{13\xspace}
\newcommand{\nPeakSessions}{8\xspace}           
\newcommand{\nSubagents}{2{,}765\xspace}
\newcommand{\nPrompts}{6{,}058\xspace}
\newcommand{\nPromptWords}{160{,}091\xspace}
\newcommand{\nPromptMedianWords}{13\xspace}
\newcommand{\nPromptMeanWords}{26.4\xspace}

\newcommand{\nFollowups}{2{,}385\xspace}        
\newcommand{\nAssistantMsgs}{489{,}448\xspace}
\newcommand{\nToolCalls}{501{,}905\xspace}
\newcommand{\nAgentHours}{2{,}153\xspace}
\newcommand{\nWallHours}{1{,}272\xspace}

\newcommand{\nOutputMTokens}{235.3\xspace}
\newcommand{\nThinkingMTokens}{57.6\xspace}

\newcommand{\nInputBTokens}{174.3\xspace}
\newcommand{\nCacheReadBTokens}{171.8\xspace}

\newcommand{\nCommits}{7{,}715\xspace}

\newcommand{\nRetentionDays}{30\xspace}         

\newcommand{\nLostShare}{9\xspace}              

\newcommand{\nXvBumps}{26\xspace}               
\newcommand{\nXvBumpsMatched}{25\xspace}        

\newcommand{\nXvBumpMedianMin}{35\xspace}
\newcommand{\nXvBumpMinMin}{2\xspace}
\newcommand{\nXvBumpMaxHours}{12.2\xspace}
\newcommand{\nXvBumpHours}{39.5\xspace}
\newcommand{\nXvBumpTimeShare}{1.8\xspace}      
\newcommand{\nXvBumpMedianPrompts}{2\xspace}
\newcommand{\nXvBumpPrompts}{157\xspace}
\newcommand{\nXvBumpGenChurn}{1{,}544{,}180\xspace}
\newcommand{\nXvBumpWrittenChurn}{118{,}420\xspace}
\newcommand{\nXvBumpWrittenMedian}{2{,}822\xspace}

\newcommand{\nSailBumps}{3\xspace}
\newcommand{\nSailAdBumpHours}{10.9\xspace}     
\newcommand{\nSailAxiomBumpHours}{1.7\xspace}   
\newcommand{\nSailTagBumpHours}{0.8\xspace}     
\xspaceaddexceptions{\%}

\xspaceaddexceptions{\%}

\begin{document}
\title{Extending concurrent separation logic to the hardware level \\
  to verify the xv6 OS kernel on RISC-V with AI agents}
\author{M. Frans Kaashoek and Nickolai Zeldovich \\ \emph{MIT CSAIL}}
\date{
  October 11, 2026 -- revision v3%
\footnote{See \autoref{sec:changelog} for a list of changes since arXiv:2609.04043v1 and v2~\cite{kaashoek:machcsl}.}
}

\maketitle

\begin{abstract}
\sys is a framework for verifying system software, such as an
OS kernel, on top of low-level semantics of a RISC-V computer,
based on the Sail RISC-V semantics.  The key idea behind \sys is to
adapt concurrent separation logic, based on Iris, to reasoning about
low-level hardware execution at the sub-instruction level: page-table
translation, TLB, privilege levels, configuration registers, instruction
fetch/decode/execute, traps and interrupts, DMA, shared memory, power
failures, etc.  Reasoning at this level of detail ensures that the system
software correctly manages all of the hardware details.

Verifying software at this low level of abstraction is tedious, but
LLM-based agents are capable of reasoning about such low-level details.
As a case study, we verify the xv6 OS kernel (\nXvKernelLines lines of
C and assembly code), which provides a traditional Unix system call
interface (processes, file system, file descriptors, and preemptive
scheduling) and has substantial internal concurrency (multi-core support
with fine-grained locking, shared memory, interrupts, DMA, etc.).  In the
verification process, we uncovered 11 bugs in the xv6 implementation,
as well as one bug in the Sail RISC-V semantics, and one bug in Sail's Lean
extraction.  The verification effort
took us \nDevDays days, including the time to develop the \sys framework.

On top of the verified xv6 kernel, we prove a theorem about console-level
behavior: e.g., if the user types \cc{echo hello world} as input on the
UART console, the only output the system can produce is \cc{hello
world} (if there are no crashes or other errors).
This theorem covers both the kernel and the \cc{init}, \cc{sh}, \cc{echo},
\cc{cat}, \cc{grep}, \cc{sync}, and \cc{seccomp} binaries in the
initial file system image.

\end{abstract}

\section{Introduction}
\label{sec:intro}

\begin{figure*}[t]
\centering
\begin{tikzpicture}[
  font=\small,
  >={Stealth[length=5pt,width=4pt]}, semithick,
  box/.style={draw, rounded corners=2pt, align=center, inner sep=4pt},
  cell/.style={box, fill=white, font=\footnotesize, minimum width=1.5cm, minimum height=0.85cm},
  tool/.style={cell, minimum width=0.85cm, minimum height=0.7cm,
               text height=5pt, text depth=1.3pt},
  inp/.style={box, fill=white, text width=2.2cm, minimum height=1.0cm},
  lbl/.style={font=\scriptsize, align=center, inner sep=2pt},
  hd/.style={font=\small\bfseries},
  thm/.style={box, fill=green!8, draw=green!45!black, double, double distance=1.2pt, thick,
              text width=15.75cm, inner sep=6pt, align=flush center},
  cor/.style={box, fill=green!4, draw=green!45!black, align=flush left, text width=5.0cm,
              minimum height=1.62cm, inner sep=6pt, font=\footnotesize},
  proofbox/.style={box, fill=white, align=left, text width=3.4cm, inner sep=9pt},
]

\node[cell, text width=2.5cm, minimum height=1.0cm] (ksrc) at (6.35,1.05) {xv6 kernel source\\(C + assembly)};
\node[cell, text width=2.5cm, minimum height=1.0cm] (usrc) at (6.35,-1.05)
  {user programs (C):\\\cc{init}, \cc{sh}, \cc{echo}, \dots};

\node[tool] (gcc)  at ( 8.75,1.05) {\cc{gcc}};
\node[tool] (as)   at ( 9.90,1.05) {\cc{as}};
\node[tool] (ld)   at (11.05,1.05) {\cc{ld}};
\node[tool] (gcc2) at ( 8.75,-1.05) {\cc{gcc}};
\node[tool] (ld2)  at ( 9.90,-1.05) {\cc{ld}};
\node[tool] (mkfs) at (11.05,-1.05) {\cc{mkfs}};
\node[hd, align=center] (bhd) at (9.90,0) {Build toolchain\\{\scriptsize\mdseries (untrusted, outside TCB)}};
\begin{scope}[on background layer]
  \node[draw, densely dashed, rounded corners=3pt, fill=white, inner sep=5pt,
        fit=(gcc)(ld)(gcc2)(mkfs)] (build) {};
\end{scope}

\node[inp] (bin) at (13.55,1.05) {xv6 kernel\\binary (ELF)};
\node[inp] (img) at (13.55,-1.05) {xv6 disk image\\(\cc{fs.img})};
\node[hd]  (ihd) at (13.55,0) {Inputs};
\begin{scope}[on background layer]
  \node[draw, densely dashed, rounded corners=3pt, fill=white, inner sep=5pt,
        fit=(bin)(img)] (inputs) {};
\end{scope}

\draw[->] (ksrc) -- (gcc);
\draw[->] (gcc) -- (as);
\draw[->] (as) -- (ld);
\draw[->] (ld) -- (bin);
\draw[->] (usrc) -- (gcc2);
\draw[->] (gcc2) -- (ld2);
\draw[->] (ld2) -- (mkfs);
\draw[->] (mkfs) -- (img);

\node[cell, text width=3.0cm] (hart) at (6.90,-4.05) {Sail RISC-V HART $\times$\,8\\(RISC-V International)};
\node[cell, text width=3.0cm] (ext)  at (10.62,-3.98) {UARTs, disk controller\\(DMA), PLIC, power on/off};
\node[cell, minimum width=1.4cm] (disk) at (13.55,-3.98) {disk};
\node[cell, minimum width=1.4cm] (mem)  at (15.38,-3.98) {shared\\memory};
\node[hd]   (mhd)  at (10.75,-3.12) {RISC-V system model (trusted)};
\begin{scope}[on background layer]
  \node[draw, rounded corners=4pt, fill=black!8, inner sep=5pt,
        fit={(hart)(mem)(mhd)($(hart.south west)+(0,-0.22)$)}] (model) {};
\end{scope}

\begin{scope}[on background layer]
  \foreach \s in {2,1} {
    \node[box, fill=white, inner sep=0pt,
          fit={($(hart.north west)+(0.14*\s,-0.14*\s)$)
               ($(hart.south east)+(0.14*\s,-0.14*\s)$)}] {};
  }
\end{scope}

\draw[->] (img.south) -- node[lbl, left, pos=0.36] {initial disk\\contents} (disk.north);
\draw[->, rounded corners=2pt] (bin.east) -| node[lbl, left, pos=0.86] {loaded at\\power-on} (mem.north);

\node[proofbox, minimum height=6.46cm] (proofs) at (1.97,-1.65)
  {\textbf{\sys}\\\textbf{specs + proofs}\\(checked by Rocq)\\[9pt]
   $\bullet$~CSL spec for every\\
   \phantom{$\bullet$~}instruction and device\\[7pt]
   $\bullet$~spec + proof for every\\
   \phantom{$\bullet$~}kernel function\\[7pt]
   $\bullet$~kernel invariants\\[7pt]
   $\bullet$~adequacy theorem};

\draw[->] (model.west) -- node[lbl, above] {lifted\\into CSL} (model.west -| proofs.east);

\node[thm] (thm) at (8.15,-6.25)
  {\textbf{Kernel theorem.}\; Starting powered off with these inputs, every execution of the
   model \\
   (under any interleaving of HARTs, devices, interrupts, and power cycles)
   maintains all invariants.};

\draw[->] (proofs.south) -- node[lbl, right] {establish} (thm.north -| proofs.south);
\draw[->] (model.south) -- node[lbl, right] {every execution satisfies} (thm.north -| model.south);

\node[cor, anchor=north west] (cor1) at (5.1,-7.95)
  {\textbf{Corollary (kernel integrity).}\;
   Kernel maintains its invariants at every instruction
   (e.g., no assertion failures, no double frees, no arbitrary
   jumps).};
\node[cor, anchor=north west] (cor2) at ($(cor1.north east)+(0.25,0)$)
  {\textbf{Corollary (process isolation).}\;
   Arbitrary user-space code cannot corrupt kernel memory,
   and can reenter the kernel only through the trap handler.};
\node[cor, anchor=north west] (cor3) at ($(cor1.south west)+(0,-0.22)$)
  {\textbf{Corollary (crash safety).}\;
   On-disk file system stays consistent under any interleaving of
   concurrent system calls and power failures.};
\node[cor, anchor=north west] (cor4) at ($(cor3.north east)+(0.25,0)$)
  {\textbf{Corollary (hardware safety).}\;
   Kernel never triggers undefined hardware behavior: no stray
   DRAM or MMIO access, no device protocol violations.};
\begin{scope}[on background layer]
  \node[draw, densely dashed, rounded corners=3pt, inner sep=5pt,
        fit=(cor1)(cor2)(cor3)(cor4)] (cors) {};
\end{scope}
\draw[->] (thm.south -| cors.north) -- node[lbl, right] {implies, e.g.} (cors.north);

\node[proofbox, anchor=north] (appproofs) at (proofs.south |- cors.north)
  {\textbf{App specs + proofs}\\(checked by Rocq)\\[9pt]
   $\bullet$~spec + proof for \cc{init},\\
   \phantom{$\bullet$~}\cc{sh}, \cc{echo}, etc.\\[7pt]
   $\bullet$~built on the kernel's\\
   \phantom{$\bullet$~}system-call specs};

\node[thm, anchor=north] (appthm) at ($(thm.south |- cors.south)+(0,-0.85)$)
  {\textbf{Application theorem (\autoref{sec:theorem}).}\;
    If the user types \cc{echo hello world} on the console, \\
    the only output the system can produce on the console is \cc{hello world}
    (and similarly for other shell commands).};

\coordinate (lane) at ($(appproofs.east)!0.5!(cors.west)$);
\draw[->] (thm.south -| lane) -- node[lbl, right, pos=0.945] {implies} (appthm.north -| lane);
\draw[->] (appproofs.south) -- node[lbl, right] {establish} (appthm.north -| appproofs.south);
\end{tikzpicture}
\caption{An overview of the proof approach for verifying the xv6 kernel, and applications running on top of it.}
\label{fig:overview}
\end{figure*}

Virtually every application uses the operating system to allocate memory,
read and write files, print output, and even to initialize the system so
that the application can start running.  As a result, applications depend
on the correctness of the operating system kernel that they are running on.
Operating system kernels, however, are hard to get right.  Kernels have to
program the underlying hardware to correctly implement user-mode isolation
for processes, such as setting up page tables, physical memory protection,
and other configuration registers.  Kernels have to implement sophisticated
abstractions, such as processes, file systems, file descriptors, etc.
Kernels also have to implement drivers that interact with peripheral
devices, such as disks.  Finally, kernels have significant concurrency:
execution on multiple cores; interrupts taken during user-space execution
and even during kernel execution itself, either from devices or from
timers; DMA from devices; hardware executing page-table walks concurrently with the
kernel's own accesses to the page table; and system crashes that can
stop the kernel's execution at any point and start again with a fresh
memory image after a reboot.

A promising approach for ensuring the absence of bugs in OS kernels
is to use formal verification.  Indeed, prior work has demonstrated
that this can be made to work for microkernels, hypervisors, and
security monitors~\cite{klein:sel4, gu:certikos, feiertag:psos,
zhou:verismo, li:armv9cca, chen:atmosphere, li:sekvm, ferraiuolo:komodo,
nelson:hyperkernel, sigurbjarnarson:nickel, karger:vmm,
walker:uclaunix, feiertag:multilevel, schiller:pdp11-kernel,
millen:kernel-validation, mccauley:ksos}.

The contribution of this work is a machine-checked proof of the
xv6 kernel~\cite{xv6} directly on a low-level model of the RISC-V
hardware~\cite{sail-riscv}, which includes the semantics of virtually
all hardware mechanisms needed for an OS kernel, including page tables,
interrupts, etc.  One advance over prior work lies in the amount
of concurrency that the xv6 kernel has, and that the proof therefore has
to handle: xv6 runs on multiple cores; uses spinlocks and sleeplocks;
uses flag variables in shared memory for some lock-free coordination;
runs with interrupts enabled in the kernel; implements preemptive
context switching even for kernel code; and handles concurrent DMA from
devices.  xv6 also includes a Unix file system and system call interface,
which it implements with fine-grained
locking, introducing further concurrency.

Another advance is that the proof covers
the interactions between the xv6 kernel and
the underlying hardware by using the
existing off-the-shelf semantics for the RISC-V hardware, namely the Sail
model from RISC-V International~\cite{sail-riscv}, combined with a model
of the interconnect, shared memory, devices, and power failures that
we developed.  This model gives a precise description of how a RISC-V
core executes, down to the sub-instruction level of instruction fetch,
decode, execute, and retire, along with page-table walking in hardware,
TLB caching, interrupt handling, configuration registers, execution in
machine/supervisor/user privilege levels, power failures and reboots, etc.
By basing the xv6 kernel proof on this model, the proof
covers many low-level details, including all of the boot
code, all of the assembly instructions required for context switching,
interrupt handling, user-kernel transitions, page-table switching,
TLB handling, device DMA, spinlocks, file system crash recovery, etc.

To eliminate the entire kernel build toolchain from our verification
assumptions, we verify the xv6 kernel by proving a kernel theorem about
the single ELF binary produced by running \cc{make} in the xv6 source
tree, as well as the \cc{fs.img} initial file system image produced by
that same \cc{make} command through running \cc{mkfs} on the initial
user-space binaries.  This theorem states that, in any execution of our
RISC-V hardware, with the memory contents at every boot being the bytes
from the kernel ELF image, and with the initial contents of the disk
at first boot being the \cc{fs.img} bytes, the execution will maintain
all kernel invariants.  The kernel invariants are strong enough
for this kernel theorem to imply many powerful corollaries: kernel
integrity, process isolation, crash safety, and hardware safety.
\autoref{fig:overview} shows an overview of this verification plan.

A powerful class of corollaries is application theorems that can be proven
on top of the xv6 kernel's own theorem.  For example, we prove a theorem
about the user typing \cc{echo hello world} on the UART serial console;
the theorem states that the only output the system can produce is
\cc{hello world} (if there are no errors).
This proof covers both the kernel and the user-level binaries
for \cc{init}, \cc{sh}, \cc{echo}, etc., that are present in the initial
file system image \cc{fs.img}.  This top-level application theorem
implicitly covers many aspects of kernel correctness, such as setting
up page tables, recovering the file system after crashes, handling
interrupts and DMA, etc.

A key challenge in realizing this verification plan is the level
of detail \emph{and} concurrency that the proof must cover.  To
address this challenge, this paper introduces \sys, which adopts ideas from
concurrent separation logic (CSL), and Iris~\cite{jung:iris-1,
jung:iris-jfp, krebbers:iris3.0} in particular, to reason about
sub-instruction-level execution of RISC-V code.  CSL enables the xv6
proof to introduce abstractions that allow its proofs to stay modular,
reasoning about one function, one core, or one device at a time, while
handling all of the concurrency present in xv6.

Another challenge is dealing with the tedious effort of verifying the
xv6 kernel at the level of the bytes in the compiled kernel image.  To
address this challenge, we made extensive use of AI agents
(Claude Code) to help us write the proofs and intermediate specs,
although we had to lay out the overall plan (\sys) in substantial detail,
and we were heavily involved in designing intermediate layers, abstractions,
and specs (e.g., our prompts to the agents added up to about 3$\times$ as
much text as this paper).
One pleasant benefit of using agents for proofs is that we do not have
to inspect the proofs because they are
machine-checked by Rocq~\cite{coq}, which will validate that the proofs imply
the desired top-level theorem, such as our \cc{echo} application theorem (\autoref{sec:theorem}).

Our main theorems and verification project have some limitations.  First,
our proofs cover safety properties (e.g., the kernel does not crash, and
a few user-level programs that we verified run correctly),
but do not yet cover liveness properties (e.g., every process will
keep making progress, or the user will eventually get an output on
the console), and
do not cover non-interference properties (e.g., one process cannot leak
data to another process or observe information from another process).
Second, the theorems are subject to the assumption that our trusted computing base (TCB) is correct,
which includes the Rocq proof checker and our model of the RISC-V system
(which includes the Sail RISC-V core model, as well as our model of
the interconnect with shared memory and peripheral devices).  Third,
the theorems apply to our model of shared memory: total store order (TSO) with load-load reordering.
Finally, the verification relies on the top-level theorem being meaningful:
what we prove is the precise theorem statement (\autoref{sec:theorem}), rather
than some vague ``xv6 is correct or bug-free''.

The overall proof of the top-level theorem about the xv6 kernel and the
user-level code running on top of it
comprises about \nTotalMLines~million lines of Rocq code; the xv6 kernel
itself, which we verified, is \nXvKernelLines lines of C and assembly, which
compile down to \nKernelInstrs RISC-V instructions.\footnote{We have recently
ported the entire development from Rocq to Lean, but this paper describes the
Rocq-based development; all of the results described in this paper were proven
in Rocq, prior to the Lean port.}  The proof
effort took \nDevDays days, starting from the bare idea and building
the \sys framework from scratch, all the way to a closed theorem about
the entire xv6 kernel.  In the process, we discovered 11 bugs in
xv6, including bugs that would crash the kernel, as well as a bug in
the Sail RISC-V model and a bug in Sail's Lean extraction (\autoref{sec:bugs}).
Other than fixing these bugs, and a few
other minor changes (such as disabling unsafe code intended for use by
students to debug their lab solutions that extend the xv6 kernel), the
xv6 kernel that we verified is the upstream xv6 implementation, which was
not designed or implemented with verification in mind.

The entire proof development is available at
\url{https://github.com/mit-pdos/xv6iris}, and the precise version of
the xv6 kernel covered by our theorems is available on the
\cc{verified} branch of \url{https://github.com/mit-pdos/xv6-riscv}.

\section{Main correctness theorem}
\label{sec:theorem}

The main correctness theorem that captures this entire verification
project is illustrated in \autoref{fig:traces}.  This is a theorem about how the
entire system will execute from the perspective of the externally visible
behavior.  The system includes the xv6 kernel, as well as the user-level
binaries in the initial file system image (\cc{init}, \cc{sh}, \cc{echo},
etc.).  The externally visible behavior is just the power-on and power-off
events, as well as input and output from the UART serial console.
The theorem, then, is a statement about the legal traces that a system
can produce.

Our specific theorem captures the fact that, if the user types \cc{echo
hello world} as input on the serial port, then the legal response
from the system is \cc{hello world} (or one of the possible errors,
as shown in \autoref{fig:traces}).  The theorem precisely states what
the console output will be: first, \cc{init} prints its message, then
the shell prints its \cc{\$} prompt, etc.  The theorem also states what
errors can occur (e.g., if the kernel runs out of memory, \cc{init}
might fail to start \cc{sh}, or \cc{sh} might fail to start \cc{echo}).
More generally, our theorem covers a range of shell commands that a
user could type, and supports several commands (\cc{echo}, \cc{cat},
\cc{grep}, \cc{sync}, and \cc{seccomp}), together with different ways
in which a user could combine them in the shell (e.g.,
redirecting output to a file or using pipes).
The theorem also covers files that persist across power cycles: after
\cc{echo again > f} and a reboot, \cc{cat f | grep a} must print
\cc{again}.  The theorem also covers running an unverified binary under
the \cc{seccomp} sandbox (\autoref{sec:user-app}): after the user types
\cc{seccomp xx}, the console output is unconstrained until the next power
cycle (because the sandboxed program keeps access to the console and could
print arbitrary output), but the kernel remains safe and the next boot is
again fully specified.
Finally, the theorem does not promise what happens if the user runs
other commands, such as \cc{echo xx > /sh}, although the theorem does
say that the kernel's own internal invariants continue to be maintained
(i.e., the user cannot corrupt the kernel).

\begin{figure*}[ht]
\centering
\begin{tikzpicture}[
  console/.style={draw, rounded corners=2pt, fill=black!4, align=left, text width=3.62cm,
                  inner sep=4pt, font=\footnotesize\ttfamily, anchor=north west},
  badconsole/.style={console, fill=red!4, draw=red!60!black},
  hd/.style={font=\small\bfseries, anchor=south west, inner sep=0pt},
]

\def\usr#1{\textcolor{blue!65!black}{\textbf{#1}}}
\def\pwr#1{{\normalfont\scriptsize\itshape\color{black!55}--- #1 ---}}
\def\off#1{\textcolor{orange!75!black}{\textbf{#1}}}
\def\any#1{{\normalfont\footnotesize\itshape\color{black!60}\dots~#1~\dots}}
\def\bad#1{\textcolor{red!70!black}{#1}~{\normalfont\scriptsize\color{red!70!black}\ding{55}}}

\node[hd] at (0,0.12) {(a) Expected session};
\node[console] (a) at (0,0)
  {\pwr{power on}\\
   init: starting sh\\
   \$ \usr{echo hello world}\\
   hello world\\
   \$ \usr{echo again > f}\\
   \$\\
   \pwr{power off, power on}\\
   init: starting sh\\
   \$ \usr{cat f | grep a}\\
   again\\
   \$};

\node[hd] at (4.17,0.12) {(b) Allowed variations};
\node[console] (b) at (4.17,0)
  {\pwr{power on}\\
   init: starting sh\\
   init: exec sh failed\\
   \pwr{power off, power on}\\
   init: starting sh\\
   \$ \usr{echo foo}\\
   exec echo failed\\
   \$ \usr{seccomp xx}\\
   \any{any output}\\
   \pwr{power off, power on}\\
   init: starting sh\\
   \$ \usr{echo foo}\\
   foo\\
   \$};

\node[hd] at (8.34,0.12) {(c) Disallowed output};
\node[badconsole] (c1) at (8.34,0)
  {\pwr{power on}\\
   \bad{init: bogus message}};
\node[badconsole] (c2) at ($(c1.south west)+(0,-0.3)$)
  {\pwr{power on}\\
   init: starting sh\\
   \$ \usr{echo foo}\\
   \bad{bar}};

\node[hd] at (12.51,0.12) {(d) Unspecified output};
\node[console, fill=orange!7, draw=orange!70!black] (d) at (12.51,0)
  {\pwr{power on}\\
   init: starting sh\\
   \$ \off{echo xx > /sh}\\
   \any{any output}\\
   \pwr{power off, power on}\\
   \any{any output}};
\end{tikzpicture}
\caption{Main correctness theorem that captures this verification
  project's result, illustrated through example traces that are
  allowed or disallowed by the theorem.}
\label{fig:traces}
\end{figure*}

The theorem statement is powerful, because it covers many important
properties of the system's execution.  For example, this theorem covers
all possible concurrent interleavings of multiple RISC-V cores, running
the kernel code as well as the user-space processes, with preemptive
scheduling of processes across cores.  The theorem requires that all kernel
code correctly implement system calls, interrupts, page faults, DMA for
disk reads/writes, etc.  The theorem requires that the kernel correctly
configure page tables, correctly flush the instruction cache when setting
up the code of a new user-level process on \cc{exec}, correctly insert
barriers to deal with non-sequentially-consistent shared memory, etc.
This theorem also in effect requires that the kernel have no concurrency
bugs, off-by-one errors, buffer overflows, use-after-free bugs, etc.
The theorem covers crashes, when the system might power off at any time,
and proves that the system will keep working correctly after it powers back
up.  This theorem also requires that the kernel safely execute arbitrary
user-level code: even though the initial user binaries for \cc{init},
\cc{sh}, \cc{echo}, etc., are verified, the user could also use the
\cc{seccomp} sandbox to run arbitrary unverified binaries at runtime
(e.g., by running \cc{echo xx > bin} and then running \cc{seccomp bin});
the theorem implies that doing so will not modify any files (except
through inherited file descriptors).

One limitation of this theorem is that it is a safety property: it says
what outputs are legal, but it does not require the system to produce any
particular output, or any output at all.  However, it is easy to check
that the xv6 system does actually produce output---e.g., by running it
under QEMU and typing in \cc{echo hello world}---which shows that the
theorem is not vacuously true.

\section{RISC-V execution model in \sys}
\label{sec:model}

To precisely reason about the execution of an OS kernel, such as xv6,
\sys starts with a formal model of the system's execution, as shown on
the left side of \autoref{fig:csl-view} (we will explain the right side
in the next two sections).  Specifically,
\sys uses the formal specification of the RISC-V ISA written in Sail, as
adopted by RISC-V International~\cite{sail-riscv}.  The model describes
the execution of a RISC-V core (more precisely, a hardware thread, or
HART in RISC-V terminology), in terms of its sub-instruction-level stages,
such as instruction fetch, instruction decode, execute, retire, incrementing
the clock cycle, checking pending interrupts, vectoring to the trap handler,
etc.  The model covers many detailed aspects of the CPU's execution,
including the precise semantics of all control/status registers (CSRs),
page tables with hardware walks, TLBs, writeback of A/D bits from the
TLB into the page table, interrupt handling, user/supervisor/machine
privilege levels, etc.  \sys translates the Sail model into Rocq using
Sail's Rocq backend, which generates a monadic encoding of the steps of
the Sail RISC-V HART model.  \sys uses the existing Sail RISC-V model,
except for one bug that we discovered and fixed (see \autoref{sec:bugs}).

\begin{figure*}[t]
\centering
\begin{tikzpicture}[
  font=\small,
  >={Stealth[length=4pt,width=3pt]}, semithick,
  st/.style={#1},
  cellbase/.style={draw, line width=0.4pt, inner sep=0pt, minimum height=0.36cm,
                   font=\tiny\ttfamily, text height=3.3pt, text depth=0.9pt},
  live/.style={fill=blue!14, draw=blue!55!black},
  own/.style={fill=blue!14, draw=blue!55!black},
  un/.style={fill=white, draw=blue!55!black, densely dotted, text=black!55},
  inv/.style={draw=orange!70!black, pattern=north east lines, pattern color=orange!60},
  grpbox/.style={draw, rounded corners=2pt, fill=white, inner sep=0pt},
  ghostbox/.style={grpbox, densely dashed},
  grptitle/.style={font=\scriptsize, inner sep=1pt, text height=1.4ex, text depth=0.4ex},
  hd/.style={font=\small\bfseries},
  lbl/.style={font=\scriptsize, align=center, inner sep=2pt},
  note/.style={font=\scriptsize\itshape, align=center, inner sep=2pt},
]

\def\grp#1#2#3#4#5#6#7#8#9{%
  \pgfmathsetmacro\gw{#6*(#5+0.05)+0.15}
  \pgfmathsetmacro\gh{#7*0.41+0.47}
  \node[#9, anchor=north west, minimum width=\gw cm, minimum height=\gh cm] (#1) at (#2,#3) {};
  \node[grptitle, anchor=north] at ($(#1.north)+(0,-0.04)$) {#4};
  \foreach \c/\r/\n/\s in #8 {
    \node[cellbase, minimum width=#5cm, st/.expand once=\s]
      at ($(#1.north west)+({0.1+#5/2+\c*(#5+0.05)},{-0.56-\r*0.41})$) {\n};
  }
}

\def\mem#1#2#3#4#5#6{%
  \node[grpbox, anchor=north west, minimum width=#4cm, minimum height=1.2cm] (#1) at (#2,#3) {};
  \node[grptitle, anchor=north] at ($(#1.north)+(0,-0.04)$) {shared memory};
  \foreach \i/\s in #5 {
    \node[cellbase, minimum width=0.5cm, st/.expand once=\s] (#1-\i)
      at ($(#1.north west)+({#6+\i*0.55},-0.56)$) {};
  }
}

\def\liveregs{0/0/pc/live, 1/0/ra/live, 2/0/sp/live,
              0/1/a0/live, 1/1/a4/live, 2/1/a5/live,
              0/2/s0/live, 1/2/t0/live, 2/2/gp/live,
              0/3/satp/live, 1/3/sip/live, 2/3/$\cdots$/live}
\def\ownregs{0/0/pc/own, 1/0/ra/own, 2/0/sp/un,
             0/1/a0/own, 1/1/a4/own, 2/1/a5/own,
             0/2/s0/un, 1/2/t0/un, 2/2/gp/un,
             0/3/satp/un, 1/3/sip/un, 2/3/$\cdots$/un}

\node[hd] (lhd) at (3.43,0.38) {RISC-V machine model: all of the state};

\grp{h0}{0}{0}{HART 0}{0.6}{3}{4}{\liveregs}{grpbox}
\grp{h1}{2.25}{0}{HART 1}{0.6}{3}{4}{\liveregs}{grpbox}
\node at (4.55,-1.05) {$\cdots$};
\grp{hN}{4.75}{0}{HART $N{-}1$}{0.6}{3}{4}{\liveregs}{grpbox}

\def\livemem{0/live,1/live,2/live,3/live,4/live,5/live,6/live,7/live,8/live,9/live,10/live}
\mem{lmem}{0}{-2.75}{6.85}{\livemem}{0.5}
\node at ($(lmem-10)+(0.5,0)$) {\scriptsize$\cdots$};
\node[note] at ($(lmem.south)+(0,0.2)$) {log of writes (TSO); per-HART read timestamps};

\def\liveuart{0/0/rx/live, 1/0/tx/live}
\def\livedisk{0/0/regs/live, 1/0/blocks/live}
\def\liveplic{0/0/pending/live, 1/0/enable/live}
\grp{luart}{0}{-4.55}{UARTs}{0.8}{2}{1}{\liveuart}{grpbox}
\grp{ldisk}{2.5}{-4.55}{disk + controller}{0.8}{2}{1}{\livedisk}{grpbox}
\grp{lplic}{5.0}{-4.55}{PLIC}{0.8}{2}{1}{\liveplic}{grpbox}

\foreach \h in {h0,h1,hN}
  \draw[<->] (\h.south) -- (\h.south |- lmem.north);
\draw[<->] (ldisk.north) -- node[lbl, right] {DMA} (ldisk.north |- lmem.south);
\draw[->, rounded corners=2pt] (lplic.east) -- ++(0.25,0) |- (hN.east);
\node[lbl, anchor=east] at ($(lplic.north east)+(0.25,0.3)$) {interrupts};

\begin{scope}[on background layer]
  \node[draw, rounded corners=4pt, fill=black!6, inner sep=0pt,
        fit={(-0.2,0.7) (7.25,-7.5)}] (lpanel) {};
\end{scope}

\begin{scope}[xshift=9.15cm]
\node[hd] (rhd) at (3.43,0.38) {CSL view: proof over partial state};

\grp{hk}{0}{0}{HART $k$}{0.6}{3}{4}{\ownregs}{grpbox}

\node[cellbase, own, minimum width=0.4cm] (lgo) at (2.73,-0.55) {};
\node[lbl, anchor=west, align=left] at (lgo.east) {owned by this proof};
\node[cellbase, un, minimum width=0.4cm] (lgu) at (2.73,-1.1) {};
\node[lbl, anchor=west, align=left] at (lgu.east) {not owned: invisible to this proof};
\node[cellbase, inv, minimum width=0.4cm] (lgi) at (2.73,-1.65) {};
\node[lbl, anchor=west, align=left] at (lgi.east) {shared, in an invariant};

\def\ownmem{0/un,1/own,2/own,3/own,4/un,5/un,6/inv,7/un,8/own,9/un,10/un}
\mem{rmem}{0}{-2.75}{6.85}{\ownmem}{0.5}
\node[text=black!35] at ($(rmem-10)+(0.5,0)$) {\scriptsize$\cdots$};
\draw[decorate, decoration={brace, mirror, amplitude=2pt}, thin]
  ($(rmem-1.south west)+(0,-0.04)$) -- ($(rmem-3.south east)+(0,-0.04)$)
  node[midway, below=1pt, font=\tiny] {$\ktext$};
\node[below=1pt, font=\tiny] at (rmem-6.south) {lock $l$};
\node[below=1pt, font=\tiny] at (rmem-8.south) {$a \mapsto v$};

\def\ownuart{0/0/rx/un, 1/0/tx/own}
\def\owndisk{0/0/regs/un, 1/0/blocks/un}
\def\ownplic{0/0/pending/un, 1/0/enable/un}
\grp{ruart}{0}{-4.55}{UARTs}{0.8}{2}{1}{\ownuart}{grpbox}
\grp{rdisk}{2.5}{-4.55}{disk + controller}{0.8}{2}{1}{\owndisk}{grpbox}
\grp{rplic}{5.0}{-4.55}{PLIC}{0.8}{2}{1}{\ownplic}{grpbox}

\draw[densely dashed] (-0.3,-5.72) -- (7.15,-5.72);
\node[lbl, anchor=west] at (-0.1,-5.98) {\textbf{ghost state} (logical; exists only in specs and proofs)};
\def\gfiles{0/0/$f_1$/un, 1/0/$f_2$/own, 2/0/$f_3$/un, 3/0/$\cdots$/un}
\def\gprocs{0/0/$p_1$/un, 1/0/$p_2$/un, 2/0/$p_3$/own, 3/0/$\cdots$/un}
\def\gmaps{0/0/$m_1$/un, 1/0/$m_2$/un, 2/0/$m_3$/own, 3/0/$\cdots$/un}
\grp{gf}{0}{-6.3}{files}{0.375}{4}{1}{\gfiles}{ghostbox}
\grp{gp}{2.5}{-6.3}{processes}{0.375}{4}{1}{\gprocs}{ghostbox}
\grp{gm}{5.0}{-6.3}{address spaces}{0.375}{4}{1}{\gmaps}{ghostbox}

\begin{scope}[on background layer]
  \node[draw, rounded corners=4pt, fill=black!6, inner sep=0pt,
        fit={(-0.3,0.7) (7.15,-7.5)}] (rpanel) {};
\end{scope}
\end{scope}

\draw[->] ($(lpanel.east)+(0,0.9)$) -- node[lbl, above] {split into\\owned\\resources}
          ($(rpanel.west)+(0,0.9)$);
\draw[<-] ($(lpanel.east)+(0,-0.9)$) -- node[lbl, below] {CSL\\soundness}
          ($(rpanel.west)+(0,-0.9)$);
\end{tikzpicture}
\caption{\sys's model of system execution (left, \autoref{sec:model})
  and the CSL representation of it (right, \autoref{sec:csl-intro}
  and \autoref{sec:csl-riscv}).}
\label{fig:csl-view}
\end{figure*}

The HART model itself does not specify how memory works, how multiple
HARTs interact through shared memory, how devices work, how DMA works, etc.
Instead, the HART is defined in terms of producing various \emph{events}
throughout its execution, most notably memory-read and memory-write events,
but also register-read and register-write events.  To build a model of a
complete RISC-V computer / SoC, \sys augments the Sail HART model with a model
of shared memory; a model of several devices,
including two UARTs and a DMA-capable disk controller; and a model of the PLIC
interrupt controller.  All of these components (multiple HARTs, the
devices, and the interrupt controller) execute concurrently, with
fine-grained interleaving at the level of individual events.

Interleaving at the level of individual memory reads and writes allows
\sys's model to capture fine-grained concurrency, such as a hardware
page-table walk interleaving with another core updating the page table,
or another core doing writeback of A/D bits to the same page-table entry.
Interleaving at the level of register reads and writes allows \sys to
model the asynchronous delivery of interrupts into the pending-interrupt
status register of individual HARTs, which can occur at any instant.
In particular, \sys's fine-grained interleaving captures not just
interleaving at the level of instructions, but interleaving at the level
of sub-instruction steps, such as individual memory reads during the
page-table walk, instruction fetch, data fetch, etc.

\sys's shared-memory model is TSO with load-load reordering: it captures TSO semantics for writes, which allows CPUs to
implement write buffering, and load-load reordering for reads, which allows CPUs
to do out-of-order loads (unlike x86-TSO~\cite{sewell:x86-tso}).
In particular, the memory state is a log
of write events, where each event specifies the address and value of
the write, and the source of the write (a specific core, or a specific
device issuing DMA writes).  Writing to memory appends a new event to
this log.  Reads under TSO can be stale: each core maintains a timestamp
(log position) that represents the prefix of the log that must be visible
to the core for purposes of memory reads.  Thus, when a core executes
a memory read, it picks some log position at or after that core's timestamp,
and gets the latest value written to that address as of that position.
There are two exceptions to this rule: first, cores see their own
writes (i.e., they check their own store buffer), and second, cores have
read coherence (i.e., once a read of some address returns a value from a
certain timestamp, future reads of that address do not roll back to earlier
timestamps).
Atomic instructions,
modeled in Sail as exclusive memory reads and writes, force the core
to increase its timestamp to the end of the global log, which means
the atomic instruction will access the latest state.  The RISC-V
\cc{fence} instruction names the kinds of accesses (reads and/or writes)
before and after it that must be ordered.  A write-to-read \cc{fence}
advances the core's timestamp to its own latest write
(to any address).  A read-to-read \cc{fence} advances the core's timestamp
to the highest read timestamp observed so far (at any address).

RISC-V allows instruction caches to be non-coherent with data accesses.
To represent this, our model includes a second timestamp associated
with every core; this timestamp determines what data can be returned
in response to an instruction fetch.  Each instruction fetch
non-deterministically reads at any log position at or after the core's
instruction timestamp, and does not advance the timestamp.
The \cc{fence.i} instruction bumps up the instruction timestamp to the
core's latest write (or to its data timestamp, if that is later).

\sys also models power failures and reboots.  It does so using a
logical ``thread'' of execution that invokes two operations in a loop:
power-on and power-off.  The power-on operation boots up the system,
and starts executing eight concurrent HARTs (xv6's \cc{NCPU}), representing all
of the cores.  The HARTs start running at the initial PC as specified by
the platform (\cc{0x80000000}, the start of DRAM on QEMU's \cc{virt} machine), in machine (M) mode.  The initial memory
contents at power-on are the bytes from the xv6 kernel ELF image, as
generated by running \cc{make} on the xv6 kernel's source code; there
is no difference between bytes representing code and data.  The initial
registers on each HART are set non-deterministically, except where
mandated by the RISC-V specification or by platform assumptions that
the xv6 kernel makes (e.g., specific bits in the \cc{misa} register,
or specific features supported by the HART).  When the system powers
off, all existing HARTs halt and do not execute any more instructions.
The next power-on creates new logical HARTs, representing the next era
of execution, with fresh memory, fresh initial register contents, etc.

The persistent on-disk state is logically preserved across power cycles;
the power-on operation starts the disk device with the state from the
previous power-off.  At the initial power-on, the disk contains
the byte-level image of \cc{fs.img} as generated by \cc{mkfs} as part
of running \cc{make} in the xv6 source directory, which builds all of
the user-space programs, initializes an empty file system image, and
writes the compiled user-space binaries to that image.  The disk
model captures the atomic-sector-write assumption, which file systems
traditionally rely on for crash safety (though the model allows larger
writes to be torn across sector boundaries), and models writeback caching
in the disk controller (although xv6 does not take advantage of it).

Finally, the system model produces a trace of events, which includes
the power-on and power-off events, as well as bytes sent and received
on the UART serial ports.  This trace allows stating a succinct theorem
about the externally visible aspects of the execution of the system,
such as what the user will see on the console, depending on what they
type as input to the console; one example is the \cc{echo} application theorem from
\autoref{sec:theorem}.

\section{\sys by example}
\label{sec:csl-intro}

The RISC-V model described above captures a large number of possible
interleavings.  For example, at any given moment, any core (HART) could take a
``step'': load or store a register, fetch some page-table entries from
memory, load some instruction bytes, load or store data accessed by the
instruction, update the TLB, etc.  Concurrently, devices can also ``step'':
either perform I/O such as reading or writing to the durable disk, or
perform memory reads or writes through DMA, or even raise interrupt
flags in some core's interrupt status register.  Directly reasoning
about these interleavings---such as explicitly considering all of these
different orders of events in the proof---is cumbersome.

Concurrent separation logic (CSL) makes it feasible to prove correctness
under all of the interleavings that can occur in the system,
but to do so by proving correctness of ``sequential'' executions.  That is,
the proof can step through the execution of code one instruction at a
time on a particular core, rather than having to step through all of
the system-wide interleavings across cores.

To achieve this, CSL requires reasoning about each individual core
in a different execution model that, instead of directly accessing
memory, registers, etc., explicitly tracks ownership of state (e.g.,
memory and registers) and other resources, as shown on the right side
of \autoref{fig:csl-view}.  The CSL framework then proves that the rules
for reasoning about instructions in this CSL model of the machine imply
a theorem statement about the execution of the same instructions in the
underlying highly concurrent execution model (e.g., the RISC-V machine
model from \autoref{sec:model}), illustrated by the ``CSL soundness''
arrow in \autoref{fig:csl-view}.

\begin{figure*}[t]
\centering
\begin{tikzpicture}[
  font=\small,
  >={Stealth[length=4pt,width=3pt]}, semithick,
  st/.style={#1},
  cellbase/.style={draw, line width=0.4pt, inner sep=0pt, minimum height=0.36cm,
                   minimum width=0.55cm, font=\tiny\ttfamily,
                   text height=3.3pt, text depth=0.9pt},
  own/.style={fill=blue!14, draw=blue!55!black},
  un/.style={fill=white, draw=blue!55!black, densely dotted, text=black!55},
  inv/.style={draw=orange!70!black, pattern=north east lines, pattern color=orange!60},
  pbox/.style={draw, rounded corners=3pt, fill=black!6, inner sep=0pt},
  rowlbl/.style={font=\tiny\sffamily, anchor=west, inner sep=0pt, text=black!70},
  hd/.style={font=\small\bfseries},
  lbl/.style={font=\scriptsize, align=center, inner sep=2pt},
  note/.style={font=\scriptsize\itshape, align=center, inner sep=2pt},
]

\def\panel#1#2#3#4#5#6#7{%
  \node[pbox, anchor=north west, minimum width=4.5cm, minimum height=2.9cm] (#1) at (#2,#3) {};
  \node[hd, anchor=north] at ($(#1.north)+(0,-0.05)$) {#4};
  \node[rowlbl] at ($(#1.north west)+(0.12,-1.035)$) {registers};
  \node[rowlbl] at ($(#1.north west)+(0.12,-1.8)$) {memory};
  \node[rowlbl] at ($(#1.north west)+(0.12,-2.53)$) {ghost};
  \foreach \n/\s [count=\i from 0] in #5 {
    \pgfmathsetmacro\cx{1.2+mod(\i,5)*0.6+0.275}
    \pgfmathsetmacro\cy{-0.83-floor(\i/5)*0.41}
    \node[cellbase, st/.expand once=\s] at ($(#1.north west)+(\cx,\cy)$) {\n};
  }
  \foreach \n/\s [count=\i from 0] in #6 {
    \pgfmathsetmacro\cx{1.2+\i*0.6+0.275}
    \node[cellbase, st/.expand once=\s] (#1-m\i) at ($(#1.north west)+(\cx,-1.8)$) {};
    \node[font=\tiny, inner sep=0pt, anchor=north] at ($(#1-m\i.south)+(0,-0.05)$) {\strut\n};
  }
  \foreach \n/\s [count=\i from 0] in #7 {
    \pgfmathsetmacro\cx{1.2+\i*0.6+0.275}
    \node[cellbase, st/.expand once=\s] at ($(#1.north west)+(\cx,-2.53)$) {\n};
  }
}

\def\fregs{pc/own, ra/own, sp/own, a0/own, a4/own,
           a5/own, s0/own, t0/un, satp/un, $\cdots$/un}
\def\fmem{text/own, stack/own, /un, \texttt{lk}/inv, \texttt{count}/inv}
\def\fghost{$g_1$/un, $g_2$/own, $g_3$/own, $g_4$/un, $\cdots$/un}
\panel{f}{5.95}{0}{proof of \cc{f}}{\fregs}{\fmem}{\fghost}

\node[anchor=north west, align=left, font=\scriptsize\ttfamily, inner sep=0pt] (code) at (0.3,-0.35)
  {void f(void) \{\\
   ~~acquire(\&lk);\\
   ~~count++;~~~~~~// body\\
   ~~release(\&lk);\\
   \}};
\node[lbl, anchor=north west, align=left, inner sep=0pt, text width=4.6cm] at ($(code.south west)+(0,-0.25)$)
  {The lock \cc{lk} protects \cc{count}: while the lock\\
   is free, the lock's invariant owns \cc{count}.};

\begin{scope}[xshift=11.9cm]
\node[cellbase, own] (lgo) at (0.4,-0.3) {};
\node[lbl, anchor=west, align=left] at (lgo.east) {owned by this proof};
\node[cellbase, un] (lgu) at (0.4,-0.8) {};
\node[lbl, anchor=west, align=left] at (lgu.east) {not owned: invisible to this proof};
\node[cellbase, inv] (lgi) at (0.4,-1.3) {};
\node[lbl, anchor=west, align=left] at (lgi.east) {shared, in the lock's invariant};
\end{scope}

\def\aregs{pc/own, ra/own, sp/un, a0/own, a4/own,
           a5/own, s0/un, t0/un, satp/un, $\cdots$/un}
\def\amem{text/own, stack/un, /un, \texttt{lk}/inv, \texttt{count}/inv}
\def\aghost{$g_1$/un, $g_2$/un, $g_3$/un, $g_4$/un, $\cdots$/un}
\panel{acq}{0}{-4.35}{proof of \cc{acquire}}{\aregs}{\amem}{\aghost}

\def\bregs{pc/own, ra/un, sp/own, a0/un, a4/un,
           a5/own, s0/own, t0/un, satp/un, $\cdots$/un}
\def\bmem{text/own, stack/own, /un, \texttt{lk}/inv, \texttt{count}/own}
\def\bghost{$g_1$/un, $g_2$/own, $g_3$/un, $g_4$/un, $\cdots$/un}
\panel{body}{5.95}{-4.35}{proof of body}{\bregs}{\bmem}{\bghost}

\def\rregs{pc/own, ra/own, sp/un, a0/own, a4/un,
           a5/un, s0/un, t0/un, satp/un, $\cdots$/un}
\def\rmem{text/own, stack/un, /un, \texttt{lk}/inv, \texttt{count}/own}
\def\rghost{$g_1$/un, $g_2$/un, $g_3$/un, $g_4$/un, $\cdots$/un}
\panel{rel}{11.9}{-4.35}{proof of \cc{release}}{\rregs}{\rmem}{\rghost}

\draw[->] (acq.east) -- node[lbl, above] {grants\\ownership\\of \cc{count}} (body.west);
\draw[->] (body.east) -- node[lbl, above] {passes\\\cc{count}\\along} (rel.west);
\node[note] at ($(acq.south)+(0,-0.3)$) {takes \cc{count} out of the lock's invariant};
\node[note] at ($(body.south)+(0,-0.3)$) {owns \cc{count}, so can update it};
\node[note] at ($(rel.south)+(0,-0.3)$) {returns ownership of \cc{count} to the lock};

\coordinate (join) at ($(f.south)+(0,-0.8)$);
\draw[rounded corners=2pt] (acq.north) |- (join);
\draw[rounded corners=2pt] (rel.north) |- (join);
\draw (body.north) -- (join);
\draw[->] (join) -- (f.south);
\node[lbl, anchor=west] at ($(join)+(0.1,0.4)$) {compose: chain continuations,\\frame the rest};

\draw[->, densely dashed] (f.north) -- ++(0,0.7)
  node[lbl, right, pos=0.55, align=left] {composes in turn into the proofs of \cc{f}'s callers, up to the\\whole-kernel proof, which starts out owning all of the state};
\end{tikzpicture}
\caption{A simplified view of how proofs of functions compose in CSL.}
\label{fig:csl-compose}
\end{figure*}

\autoref{fig:csl-compose} shows an example of the benefit of the CSL
approach, illustrating how CSL can be used to prove a simple function
\cc{f} that acquires a lock in order to increment a shared counter.
The proof of the overall function is built on top of proofs for its three
constituent pieces: calling \cc{acquire}, incrementing \cc{count}, and
calling \cc{release}, as if executing sequentially.
Each of these three blocks requires ownership of
some subset of resources (e.g., \cc{acquire} requires ownership of the CPU
registers touched by its implementation; the body requires ownership of
the \cc{count} variable in memory; and all of the pieces require
knowing that the kernel text instructions are present in memory).
The proof of the overall function, then, requires ownership of all of
the resources needed for the combination of those three blocks.

The key benefit of CSL in this example is that the proof of \cc{f}
ensures that \cc{f} executes correctly, even though the proof never had
to explicitly consider the possibility of another core trying to acquire
the lock while \cc{f} runs, an interrupt arriving just as \cc{acquire}
starts executing, devices performing DMA accesses to memory, etc.
The correctness of all these concurrent events is a consequence of all
of the proofs (e.g., concurrent \cc{acquire} calls, interrupt handlers,
device proofs, etc.) following the same consistent set of CSL rules,
which make it possible for a top-level adequacy theorem to compose all
of these individual proofs into an overall whole-system theorem that
covers all possible interleavings.

\subsection{Spinlock example}

To explain how \sys uses CSL to reason
about kernel execution at the RISC-V instruction level, this section will
present a simple implementation of a spinlock as an example (the
\cc{acquire} and \cc{release} functions that showed up in
\autoref{fig:csl-compose}), along with the
rules that CSL provides for reasoning about such code.  \sys in particular
builds on top of the Iris concurrent separation logic~\cite{jung:iris-1,
jung:iris-jfp, krebbers:iris3.0}, as mechanized in Rocq~\cite{coq}.
The C code of the example spinlock implementation that we will consider
is shown in \autoref{fig:spinlock-c}.
The simplified example we present in this section assumes a sequentially
consistent memory model.

\begin{figure}[ht]
\input{code/spinlock.c}
\caption{Simplified C implementation of a spinlock in the xv6 kernel.}
\label{fig:spinlock-c}
\end{figure}

All \sys specifications and proofs are built on top of the RISC-V hardware
model described in the previous section.  To specify and reason about this
C implementation, the developer first compiles the C code into machine
code; the resulting assembly for this spinlock implementation is shown in
\autoref{fig:spinlock-asm} (the \cc{fence} instruction in \cc{release}
matters only under a weak memory model, so this section ignores it).
Grounding all specifications and proofs in
machine code allows \sys to factor out the compiler, linker, assembler,
and other parts of the build process from the trusted computing base;
the theorems proven in \sys are statements about executing the machine
code on the RISC-V model, which avoids reasoning about C semantics,
compiler correctness, build flags, etc.

\begin{figure}[ht]
\input{code/spinlock.asm}
\caption{Compiled assembly of the spinlock from \autoref{fig:spinlock-c}.}
\label{fig:spinlock-asm}
\end{figure}

\subsection{Resources}

\sys specifications are formulated in terms of concurrent separation logic.
A key idea in concurrent separation logic is the notion of ownership
of resources.  For instance, a specification can represent ownership
of memory location $a$ using a resource $\memptsto{a}{v}$, which says
two things: first, memory location $a$ contains value $v$, and second,
no other thread owns this resource.  The second point implies that no
other thread can read or write memory location $a$, and therefore, the
proof can reason about its own accesses to $a$ without worrying about
concurrency.  As an analogy, ownership of $\memptsto{a}{v}$ is similar
to Rust's notion of some function owning a mutable reference to $a$.
Much as in separation logic, Rust's borrow checker guarantees that if
one function holds a mutable reference, no other function can hold a
mutable or shared reference to the same variable.

The specification for executing code, such as the spinlock \cc{acquire}
and \cc{release} functions, is stated in terms of separation-logic
implication---that is, defining what resources are required to invoke
the function.  For example, \autoref{fig:spinlock-spec} shows the
specification for the example spinlock.  The $\wand$ operator used in
this specification is separation logic's implication: it says that, given
resources on the left of the $\wand$, a proof can obtain the resources on
the right of the $\wand$.

The $\wand$ operators chain, so the specification of \cc{acquire}, for
example, requires the caller to establish all of the facts required by the
chain of $\wand$ operators to invoke \cc{acquire}.  $\pc{\cc{acquire}}$
is a requirement that the \cc{pc} register contain the address of
\cc{acquire}, which is where \cc{acquire}'s first instruction lives,
as shown in \autoref{fig:spinlock-asm}.  $\regptsto{\cc{a0}}{l}$
requires that the \cc{a0} register---the first argument in the
RISC-V calling convention---contain the address of the lock, $l$.
The next two preconditions require passing ownership of both the
\cc{a4} and \cc{a5} registers, containing any value; \cc{acquire}
does not care what the existing values are, but it will overwrite them.
The $\regptsto{\cc{ra}}{r}$ resource passes ownership of the return-address
\cc{ra} register, and says that it contains some return address $r$.

Resources can be more general than just ownership of specific registers or
memory locations.  For example, $\ktext$ is a resource representing the
kernel's text section (instructions) in memory (which happens to include
the code of \cc{acquire} and \cc{release}).  This $\ktext$ resource is
non-exclusive (``persistent''), meaning that multiple proofs (e.g., concurrent HARTs)
can own it at the same time, because it states that the kernel text is
read-only.  As a result, this resource is sufficient as a precondition
for reading memory (during instruction fetch), but is not sufficient as a
precondition for memory write instructions.  Thus, it would not be possible
to prove correctness of code that performs a store into the kernel's text.
The other two resources in this specification, $\textlog{is_lock}$ and
the parenthesized continuation, are also more than just simple memory
ownership, and we will discuss them later in this section.

\begin{figure}[ht]
\centering
\subfigure[Specification for \cc{acquire}.]{%
\begin{minipage}[t]{0.48\columnwidth}
\begin{specblockt}
\specline{\pc{\cc{acquire}} \wand} \\
\specline{\regptsto{\cc{a0}}{l} \wand} \\
\specline{(\exists~v_4.~~\regptsto{\cc{a4}}{v_4}) \wand} \\
\specline{(\exists~v_5.~~\regptsto{\cc{a5}}{v_5}) \wand} \\
\specline{\regptsto{\cc{ra}}{r} \wand} \\
\specline{\ktext \wand} \\
\specline{\islock{l}{P} \wand} \\
(~ & \pc{r} \wand \\
   & \regptsto{\cc{a0}}{l} \wand \\
   & (\exists~v_4.~~\regptsto{\cc{a4}}{v_4}) \wand \\
   & (\exists~v_5.~~\regptsto{\cc{a5}}{v_5}) \wand \\
   & \regptsto{\cc{ra}}{r} \wand \\
   & P \wand \\
   & \wpcycle ~) \wand \\
\specline{\wpcycle}
\end{specblockt}
\end{minipage}%
\label{fig:spec-acquire}}
\hfill
\subfigure[Specification for \cc{release}.]{%
\begin{minipage}[t]{0.48\columnwidth}
\begin{specblockt}
\specline{\pc{\cc{release}} \wand} \\
\specline{\regptsto{\cc{a0}}{l} \wand} \\
\specline{\regptsto{\cc{ra}}{r} \wand} \\
\specline{\ktext \wand} \\
\specline{\islock{l}{P} \wand} \\
\specline{P \wand} \\
(~ & \pc{r} \wand \\
   & \regptsto{\cc{a0}}{l} \wand \\
   & \regptsto{\cc{ra}}{r} \wand \\
   & \wpcycle ~) \wand \\
\specline{\wpcycle}
\end{specblockt}
\end{minipage}%
\label{fig:spec-release}}

\caption{Specifications for the example spinlock in \autoref{fig:spinlock-asm}.}
\label{fig:spinlock-spec}
\end{figure}

\subsection{Reasoning about machine execution}

One important aspect of \sys lies in how it uses CSL-style resources to
specify the execution of code in a system.  In a traditional Hoare-logic
style, specifications take the form of pre- and postconditions associated
with some function.  In this style, one might expect that there is a
clear pre- and postcondition associated with running \cc{acquire}.

Having explicit pre- and postconditions in a specification implicitly
requires that the execution semantics be defined in a compatible way,
in terms of executing functions, where functions eventually return to
their callers.  This is not how hardware semantics are defined: there
is no notion of functions, or returning to a caller, and there is not
even a notion of the PC necessarily advancing sequentially through code.
For instance, at any time, an interrupt might come in and redirect the
CPU's execution to the interrupt handler, rather than advancing the PC to
the next instruction.  Moreover, even if an interrupt does not come in,
the CPU will fetch the next instruction based on the state of the TLB,
page table, and the physical memory at the PC address, which might not
even be a valid DRAM address; all these are facts that the proof must
explicitly consider.

To address this, all specifications in \sys are ultimately statements
about the low-level execution of the RISC-V CPU, which is defined based on
the Sail semantics: the CPU keeps fetching instructions from the PC (through
virtual memory translation), decoding them, executing them, retiring them,
possibly vectoring to the interrupt handler, etc., in an infinite loop.
We formally capture the notion that it is safe to run the CPU in this
way using the resource $\wpcycle$.\footnote{``Safe'' here means that
the execution proceeds based on the underlying operational semantics,
maintains all invariants, and does not trigger any undefined behavior.
This is a somewhat technical statement, but it turns out to allow the \sys
proof to conclude an adequacy theorem that implies strong corollaries
about the xv6 kernel, as illustrated in \autoref{fig:overview}.} Here,
$\textlog{wp}$ stands for ``weakest precondition'', meaning that $\wpcycle$ is
a resource that is sufficient to satisfy the precondition of running
$\textlog{CpuLoop}$, which is our infinite loop of RISC-V CPU execution.

One implication of the $\wpcycle$ encoding is that there are no
postconditions, because the CPU is never done.  Instead, $\wpcycle$
says that the CPU can start running, and from that point onwards,
it can continue running safely.  To illustrate what $\wpcycle$ means,
consider a simple example where the CPU is stuck in an infinite loop:
the PC points at an instruction that jumps back to the same exact PC value,
and there is nothing that could possibly interrupt the CPU (no interrupts
enabled, no shared page tables, etc.).  In this case, a proof can directly
establish that $\wpcycle$ holds, by inductively arguing that the CPU
will forever keep executing this one instruction safely.

In the general case, \sys views the safety of the xv6 kernel in a similar
way: it effectively treats the OS kernel, along with any user-space
programs it executes, interrupts it takes, etc., as one big complicated
infinite loop.  The kernel boots up, initializes its state, and then enters
an infinite loop of running user-space processes, switching page tables,
handling system calls, handling interrupts, context-switching between
processes, etc.  Thus, proving the safety of the xv6 kernel's implementation requires
establishing $\wpcycle$ at the initial CPU starting configuration---that
is, it is safe to enter this ``infinite loop'' starting from whatever
state the CPU happens to have when it first powers up.

The remaining challenge lies in proving this ``infinite loop'' $\wpcycle$
statement for the entire kernel.  We do not want the proof to consider
this entire complex loop all at once; that would not
be feasible.  Instead, we want to decompose the proof along modular
abstraction boundaries, such as function calls and interrupt handlers:
for instance, the proof of a function like \cc{acquire} should be
an argument about the execution of that function's instructions.
What makes this decomposition tricky is that a proof about a single
function does not know why (or whether) it is going to be safe to
continue executing once the function returns.  For example, the proof of
\cc{acquire} does not know whether, after \cc{acquire} jumps back to the
return address in the \cc{ra} register, the caller's code might write
garbage into kernel memory in a way that violates the kernel's invariants.

To enable this decomposition, most specifications in \sys are formulated
in terms of \emph{continuations}.  For example, the specification of
\cc{acquire}, shown in \autoref{fig:spinlock-spec}, says that
$\wpcycle$ holds, as long as the caller satisfies the
precondition, and also supplies a proof about ``the rest of the infinite
loop'' when \cc{acquire} returns, represented by the last premise of
the specification.  That last premise says that it is the job of the
caller's proof to establish $\wpcycle$ starting from the state at which
\cc{acquire} is going to return.  In this state, the PC will be at $r$,
which is the return-address register's contents; ownership of the \cc{a0},
\cc{a4}, \cc{a5}, and \cc{ra} registers is given to the continuation;
and the continuation also receives the $P$ predicate that is the result
of acquiring the lock (we will describe this $P$ predicate shortly).

This formulation allows \cc{acquire}'s proof to establish that it will
be safe to forever execute the machine: \cc{acquire} itself maintains
safety while executing \cc{acquire}'s instructions, and the caller proved
that it will continue to be safe to execute after \cc{acquire} returns.
In this example, there is just one continuation shown, for returning
to \cc{acquire}'s caller, but in practice in xv6 proofs there are
often multiple continuations (some hidden inside of abstract predicates),
such as a continuation for proving the safety of jumping to the interrupt
handler at any point (instead of continuing on to execute the current PC),
a continuation for the safety of context-switching to another kernel
thread, etc.

\subsection{Lock specification}

Returning to the $P$ predicate, when \cc{acquire} returns, the caller
can assume that the lock has been acquired, which in CSL is represented
by gaining ownership of some predicate $P$ associated with the lock.
The $\islock{l}{P}$ premise in \cc{acquire}'s spec ties the predicate $P$
to the lock address $l$ that is being acquired.  The specification of
the spinlock allows the developer to associate any predicate $P$ with
the spinlock.  For instance, this $P$ likely contains memory points-to
facts for shared memory locations protected by this lock.

In our shared-counter example from \autoref{fig:csl-compose}, the $P$ predicate
would be $\exists~v.~~\memptsto{\cc{count}}{v}$, granting ownership of the \cc{count}
variable in memory, with some value $v$.  Owning $\memptsto{\cc{count}}{v}$ allows the body
of \cc{f} to assume exclusive ownership of that memory location; the
CSL rules mean that no other HART can be reading or writing that memory
location at the same time.  As a result, the proof of \cc{f}'s body can
proceed without having to consider the possibility of concurrent accesses
to the same memory.  In particular, \cc{count++} could be compiled down
to three non-atomic instructions (load, increment, and store) and the
proof need not worry about the possibility of another HART modifying
the memory between the load and store.

The specification for \cc{release} does the opposite operation with
$P$: it is now the caller's obligation to return ownership of $P$ to
the spinlock in order to release it.  This ensures that the predicate
$P$ holds for the next lock holder's acquisition.  In the shared-counter
example, calling \cc{release} requires the proof to return ownership
of $\exists~v.~~\memptsto{\cc{count}}{v}$.

\subsection{Instruction specifications}

To prove the correctness of the spinlock implementation, a proof in
\sys must reason about the individual instructions that make up the
implementation, chaining them together to conclude the specification
of the entire \cc{acquire} or \cc{release} function.  For example,
\autoref{fig:instr-spec} shows specifications for several individual
RISC-V instructions that appear in \cc{acquire}'s machine code.  Each of
these specifications requires the caller to prove the premises of the
spec based on the resources the caller already holds.

\begin{figure}[ht]
\centering
\subfigure[Specification for \cc{LI}.]{%
\begin{minipage}[t]{0.24\columnwidth}
\begin{specblockt}
\specline{\pc{i} \wand} \\
\specline{\instr{i}{(\cc{LI}~r,x)} \wand} \\
\specline{(\exists~v.~~\regptsto{r}{v}) \wand} \\
(~ & \pc{(i+2)} \wand \\
   & \regptsto{r}{x} \wand \\
   & \wpcycle ~) \wand \\
\specline{\wpcycle}
\end{specblockt}
\end{minipage}%
\label{fig:spec-li}}
\hfill
\subfigure[Specification for \cc{RET}.]{%
\begin{minipage}[t]{0.24\columnwidth}
\begin{specblockt}
\specline{\pc{i} \wand} \\
\specline{\instr{i}{(\cc{RET})} \wand} \\
\specline{\regptsto{\cc{ra}}{r} \wand} \\
(~ & \pc{r} \wand \\
   & \regptsto{\cc{ra}}{r} \wand \\
   & \wpcycle ~) \wand \\
\specline{\wpcycle}
\end{specblockt}
\end{minipage}%
\label{fig:spec-ret}}
\hfill
\subfigure[Specification for \cc{AMOSWAP}.]{%
\begin{minipage}[t]{0.48\columnwidth}
\begin{specblockt}
\specline{\pc{i} \wand} \\
\specline{\instr{i}{(\cc{AMOSWAP}~r_v,r_v,(r_a))} \wand} \\
\specline{\regptsto{r_a}{a} \wand} \\
\specline{\regptsto{r_v}{v} \wand} \\
\specline{(\textlog{True} \vs \exists~x.~~\memptsto{a}{x} \ast ( \memptsto{a}{v} \vs Q~v~x )) \wand} \\
(~ & \pc{(i+4)} \wand \\
   & \regptsto{r_a}{a} \wand \\
   & \regptsto{r_v}{x} \wand \\
   & Q~v~x \wand \\
   & \wpcycle ~) \wand \\
\specline{\wpcycle}
\end{specblockt}
\end{minipage}%
\label{fig:spec-amoswap}}

\caption{Specifications for individual RISC-V instructions.}
\label{fig:instr-spec}
\end{figure}

For instance, the first instruction of \cc{acquire} is a load-immediate
(\cc{LI}).  The proof of \cc{acquire} uses the specification of the \cc{LI}
instruction shown in \autoref{fig:spec-li}.  The \cc{acquire}
proof holds the resource $\pc{\cc{acquire}}$, as well as the resource
$\regptsto{\cc{a4}}{v_4}$, which it uses to satisfy \cc{LI}'s spec.
The instruction specification also requires the caller to prove that
this instruction is actually present in memory at the PC's address (the
$\textlog{instr}$ resource).  This fact is implied by $\ktext$, which
is in the precondition of \cc{acquire}.
The only remaining fact that \cc{acquire}'s
proof has to supply in order to use \cc{LI}'s specification is the
continuation.  This effectively reduces \cc{acquire}'s remaining proof
obligation to proving a $\wpcycle$ while holding $\pc{(\cc{acquire}+2)}$
as well as $\regptsto{\cc{a4}}{1}$.  This allows the proof of \cc{acquire}
to keep stepping through its instructions until it hits the return
instruction, \cc{RET}.  The specification of \cc{RET} requires proving a
$\wpcycle$ for $\pc{r}$, where $r$ is the return address in the \cc{ra}
register.  The proof of \cc{acquire} already has this $\wpcycle$,
provided by \cc{acquire}'s caller, which allows the proof to conclude.

\subsection{Framing ownership}

The specifications for individual instructions, such as \cc{LI}
and \cc{AMOSWAP}, mention ownership of specific registers and memory
locations, but say nothing about ownership of other registers and the
rest of the memory.  This is a key aspect of CSL's approach to modularity.
If consumers of these specifications (e.g., proofs that use the specs of \cc{LI}
and \cc{AMOSWAP}) own some resources that are disjoint from the ones
mentioned in the specification, those consumers can \emph{frame} their
ownership over the invocation of this spec.

For example, consider the proof of \cc{acquire}.  The proof starts out
by owning many resources, stemming from the precondition of \cc{acquire}'s
specification:

\[
  \pc{\cc{acquire}} \ast
  \regptsto{\cc{a0}}{l} \ast
  \regptsto{\cc{a4}}{v_4} \ast
  \regptsto{\cc{a5}}{v_5} \ast
  \regptsto{\cc{ra}}{r} \ast
  \ktext \ast
  \islock{l}{P}
\]

When the proof invokes the specification of \cc{LI}, which is the first
instruction of \cc{acquire}, it chooses the $i$ variable in \cc{LI}'s
spec to be the address $\cc{acquire}$, which indeed contains the \cc{LI}
instruction.  The proof then derives the $\instr{i}{(\cc{LI}~r,x)}$
fact from $\ktext$, with $i=\cc{acquire}$, $r=\cc{a4}$, and $x=1$.
The proof then passes ownership of $\instr{i}{(\cc{LI}~r,x)}$, $\pc{i}$,
and $\regptsto{\cc{a4}}{v_4}$ to \cc{LI}'s specification.  The remaining
resources held by \cc{acquire}'s proof at this point get passed to
the continuation.  The continuation's proof gets to combine these
passed resources (unchanged) with the resources returned from \cc{LI},
which have been updated.  In particular, the resources handed back by
\cc{LI}'s specification include $\pc{(i+2)}$ and $\regptsto{\cc{a4}}{1}$,
which allows the rest of the proof to proceed.  The key modularity property
here is that \cc{LI}'s specification and proof never had to mention these
other resources, or even know what kinds of other resources might exist.

Framing also enables composition at the level of function proofs.  In the
shared-counter example from \autoref{fig:csl-compose}, the proof of \cc{f}
owns the \cc{s0} register, and uses framing to preserve ownership of
\cc{s0} across its calls to \cc{acquire} and \cc{release} (which do not
require ownership of that particular resource).

\subsection{Invariants}
\label{sec:csl-intro:inv}

Memory locations that are accessed by multiple threads, such as the
spinlock's lock bit itself, cannot be owned by any one thread, because
all threads that are trying to acquire the spinlock need to be able to
access it at the same time.  In CSL, this is represented using the notion
of an invariant, which is a separation-logic resource (predicate) that
is true at all times.  In the example from \autoref{fig:csl-compose}, the
\cc{lk} and \cc{count} memory locations are owned by the lock invariant,
as indicated by the orange-striped cells; when the lock is acquired, ownership
of \cc{count} passes to the proof of the function that acquired the lock.

For example, the $\textlog{is_lock}$ resource that showed up in the
specifications of \cc{acquire} and \cc{release} is precisely such an
invariant, and its definition is shown in \autoref{fig:spinlock-inv},
where the box denotes an invariant.
The definition says that, at every instant, the lock invariant owns the
memory location $l$, and it has some value $v$.  Moreover, if $v$ is 0,
the lock invariant also owns the lock predicate $P$.  Otherwise, $v$
must be 1, and there is no $P$ in the lock invariant (it has been handed
out to the acquiring thread).  Like $\ktext$, an invariant is persistent:
any number of proofs can hold $\islock{l}{P}$ at the same
time.  This is why the specifications of \cc{acquire} and \cc{release}
do not return $\ktext$ or $\islock{l}{P}$ to their continuations; the
caller simply keeps its own copy.

\begin{figure}[ht]
\[
\islock{l}{P} \triangleq \knowInv{}{\exists~v.~~\memptsto{l}{v} \ast ( ( v = 0 \ast P ) \lor ( v=1 ) )}
\]
\caption{Invariant for the spinlock example.}
\label{fig:spinlock-inv}
\end{figure}

Interacting with the invariant requires a special kind of specification
for an instruction.  Instead of passing in full ownership of the memory
location in the precondition (which would preclude concurrent access
by multiple threads), the specification of the atomic \cc{AMOSWAP}
instruction, as shown in \autoref{fig:spec-amoswap}, allows the caller to
open the invariant for an instant during the instruction's execution.
This is encoded in \cc{AMOSWAP}'s precondition using the $\vs$
resource, as we will now explain.

The $\vs$ notation encodes a proof callback (in Iris, this object is
called a view shift, and is defined in terms of the update modality or
fancy-update modality).  Specifically, $A \vs B$
is a resource that allows the holder to ``invoke'' the callback by giving
up ownership of $A$, and receive ownership of $B$ in return; a callback that
requires no resources to invoke is written $\textlog{True} \vs B$.
The precondition of \cc{AMOSWAP} uses $\vs$
to encode the fact that, at some instant during \cc{AMOSWAP}'s execution,
the caller will be required to give ownership of $\memptsto{a}{x}$, for
some value $x$, and also give another $\vs$ resource: a proof callback
that, given $\memptsto{a}{v}$, returns ownership of some resource $Q~v~x$,
where $Q$ is specified by the caller of this specification.  Not shown
in the specification is the coupling between the two $\vs$ callbacks,
requiring them to be executed together at the same point in the proof
(invoking the outer $\vs$ obligates the invoker to also invoke the
inner $\vs$ before continuing with more execution steps).

This callback machinery allows the caller to open an invariant, such as
$\textlog{is_lock}$, to temporarily obtain ownership of resources stored
in the invariant, for an instantaneous point in time.  The outer $\vs$
corresponds to allowing the caller to open the invariant, and the inner
$\vs$ allows the caller to close the invariant back up with a new value
stored at the lock's memory address.  The $Q$ predicate allows the caller
to pass along any resources that it may have taken out of the invariant
while it had it open.

For example, in the proof of \cc{acquire}'s invocation of \cc{AMOSWAP},
the $Q~v~x$ predicate is defined as holding no resources if $x$ is 1,
and holding $P$ if $x$ is 0.  In \cc{acquire}'s use of \cc{AMOSWAP},
the $v$ argument to $Q$ does not matter.  $Q$ encodes what must be true
when \cc{AMOSWAP} finishes: if the atomically exchanged old value was 1,
then \cc{acquire} did not acquire the lock, and as a result, it got no
resources.  But, if \cc{AMOSWAP} observed that the old value was 0
(and it was replaced by 1), then the caller acquired the lock, and
$Q$ contains $P$.  Specifically, once the outer
$\vs$ callback opens the lock invariant and gives up $\memptsto{a}{x}$,
and the inner callback receives $\memptsto{a}{v}$ as the new contents of
the memory location, the proof can consider two possible cases: either
$x$ was 0 or it was 1.  If $x$ was 1, this
means the lock was already held, and $Q$ is empty.  On the other hand,
if $x$ was 0, then the lock invariant previously held $P$, but now does
not need to hold $P$, because the location is now 1.  This allows the
callback to keep $P$ in $Q$ and reclose the invariant without $P$.

\subsection{Discussion}

An advantage of \sys is that it reasons about the execution of code
based on precise semantics for RISC-V hardware.  This is important
for an OS kernel, because the OS kernel is responsible for setting up
the hardware mechanisms needed to ensure its own correctness, such
as configuring page tables, managing the TLB, setting configuration
registers, setting up interrupt trap handlers, enabling and disabling
interrupts, etc.  \sys makes reasoning about all of these operations
explicit and consistent with reasoning about all other parts of the
kernel's implementation.  This allows a \sys proof to make relatively
few assumptions about the kernel's correctness: we do not need to assume
that memory accesses are properly synchronized, that interrupts are
properly handled, that page tables are correctly configured, that the
kernel correctly flushes the TLB when switching between user and kernel
page tables, etc.

One cost of \sys is that it requires the proof to be explicit about the
execution of each instruction, and to consider many details.  This level of
tedium would be prohibitive if the proofs (and even the intermediate specs
for each function inside xv6) had to be written by hand.  Using agents
makes it possible to automate large parts of the proof effort required
to verify xv6 in \sys.

A second cost of \sys is that the proof is tied specifically to the
precise binary image of the kernel.  As a result, small changes to the
kernel's source code can lead to pervasive changes throughout the kernel
image (such as shifting relative offsets or instruction alignments),
thereby requiring many proofs to be fixed.  This again is a tedious
task that would be impractical to do by hand at each kernel recompile,
yet we find that agents are able to perform these tedious proof updates
without substantial developer input.

Although \sys is based on low-level RISC-V semantics, higher-level proofs
need not reason directly about the precise bit-level representation
of all system state.  Instead, developers introduce abstractions and
invariants that connect the low-level machine state, such as registers
and memory, to higher-level abstract states and predicates.  For example,
the xv6 proof introduces abstractions such as a logical mapping from
virtual to physical addresses (abstracting away
the 3-level page-table tree structure and the TLB); a file on disk containing
some sequence of bytes; a runnable process with a PID and name; etc.
Furthermore, CSL allows splitting up ownership of these abstractions
at the level of abstract state, rather than physical state.  That is,
a proof of some function, such as \cc{sys_write}, can own the bytes of
a file, regardless of whether those bytes are currently stored in the
buffer cache, in the write-ahead log, or installed in the file inode's
actual data blocks.  This is implemented by introducing \emph{ghost
state}, which is logical state that appears only in the specs and proofs
(as shown in \autoref{fig:csl-view} and \autoref{fig:csl-compose}).
Ghost state can be connected to physical state through explicit predicates
and invariants, and ghost state can then be owned by different threads
much like physical resources such as registers and memory locations.

Finally, \sys inherits the key benefit of CSL-style local reasoning.
In our example spinlock specification and proof, CSL allows the caller to
reason about its own execution as if no other threads could interfere:
the caller gains complete ownership of the locked state,
as described by $P$, and the proof implicitly promises that no other CPUs
can modify memory in a way that would invalidate $P$, that no concurrent
device DMA could occur that would overwrite $P$'s memory, and that no
interrupts can happen that could violate $P$.  This allows the formal proof
to avoid explicit reasoning about all of these possible interleavings.
At the same time, the theorem proven using CSL does establish the fact
that the kernel is correct for all of these possible interleavings.

\section{CSL for RISC-V}
\label{sec:csl-riscv}

This section explains how \sys provides reasoning principles for the
Sail RISC-V model using CSL.

\subsection{Ownership of state}

\paragraph{CPU state.}

\sys represents the ownership of every CPU register using a points-to
fact, $\regptsto{r}{v}$.  Each HART has its own set of registers, so
the points-to resource is indexed by the HART ID (we omit this index
for simplicity in this paper's presentation).  The register set includes
the 32 general-purpose RISC-V registers, which showed up in the previous
section's spinlock example, as well as many control and status registers,
which the Sail model folds into the notion of a ``register''.
For example, the TLB is modeled as a separate register in the Sail
model, storing all of the TLB entries; \sys represents it as ownership of
$\regptsto{\cc{tlb}}{\textit{tlb-entries}}$.  The RISC-V PMA and PMP tables
are also represented by registers.  The \cc{satp} and \cc{stvec} registers, storing
the page-table root pointer and the trap vector, are similarly represented.

Representing ownership of each register using a separate points-to resource
provides two advantages.  First, it allows \sys to accurately model
registers that are concurrently modified by other parts of the system,
such as the interrupt controller asynchronously and non-deterministically
deciding to set the interrupt-pending bit.  This ensures that we
correctly model how the CPU handles interrupts, down to the precise
point within the execution of a single instruction at which it checks
for pending interrupts.

Second, separate ownership of each register allows the xv6 proof to
distribute each register to the subsystem responsible for managing it.
For instance, the xv6 proof defines one predicate that controls virtual
address translation; that predicate, and all of the proof machinery around
it, owns the \cc{satp} and \cc{tlb} registers.  The proof also defines
another predicate in charge of interrupt handling; that predicate owns
the \cc{stvec} register, the interrupt-enable bit of the \cc{sstatus}
register (owned through ghost state that subdivides the \cc{sstatus}
register into individual ghost bits that can be owned by different
predicates), and sufficient memory locations to execute the interrupt
handler, as well as other resources.  Finally, the proof also has a
separate invariant that owns the cycle-counter/timer register, and
is used to enable timer interrupts.  Distributing ownership of these
registers among different predicates and invariants avoids the need to
explicitly consider them in the rest of the proof.  For example, the
rest of the xv6 proof largely ignores the fact that each instruction
involves one or two virtual memory translations for instruction fetch,
and perhaps more translations for data loads and stores.  Similarly,
most of the proofs are independent of the fact that an interrupt can
occur at any time (when interrupts are enabled).

In addition to full ownership of resources, CSL has a notion of a
resource being persistent, which means that any number of threads can
hold it at the same time.  This is useful for representing read-only
state that is accessed by multiple threads.  For example, when the system
is booting up, it initializes certain configuration registers, such as
RISC-V's \cc{medeleg} and \cc{senvcfg} registers, and they are never
modified again.  Thus, after writing these registers, the kernel proof
transforms its exclusive ownership of the register into a persistent fact.
This simplifies reasoning about these registers, since many parts of the
kernel proof can now own this persistent fact stating the register's value.

\paragraph{Shared memory.}

To represent ownership of shared memory under TSO with
load-load reordering, \sys provides
an $\tsoptsto{a}{H}$ resource, which states that the timestamped history
of writes to address $a$ is exactly $H$.  \sys also provides a per-CPU
resource representing the CPU's TSO timestamp, $\tsots{t}$, which can
grow monotonically but never decrease.  The combination of these two
resources allows reasoning about the effects of reads and writes: a write
extends the history, while a read returns some value from the history
subject to the CPU's timestamp (for writes from other cores), or this
core's own latest write.

Much of the xv6 kernel uses spinlocks to protect shared memory, which
allows reasoning about memory as if it were sequentially consistent.
To model this regime,
\sys provides a points-to fact to represent ownership of a consistent
physical memory location, $\memptsto{\textit{pa}}{v}$.  Underneath,
this is defined as $\tsoptsto{\textit{pa}}{H}$ together with an assertion
that the local core's timestamp forces reads to observe the latest write.

The xv6 kernel runs with paging enabled, which
means that instructions and data are accessed at virtual addresses, rather
than physical addresses.  The xv6 proof represents this using a
$\vmemptsto{\textit{va}}{v}$ resource, which combines ownership of some (existentially quantified)
physical memory address $\textit{pa}$, $\memptsto{\textit{pa}}{v}$, together with a fact about
the current virtual memory configuration, saying that virtual address
$\textit{va}$ currently translates to physical address $\textit{pa}$.  This translation
statement is represented using ownership of ghost state: the proof
stores the current virtual memory mapping in a ghost variable, and each
translation is a fact about a single element in this ghost map.  The page-table
predicate, mentioned above, connects the ownership of \cc{satp}
and \cc{tlb}, as well as the physical pages comprising the 3-level page
table, to the logical view established by that page table and TLB.

The xv6 kernel does not modify its own code.  As a result, the xv6
proof makes all of the memory points-to facts for kernel code
persistent.  This allows freely sharing them between all proofs,
so there does not need to be exclusive ownership of instruction memory.
This also ensures that nothing in the kernel can modify the instructions
themselves (since doing so would require the proof to obtain exclusive
ownership of instruction memory).

\paragraph{Devices.}

RISC-V exposes devices by mapping their control registers in a separate
part of the physical address space (MMIO).  Reads and writes to these
physical addresses do not work through memory points-to facts, because
these registers do not behave like memory.  \sys introduces explicit
state for each device, in its device model, together with resources
defining ownership of this device state.  Device state is modified both
by the device itself (e.g., a UART receiving and sending bytes in the
background, or the disk controller executing DMA requests to read/write
disk blocks) and in response to reads/writes to the device's
memory-mapped control registers.

This device state in the xv6 proof is owned by an invariant, one for each
device (UART, PLIC, and virtio disk in xv6, plus one for the
external-interrupt pins that the PLIC drives into each HART).  \sys requires the developer to
prove that this invariant is maintained by the device's own operation,
as well as by MMIO accesses from kernel driver code.  To this end, the
developer must prove that the logical device thread (representing the
behavior of the device according to the \sys model) correctly executes
against this invariant, in a style similar to what we described in
\autoref{sec:csl-intro} for kernel software itself.  The proof for the
kernel device driver follows exactly the same steps as any other kernel
code that uses invariants.  For example, in the case of the DMA-capable
disk controller, the invariant associated with the disk owns both the
device state and the physical memory used for the command queue,
buffers of pending block reads/writes, etc.  Ownership of command slots
in the ring buffer is handed off from kernel to device and back, through
the invariant, as a result of command submission and completion.  Proving
that the device upholds the invariant ensures that the kernel's driver
code can safely reason about its interactions with the device through
the abstraction of a CSL invariant over the device state, even in the
presence of concurrent DMA operations to shared memory, interrupts, etc.

Load and store operations to device MMIO addresses follow a different
specification from memory loads and stores.  \sys proves separate
specifications for every MMIO address (e.g., one specification for
accessing the UART status register, another specification for storing
into the UART TX FIFO, etc.).  The specification uses $\vs$ to open the
device invariant and describe what effect that particular MMIO access
has on the device state.

\subsection{Execution}

At the whole-instruction level, execution is captured in \sys by the
$\wpcycle$ resource, representing the fact that the HART can safely run
$\textlog{CpuLoop}$, the infinite loop of RISC-V \emph{cycles}; each cycle
of the Sail model fetches and executes one instruction, or takes an
interrupt or trap.  The $\textlog{CpuLoop}$ operation is indexed by a HART
ID, which determines the set of registers that are used in executing
each cycle; much like register points-to facts, we usually do not explicitly
show this HART ID index of $\textlog{CpuLoop}$, although it is important when
reasoning about preemption of kernel threads and the scheduler resuming
their execution on a different HART (\autoref{fig:swtch-spec} shows it).

To reason precisely about the sub-instruction-level execution of a RISC-V
HART based on the Sail model, \sys defines $\wpcycle$ as a statement
about executing the individual steps of the Sail RISC-V HART model.
Specifically, \sys has a lower-level notion of $\wpsail{m}{\Phi}$, where
$m$ is the functional description of how the Sail RISC-V HART model
executes, represented using a monadic encoding in Rocq.  Here, $\Phi$
is the explicit postcondition that will be true after $m$ executes.
One notable difference between $\wpcycle$ and $\wpsail{m}{\Phi}$
is that the control flow is now made explicit, and there is also an
explicit postcondition, in contrast to the continuation-based $\wpcycle$.
The reason is that, in our RISC-V operational semantics, the steps of
a single cycle are statically defined by Sail's model: $m$ is a fixed
monadic program that terminates, so $\Phi$ can describe its result.
In contrast, which instruction the next cycle executes is determined by
dynamic state, such as page tables, memory contents, interrupts, etc.,
which is why instruction-level specifications take a $\wpcycle$
continuation instead.  \sys defines
$\textlog{CpuLoop} \triangleq \textlog{forever} \left\{\textit{cycleModel}\right\}$,
where \textit{cycleModel} is the entire definition of the Sail model for
a single cycle of RISC-V execution.
$\wpcycle$ has no postcondition because the processor executes cycles
indefinitely, until powered off; all of the facts at the end of each
instruction's execution are represented using CSL resources and passed
into the proof of the next instruction's execution.

\sys uses standard sequencing combinators to reason about the steps of
the Sail model.  In particular, proving $\wpsail{a; b}{\Phi}$ reduces to proving
$\wpsail{a}{\wpsail{b}{\Phi}}$, meaning that, to prove that $\Phi$
holds after running $a; b$, it suffices to prove that, after running
$a$, there is a $\wpsail{b}{\Phi}$ resource that allows running $b$
to establish $\Phi$.  A similar combinator works for sequencing with
return values (i.e., $a$ could return some value that $b$ depends on,
and in that case, $\Phi$ is a function of that return value).

This encoding allows \sys to state individual specifications for
sub-instruction stages of execution.  For example, in the Sail model
as extracted to Rocq, there is a separate function used to perform
address translation, \cc{translateAddr}.  The xv6 proof establishes
$\wpsail{\cc{translateAddr}}{\Phi}$, with a $\Phi$ that specifies how
address translation works (i.e., address translation
through a 3-level page table produces a result consistent with the
abstract-state view of that page table according to the xv6 kernel's
page-table invariant).  Similarly, \sys proves sub-instruction
specifications for the execute stages of different instructions, such
as ALU operations, bitwise operations, etc.  \sys also proves specs for
all other stages of the Sail model, including instruction fetch, decode,
trap handling, computing the next PC, etc.

Each individual $\wpcycle$ specification, such as the ones shown in
\autoref{fig:instr-spec}, is then proven on top of these lower-level
sub-instruction stage specifications.  For instance, the spec for
\cc{AMOSWAP} combines a \cc{translateAddr} spec for doing a virtual
memory lookup of the program counter, together with an instruction fetch
from physical memory, decode, another \cc{translateAddr} to look up
the physical address for the virtual address being swapped, then some
register operations, then an exclusive memory write (to ensure atomic
read-modify-write semantics) to perform the actual swap, etc.  This ensures that the
instruction-level specs at the $\wpcycle$ level are completely faithful
to the Sail model at the sub-instruction level.  This includes both
fine-grained concurrency between HARTs and devices, at the individual
memory event level, and fine-grained power failures, which can
occur between any two sub-instruction steps.

\paragraph{Non-HART execution.}

As mentioned above, devices are also represented by a logical thread
of execution in \sys's semantics, and the developer is responsible for
proving a WP-style statement for the execution of the device hardware
model against the developer's stated invariant over the device state.
The device WP is similar to the $\wpcycle$ and $\wpsail{m}{\Phi}$
specifications we have already described, except that the low-level
operations represent the steps that a device can take.  Finally, there
is a global power-cycling thread, also with its own set of operations
(power-on and power-off); the developer also proves a WP for this
power-cycling thread, which requires the developer to prove that any
preconditions of starting execution of the kernel code on each HART are
satisfied when that HART is started by power-on.

\subsection{Adequacy}

CSL specifications can be complex and can have complex implications.
For example, if a specification inadvertently requires ownership of
the same object twice in its precondition, that precondition can never be
satisfied by any caller under CSL's rules, and the specification is
vacuously true.
Similarly, if a CSL specification promises the same object twice in its
postcondition, it is unsatisfiable.  The presence of proof callbacks
($\vs$) further complicates the meaning of a specification.

To ensure that \sys specifications are meaningful, \sys provides an
adequacy theorem, which turns
a CSL-style specification about WP execution of the kernel into a
lower-level statement purely about the execution of the system model
(i.e., Sail's model plus \sys's shared-memory and device models),
with no separation logic, WPs, etc.  The adequacy theorem states that,
for any execution of the system according to the operational semantics
of our system model, as described in \autoref{sec:model}, the state
of the system at each step will satisfy the invariants established in
the proof.  For instance, one implication of this adequacy theorem is
that the durable disk state will remain consistent at all times.

One important job of the adequacy theorem is to ensure that all invariants
used in the CSL proofs can be instantiated.  In the absence of an
adequacy theorem, it is easy for developers to assume some invariant
holds, and prove functions with respect to that invariant, while not
realizing that this invariant cannot be instantiated, and in fact the
existence of the invariant might subtly imply false.  To rule this out,
the adequacy theorem must explicitly initialize every invariant.

The adequacy theorem also connects the CSL specifications and proofs
to concrete states over which the system operates.  For example, the
file system in xv6 is proven correct using an invariant that holds over
the contents of the durable disk state.  This invariant states that
the block free bitmap does not mark as free any blocks that are currently
in use by some file, that the inode link counts are consistent, etc.
The adequacy theorem provides a more concrete statement, about the system
starting with the literal contents of the bytes from \cc{fs.img} on disk.
To this end, the adequacy proof shows that these initial bytes produced
by \cc{mkfs} satisfy the file system's invariant, bootstrapping the
inductive reasoning across arbitrarily many power cycles.

\section{Proving xv6 in \sys}
\label{sec:proof}

This section describes the abstractions and specifications used in the
\sys-based proof of xv6, as well as particularly interesting aspects of
the proof itself.

\subsection{Interrupt handling}

The xv6 kernel enables interrupts during parts of its supervisor-mode
execution.  In particular, the kernel enables interrupts while executing
system calls, so that long-running system calls can be preempted by a
timer interrupt, and so that device interrupts can be serviced while
executing long-running system calls.  Similarly, the kernel scheduler
on each core (HART) periodically enables interrupts while scanning for
runnable processes, to check for timer interrupts or device interrupts
that might be needed to wake up a process that might be waiting for UART
input or disk I/O.

Interrupts introduce a tricky source of concurrency, because when the
kernel is executing with interrupts enabled, it can be preempted by
an interrupt at any time.  The interrupt handler executes on the same
kernel stack as the interrupted thread.  Furthermore, the interrupt
handler could put the thread to sleep (e.g., calling \cc{yield} in the
timer interrupt to schedule another process), and the same process could
be resumed on a different HART instead.

The xv6 proof addresses these challenges by introducing a resource
responsible for handling interrupts and CPU migration.  In particular, the
xv6 proof defines a resource $\textlog{sie_cap_gpr}$, which exposes several
arguments.
One of these arguments is $b$, a boolean flag indicating whether
interrupts are currently enabled.  Another argument is $n$, the number
of free stack locations on the current stack.  Another argument is $m$,
a map giving the values of the general-purpose registers on the current HART.

Every developer-visible instruction specification in \sys takes
$\textlog{sie_cap_gpr}$ as a precondition and returns it to the continuation.
This allows the proof of every instruction to do an inductive loop over
an arbitrary number of interrupts that could arrive at that instruction
boundary (corresponding to $\wpsail{\cc{dispatchInterrupt}}{\Phi}$,
where \cc{dispatchInterrupt} is the Sail functional model of checking
for pending interrupts and jumping to the trap handler).

\paragraph{Register migration.}
Instruction specifications use the $m$ argument of $\textlog{sie_cap_gpr}$
to refer to the values of general-purpose registers, rather than the
raw $\regptsto{r}{v}$ resource shown earlier.  The reason is that the
raw register points-to resource is tied to that register on a specific
HART ID.  If the proof of some function owned this resource, but was
preempted by an interrupt and migrated to another HART, it would not be
able to meaningfully use that resource on the new HART ID, because the
ID of the HART on which the code is executing might differ from the ID
of the register points-to fact.  When the interrupt handler resumes a
thread on a different HART, it ensures that all of the saved registers
are restored, and therefore the preempted thread resumes execution with
the same exact $m$.

\paragraph{Stack reservation.}
The number of available stack locations $n$ takes into account the total
stack size, and the number of stack locations that need to be available
to execute the trap handler.  When the kernel executes an instruction
to disable interrupts (writing to the \cc{sstatus} register), the $b$
argument of $\textlog{sie_cap_gpr}$ flips to false, and the number of available
stack locations increases by the trap handler's reserve amount.  When the
kernel re-enables interrupts, the precondition of the instruction spec
that writes this bit to \cc{sstatus} requires the caller to prove that
the current $n$ stack budget is large enough to accommodate the trap
handler's stack use, and subtracts that stack use from $n$ (reserving it
for the trap handler).

\paragraph{Trap handler address.}
The $\textlog{sie_cap_gpr}$ resource owns the \cc{stvec} register, and
existentially quantifies over the trap handler's address.  In particular,
$\textlog{sie_cap_gpr}$ states $\exists~h.~~\regptsto{\cc{stvec}}{h} \ast
(\pc{h} \wand \ldots \wand \wpcycle)$.  This says that \cc{stvec} has
some address $h$ such that, if the PC were to jump to $h$, along with some
other conditions, it would be safe to keep executing.

When interrupts are disabled, $\textlog{sie_cap_gpr}$ gives up ownership of
\cc{stvec}, allowing the rest of the kernel code to manipulate \cc{stvec}.
This is important when the kernel jumps to executing user-mode code.
In xv6, there is a different trap handler for user-mode traps, which,
among other things, implements demand allocation of zero pages for
process memory through page faults, and executes system calls in response
to an \cc{ECALL} instruction.  Once the kernel disables interrupts,
the proof gets back ownership of \cc{stvec}, allowing it to install
a different handler address by writing to that register, and use that
to establish a different interrupt-safety invariant that holds while
executing instructions in user mode.

\subsection{Page table and TLB}

To factor out virtual memory translation from the rest of the kernel
proof, the xv6 proof introduces a resource that owns the page-table
machinery.
In particular, $\textlog{strans_inv}$ owns the \cc{satp} register, which points
to the root of the page table, as well as the physical pages comprising
the 3-level page-table structure, and the Sail model's \cc{tlb} state.  The proof
also maintains a map in ghost state that represents the logical mapping
implemented by the kernel's page table.  The $\textlog{strans_inv}$ resource
states that the TLB must be consistent with the mappings in the page table.

\paragraph{A/D writeback.}
One complication is that the page-table hardware can write back accessed
(A) and dirty (D) bits to page-table entries, if those pages were in fact
accessed or modified by some HART.  At the same time, the kernel uses a
single page table shared across all HARTs, so the A/D writeback on one
HART can modify the page table that another HART is walking.  To prove
the correctness of this, the xv6 proof places ownership of the physical
page-table pages into an invariant.  For every instruction, the execution
of the page-walk and TLB phases of the Sail model must be able to open the
invariant and modify the physical page table in memory, updating its
A and D bits.  To allow this, $\textlog{strans_inv}$ existentially quantifies
over A and D bits for every page, allowing arbitrary A/D bit updates.

\paragraph{Translation tiers.}
In xv6, the kernel can operate in three different address-translation
regimes.  When the system first boots up, every HART is running in
machine (M) mode, which by definition does no address translation.
The xv6 initialization code then switches to supervisor (S) mode, but
running without a page table (``Bare'' mode in RISC-V terminology), which
appears as an identity mapping between virtual and physical addresses.
Later, xv6 allocates a page table, containing both identity and
non-identity mappings (a kernel stack for every kernel thread,
and a trampoline page containing code that switches between kernel and
user page tables, at the top of the virtual address space).  Once the
kernel installs this page table, it can start accessing non-identity
virtual addresses, such as per-kernel-thread stacks, and the trampoline
to switch to user mode.

One complication of this bootup sequence is that, while one HART may
have already initialized and switched to using the kernel page table
(where non-identity mappings are present), other HARTs might still be
executing early in the boot sequence, still in the identity-mapping regime
without the non-identity mappings.  As a result, a kernel virtual memory
points-to fact ($\vmemptsto{\textit{va}}{v}$) that is valid for accessing memory
on one HART is not necessarily valid for accessing memory on another HART.

To define sound rules for accessing memory through a
$\vmemptsto{\textit{va}}{v}$ resource, the xv6 proof introduces the notion of
a \emph{kernel translation tier}, which takes advantage of the fact that
kernel mappings grow monotonically in xv6.  Tier 0 represents identity
mappings that are valid in all three regimes (M mode, Bare mode, and
with the kernel page table).
Tier 1 represents mappings that are present in the kernel page table.
Tier 0 mappings are a subset of tier 1 mappings.  Every kernel virtual memory
points-to fact is indexed by a tier (0 or 1), even though we have
not explicitly shown this tier index in this paper's presentation.
The state of the current HART also keeps track of the tier that this
HART has reached; this is an argument \textit{kt} of $\textlog{sie_cap_gpr}$.
The specifications for memory load and store instructions require that
the tier of the memory points-to fact be less than or equal to the
tier of $\textlog{sie_cap_gpr}$.

\paragraph{Switching user and kernel page tables.}

When xv6 switches to running user code, it installs the user process
page table, which does not map the kernel's code or data.  The user
page table contains the user's own memory, plus two special pages: a
trampoline page, containing kernel code to jump in and out of the kernel
(which is responsible for switching page tables), and a trapframe page,
which is used to save the user's registers when
trapping into the kernel.

To switch between user and kernel page tables, the trampoline assembly
code executes a conservative but safe sequence: \cc{sfence.vma} to flush
the TLB, then load the \cc{satp} register with the new page-table root,
then \cc{sfence.vma} to flush the TLB again.  One complication is that,
while the HART is executing between the two TLB flushes, the TLB might
contain mappings from \emph{either} of the two page tables.  To pin down
the precise reasoning for why this sequence is safe, the xv6 proof states
that, in this page-table-switching window, the TLB must contain only
the mappings that are present in both the user and kernel page tables
(namely, the trampoline code page).

\subsection{Context switching}

The kernel uses \cc{swtch} to implement context switching between
kernel threads.  In particular, when a kernel thread decides to
\cc{yield} or go to sleep, it uses \cc{swtch} to switch to executing
a per-HART scheduler thread.  The scheduler thread, in turn, loops over
all processes, and when it finds a \cc{RUNNABLE} process, it uses \cc{swtch}
to switch to executing that kernel thread.  Timer interrupts can preempt
execution of a process, either in user mode or in kernel mode, and force
that process to give up the CPU by calling \cc{yield}.  The function
signature of \cc{swtch} is shown in \autoref{fig:swtch-h} in C syntax,
although the actual implementation of \cc{swtch} in xv6 is in assembly.

\begin{figure}[ht]
\input{code/swtch.h}
\caption{Signature of the \cc{swtch} context-switching function.}
\label{fig:swtch-h}
\end{figure}

The specification of \cc{swtch} needs to capture the fact that a scheduler
can resume executing a kernel thread by using \cc{swtch}, regardless of
what the kernel thread was doing at the time it was preempted or decided to
sleep on its own.  The xv6 proof captures this notion in the specification
of \cc{swtch} by storing a $\wpcycle$ inside the resource that defines
a valid \cc{struct context}.  $\textlog{own_context}~\textit{ptr}~c$ says
that the \cc{struct context} at address $\textit{ptr}$ stores the register
values represented by tuple $c$.  $\textlog{valid_context}~\textit{ptr}$
says that there is some tuple $c$ of registers saved at $\textit{ptr}$,
and, if any HART $h$ restores the callee-saved registers (\cc{sp} and
\cc{s0}--\cc{s11}) from $c$, that HART can
resume execution at $c.\textit{RA}$.  The HART ID $h$ is universally
quantified, because the HART that resumes the kernel thread need not
be the one that saved its context; the resulting $\wpcycleh{h}$ states
that it is HART $h$ that continues executing the thread.

\begin{figure}[ht]
\centering
\begin{minipage}[t]{0.46\columnwidth}
\begin{specblockt}
\specline{\textlog{own_context}~\textit{ptr}~c \triangleq} \\
\quad
  & \vmemptsto{\textit{ptr}+0}{c.\textit{RA}} \ast \\
  & \vmemptsto{\textit{ptr}+8}{c.\textit{SP}} \ast \\
  & \vmemptsto{\textit{ptr}+16}{c.\textit{S0}} \ast \\
  & \ldots \\
  & \vmemptsto{\textit{ptr}+104}{c.\textit{S11}} \\
& \\

\specline{\textlog{valid_context}~\textit{ptr} \triangleq} \\
  & \exists~c.~~\textlog{own_context}~\textit{ptr}~c \ast \\
  & \forall~h~m.~~\textlog{sie_cap_gpr}~m~\ldots \wand \\
  & \phantom{\forall~h~m.~~} \textlog{callee_saved}~m = (c.\textit{SP}, c.\textit{S0}, \ldots, c.\textit{S11}) \wand \\
  & \phantom{\forall~h~m.~~} \pc{c.\textit{RA}} \wand \\
  & \phantom{\forall~h~m.~~} \wpcycleh{h}
\end{specblockt}
\end{minipage}%
\hfill
\begin{minipage}[t]{0.50\columnwidth}
\begin{specblockt}
\specline{\ktext \wand} \\
\specline{\pc{\cc{swtch}} \wand} \\
\specline{\textlog{sie_cap_gpr}~m~\ldots \wand} \\
\specline{(\exists~c.~~\textlog{own_context}~m[\cc{a0}]~c) \wand} \\
\specline{\textlog{valid_context}~m[\cc{a1}] \wand} \\
(~\forall~h'~m'.~~
   & \pc{m[\cc{ra}]} \wand \\
   & \textlog{sie_cap_gpr}~m'~\ldots \wand \\
   & \textlog{callee_saved}~m = \textlog{callee_saved}~m' \wand \\
   & (\exists~c.~~\textlog{own_context}~m[\cc{a0}]~c) \wand \\
   & \textlog{valid_context}~m[\cc{a1}] \wand \\
   & \wpcycleh{h'} ~) \wand \\
\specline{\wpcycleh{h}}
\end{specblockt}
\end{minipage}

\caption{Simplified specification for \cc{swtch}.}
\label{fig:swtch-spec}
\end{figure}

The spec for \cc{swtch} itself (\autoref{fig:swtch-spec}) says that, as long as the caller
can prove that it owns the memory locations for storing the old context
(first argument, passed in \cc{a0}), and there is a valid context at
the new pointer (second argument, passed in \cc{a1}), and the caller
can prove that it is safe to resume execution at the return address with
the callee-saved registers restored, then the caller can provably invoke \cc{swtch}.
The spec makes the HART ID argument of $\textlog{CpuLoop}$ explicit
(\autoref{sec:csl-riscv} otherwise omits it): \cc{swtch} is invoked on
HART $h$, but the continuation must hold for any HART $h'$, since
\cc{swtch} might return on another HART after the scheduler there
resumes the thread.  Not shown in this spec is that, if the caller knows the scheduler will never resume this
process again (e.g., a killed process exiting), the caller's proof need
not supply this continuation WP.

\subsection{Locks}

The xv6 kernel uses spinlocks to coordinate access to shared data
structures between concurrent threads.  \autoref{sec:csl-intro} showed
a simplified version of the xv6 spinlock, along with basic CSL-style
specifications for the spinlock.  This subsection describes several
complications of how xv6 implements and uses spinlocks, and how the
proof is designed to address them.

\paragraph{Disabling interrupts for spinlocks.}

xv6 couples spinlocks with interrupt handling, because it is important that
a CPU holding a spinlock not be preempted by an interrupt; otherwise, the
kernel could deadlock.  In the spinlock implementation, this is done
using two functions, \cc{push_off} and \cc{pop_off}.  \cc{push_off}
increments a per-CPU interrupt-disable counter, and \cc{pop_off}
decrements it.  \cc{push_off} saves the original interrupt-enable bit,
so that when \cc{pop_off} brings the counter back to zero, it correctly
restores the state (either enabled or disabled).

This shows up in the specifications through two arguments to $\textlog{sie_cap_gpr}$:
the current depth $d$ of \cc{push_off} calls, and the original
interrupt-enable bit $\textit{eb}$ when the first \cc{push_off} ran.
When $d$ is non-zero, the interrupt-enable bit $b$ is false.  When $d$
goes from 1 to 0, $b$ becomes $\textit{eb}$.  When $d$ goes from 0 to 1,
$\textit{eb}$ becomes $b$.  When there is no active \cc{push_off} on the
current CPU (i.e., $d=0$), the kernel can explicitly toggle the interrupt-enable
bit by modifying the \cc{sstatus} register; this toggles the $b$ argument of
$\textlog{sie_cap_gpr}$.

One side effect of this design is that spinlocks must be released on
the same CPU on which they are acquired.  This is represented by the
\cc{acquire} function returning an additional $\textlog{lock_held}$
resource that is tied to the specific CPU on which the \cc{acquire}
executed.  This might not be the CPU on which \cc{acquire} started, because
\cc{acquire} might have been preempted before it disabled interrupts,
but it is the CPU on which \cc{acquire} returned.  \cc{release} requires
passing this $\textlog{lock_held}$ resource for the current CPU.

\paragraph{Avoiding deadlocks.}

The spinlock implementation in xv6 checks whether the caller is already
holding the spinlock it is trying to acquire, and if so, panics.  Moreover,
xv6 tries to avoid deadlock through spinlocks by maintaining a consistent
locking order, so that locks are never acquired in conflicting order
by different CPUs.  The spinlock specification addresses both of
these problems by maintaining the current lock-held set for each CPU,
as the $\textit{lkset}$ argument of $\textlog{sie_cap_gpr}$.  The precondition
of \cc{acquire} requires the caller to prove that the lock it is about
to acquire is ranked higher than any other lock in the current
lock-held set.  This ensures both that the lock it is about to acquire
cannot already be held by the same CPU (preventing the panic) and that
locks are never acquired in a conflicting order (preventing deadlock).

To be precise, the \sys top-level adequacy theorem currently does not
guarantee the absence of deadlock (since that is a liveness property),
but the xv6 specification for \cc{acquire} forces the rest of the kernel
code to follow the lock-ordering discipline, and in practice eliminates
the possibility of spinlock-based deadlock.

\paragraph{Dynamic allocation.}

In xv6, pipes are implemented by dynamically allocating a page of kernel
memory to store the state of a \cc{struct pipe}.  The implementation of
pipes allows concurrent reads and writes from multiple cores, and uses a
spinlock to protect the pipe state.  This spinlock itself also lives in the
dynamically allocated page of memory.  This complicates the spinlock
specification because the spinlock's invariant is no longer true at all
times: at some point, the page of memory will be deallocated, when the pipe
is fully closed, and from that point on, the page might be used for something
else, which is inconsistent with the pipe's locking discipline.  The xv6
proof solves
this by making the lock invariant \emph{cancellable}.  Namely, in order to
acquire the lock, the caller must prove that the pipe's reference count has
not reached zero.  This is established for each caller that reads from or
writes to a pipe by the fact that the \cc{struct file} referring to either
end of the pipe holds a reference on the \cc{struct pipe}.  When the reference
count reaches zero, the proof can conclude that no future caller can ever
satisfy the precondition to acquire the lock, and it is safe to cancel
the pipe's lock invariant, which allows reclaiming full ownership of the
pipe's memory page, which in turn allows freeing it.

\paragraph{Sleeplocks.}

In addition to spinlocks, xv6 implements sleeplocks for cases when the
thread holding a lock may need to go to sleep (i.e., yield the CPU to
other processes), which shows up in the file system when it needs to hold
a sleeplock while waiting for disk I/O.  For example, the file system locks
an inode, to ensure other kernel threads do not access or modify the state
of that inode, while reading the inode in from disk or writing it back.

One requirement for sleeplocks is that the caller must not hold any
spinlock while trying to acquire a sleeplock.  Otherwise, the sleeplock
acquire might switch to the scheduler while holding a spinlock, which could
result in deadlock.  To ensure this does not happen, the specification
for the sleeplock acquire requires that the caller have an empty lock-held set,
as specified by the $\textlog{sie_cap_gpr}$ $\textit{lkset}$ argument.

An exception to this rule shows up in the file system code for
releasing the last reference on an inode.  When the caller drops the
last reference to an inode, and it observes that the inode now
has a zero link count, it needs to then free the inode on disk.  The xv6
kernel implementation acquires the sleeplock while holding the spinlock
protecting the inode's in-memory reference count.  The reason this is safe
is that there cannot be any other concurrent threads trying to acquire
this sleeplock, since the caller holds the only in-memory reference to
the inode (the inode's refcount is 1).  To prove this
is safe, the sleeplock specification introduces a resource to represent
any active sleeplock holder.  Trying to acquire a sleeplock requires
supplying such a resource (which, in practice, is a share of an
in-memory reference to the inode).  Any acquired sleeplock holds this
resource until the sleeplock is released.  When the caller knows that
it holds the only reference, it can prove that there cannot be any such resource held
by the sleeplock, and therefore, acquiring the sleeplock will succeed
immediately, and will not go to sleep.

\subsection{Memory allocation}

xv6 implements a page allocator: \cc{kalloc} returns a fresh page of
memory, and \cc{kfree} puts it back on the free list.  The essence of the
specification for \cc{kalloc} and \cc{kfree} is a standard CSL encoding of
memory allocation: \cc{kalloc} returns ownership of 4096 bytes of memory,
at the returned address, and \cc{kfree} requires the caller to give up
ownership of all 4096 bytes.  \cc{kalloc} can non-deterministically fail
if there is no memory available, in which case it returns 0.

One complication is that xv6 uses \cc{kalloc} during bootup to initialize
system data structures like the page table, the per-process stacks, etc.
This initialization does not check for \cc{kalloc} returning 0 because
it will always succeed, since no other processes are running and there
must be memory still available.  However, the standard \cc{kalloc} spec
does not allow the caller to conclude that allocation will succeed,
because it is aimed at running in a concurrent setting.

The xv6 proof addresses this by having a single spec cover both modes of operation:
\cc{kalloc} starts out operating in a \emph{counted} regime, represented by a
$\textlog{kalloc_env}~(\textlog{Some}~n)$ resource, which is \emph{exclusive}: only one
CPU can own it at a time.  The $n$ argument represents the number of
free pages available for allocation.  The spec for \cc{kalloc} says that, if
the caller holds $\textlog{kalloc_env}~(\textlog{Some}~n)$, then allocation must succeed
if $n>0$.  This allows the bootup sequence to prove that all allocations
will succeed, by precisely tracking the number of free pages.  Once
CPU~0 finishes the boot-time allocations, the proof exchanges
$\textlog{kalloc_env}~(\textlog{Some}~n)$ for $\textlog{kalloc_env}~\textlog{None}$,
which represents the \emph{concurrent}
regime of \cc{kalloc}.  In this regime, allocation can non-deterministically
return 0.  The benefit of this regime, however, is that
$\textlog{kalloc_env}~\textlog{None}$ is persistent, which allows all cores to own it and allocate
memory concurrently.  Once the allocator enters the concurrent regime,
it cannot go back to the counted regime.

The traditional specification for a memory allocator requires the caller
to give up ownership of $\memptsto{a}{v}$ for every memory location that
they are freeing.  This is subtly too strong in a TSO shared-memory model,
because $\memptsto{a}{v}$ implicitly says that address $a$ has a stable
value if read on the current CPU.  This is not strictly necessary to
free a page of memory: the allocator does not care if the current CPU
sees a stable value for the memory being freed.  We had to change our
\cc{kfree} specification to accept a weaker precondition of owning every
address being freed, without requiring it to be stable.

\subsection{File descriptors}

xv6 uses reference counts in \cc{struct file} to keep track of how many
outstanding references exist to a particular open file.  This reference
count is incremented every time a user process calls \cc{dup} or \cc{fork},
and decremented when a file descriptor is closed.  The kernel implementation
does not do overflow checking when incrementing the reference count.
The reason this is safe is that there can be only a bounded number of open
file descriptors that can hold such a reference.  To formally prove this
argument, the xv6 proof introduces a notion of a file reference slot,
which is a token representing the right to increment a file's refcount.
There is a fixed number of file reference slots in each era (the period
between one power-on and the next), distributed at boot
time across all \cc{struct proc} entries, each of which receives one
slot per entry in its file descriptor table (\cc{NOFILE} pointers to \cc{struct file}).
When a process opens a file descriptor, the proof transfers ownership
of the slot resource to the \cc{struct file}; when the file descriptor is
closed, the proof moves the unused reference slot back into the process
invariant.  Each \cc{struct file}
keeps a list of these slots, and maintains an invariant that its
refcount is equal to the number of slots in its list.  Since only a bounded
number of slots are distributed at system boot time, this allows
the proof to conclude that the refcount increment cannot overflow.

A second complication with file descriptors is that multiple processes
might have a file descriptor referring to the same \cc{struct file},
yet the \cc{struct file} has a single reference on an underlying
\cc{struct pipe} or inode.  As a result, when processes concurrently
read from file descriptors that actually represent the same underlying
\cc{struct file}, they must share the fact that they are holding
a reference on the \cc{struct pipe} or inode between them.  For example,
recall that acquiring a spinlock on \cc{struct pipe} requires proving that
the pipe's refcount is non-zero, as part of the pipe's cancellable spinlock.
This is represented using fractional ownership of a resource representing
a unit refcount on either a \cc{struct pipe} or inode.

\subsection{File system}

xv6 has a traditional Unix file system with directories and files,
and system calls for creating, reading/writing, and removing them.  The
file system---without the disk driver, pipes, and file descriptor system
calls---constitutes $\sim$30\% of the xv6 kernel in terms of lines of code.
Formally reasoning about the file system code is tricky because the file
system handles crash recovery (by executing each file system operation as
a transaction), has substantial internal concurrency (system calls involving
different inodes may run in parallel on different CPUs, and commit as
a group), and must handle tricky Unix semantics (e.g., a file can be
unlinked from a directory but still be in use by another process that has
the file open).  This subsection highlights a few interesting issues
in the xv6 proof of the file system.

\paragraph{File system durability.}

A key property of the file system is that it must be consistent across
crashes.  This is formulated in the xv6 proof using an invariant about the
durable on-disk state; this invariant persists across reboots.  xv6 uses a
write-ahead log to update the disk contents in a way that is atomic with
respect to crashes.  The xv6 proof represents this by introducing two views
of the disk contents: $D$, which represents the durable state on disk,
and $M$, which represents the in-memory state that will be written to
the log and committed.

The $D$ state reflects what the on-disk state will be after recovery
finishes (i.e., after applying the log).  This means that $D$ is
updated to match $M$ when the log header recording the commit is written to the disk.
The \cc{recover_from_log} function installs the log after a crash and
reboot; its specification says that it keeps $D$ unchanged.

The $M$ state reflects the state of the file system present in the
buffer cache.  For instance, when a system call executes, it updates
blocks containing the relevant file data, inodes, bitmaps, etc., and hands
the blocks to the write-ahead log for eventual commit.  The $M$ state
is updated at the time blocks are handed to the write-ahead log layer.
When the log commits transactions, it updates $D$ to be equal to $M$.
This allows the log to prove that the on-disk state, which is now $D=M$,
satisfies the file system consistency invariant, because the in-memory
state $M$ satisfied the consistency invariant before commit.

One complication with the in-memory $M$ state is that it is not always a
consistent state of the file system.  For example, during the execution
of \cc{sys_unlink}, the kernel first updates the parent directory,
removing the directory entry, and then updates the child, dropping its
link count.  Between these two operations, the $M$ state is not consistent.
Furthermore, ownership of different parts of $M$ is distributed among many
locks, because the implementation is concurrent: different CPUs may be
running system calls that have locked different parts of the file system.

To prove that the file system is consistent at commit time, the xv6
proof maintains an invariant that every write-locked inode (i.e., an
inode locked by a system call that may modify it) must be part
of a transaction.  At commit time, the write-ahead log waits until all
of the in-flight transactions have finished before writing the set of
all modified blocks to the log.  This ensures that, at commit time,
no inode can be write-locked, because there are no pending transactions.
This allows the commit proof to observe that all inodes are in a consistent
state, because all of them satisfy the inode's lock invariant.

\paragraph{Writing the log header.}

The xv6 kernel commits a transaction by writing the \emph{log header}
block, which lists the blocks in the log, to disk.  If xv6 experiences a crash, xv6 will check the log
header on reboot.  If there is a committed transaction in the log, it
will install the blocks modified by the transaction into their home
locations on disk.  The implementation of xv6 ensures commits are atomic
by writing a single block and waiting until the disk acknowledges that
the write has completed.

There are two complications that the proof must handle: (1) an xv6 disk
block is two disk sectors, and only single-sector writes are atomic;
(2) unless explicitly disabled, the virtio disk uses a writeback cache
and writes sectors asynchronously: if the write operation completes,
the disk guarantees that the sector is in its writeback cache but it
may not be on the disk itself, and thus a power failure may cause the
acknowledged sector writes to be lost.  The reason that commits are
synchronous and atomic in xv6 is that (1) xv6's log header
fits in a single sector, and (2) the xv6 disk driver does not negotiate
\cc{VIRTIO_BLK_F_FLUSH}, making disk operations
write-through.

The proof establishes write-through behavior as a property of the xv6
driver.  The proof of \cc{virtio_disk_init} computes the feature bits
that the driver writes to the device, and the device protocol
invariant records that \cc{VIRTIO_BLK_F_FLUSH} was not negotiated, from
which the proof concludes that the cache is empty whenever a request
completes.

Reasoning about what is on disk after a crash requires setting up
ownership carefully, because a crash destroys all threads and every
resource they own; the persistent state of the disk is therefore owned
by a crash invariant that survives power cycles.
A disk write is specified
through a \emph{write permit}: a proof callback (a $\vs$ resource,
\autoref{sec:csl-intro}) over the crash invariant
that the proof constructs when the driver issues the request and that the
proof applies at the moment a sector lands on disk.  Since a block
write consists of two sector writes that may be written in either
order to the disk, and
both need to update the log's ghost state describing the disk, the
proof cannot hand out two independent permits, one per sector.  Instead,
a request carries a single permit for a set of sectors;
the device picks which sector
to write first, and applying the permit to that sector returns the permit for the remaining sectors,
threading the ghost state through from one sector write to the next.  The
final permit, when no sectors remain, delivers the postcondition that
the caller receives when the write completes.

Atomicity of the commit follows from the layout of the xv6 log header, which
fits in the first 124 bytes of its block, and from the fact that
recovery reads only those bytes.  The proof builds the header write's
permit for both sectors.  When the disk writes the block's first sector
(which holds the 124-byte header),
the proof switches the log header from the old transaction to the new
one in one step; writing the other sector, whether it happens before
or after the first, changes nothing that recovery reads.
Together with write-through completion, this shows
that when the write of the log header (\cc{write_head}, called by
\cc{commit} from \cc{end_op}) completes, the new log
header is on disk, and that a crash in the middle of the write leaves
either the old or the new log header, but never a mix.

\paragraph{Log budget for transactions.}

The xv6 log is bounded in size, and the
code is carefully written to ensure that any transaction writes no more
than \cc{MAXOPBLOCKS} blocks, which in turn ensures that
every transaction will be able to commit.

To prove that each transaction stays within its budget of log blocks,
we originally used a worst-case bound by counting the number of times
the kernel logs a write in a transaction, and requiring that this
never exceed \cc{MAXOPBLOCKS}.  This turned out to be insufficient
because the kernel relies on log absorption to stay below this bound.
In particular, the kernel might update the free bitmap block multiple
times within a single transaction (e.g., freeing multiple disk blocks
when deleting a file), or update the same data block multiple times
(e.g., writing the directory entries for \cc{.} and \cc{..}); those
writes are absorbed by the log and do not count separately towards the
transaction's budget.

We extended the spec to also track the set of blocks logged
by the transaction so far; a \cc{log_write} of a block already in that set
is absorbed and does not consume one of the transaction's
\cc{MAXOPBLOCKS} units.
The function
\cc{begin_op} that starts a transaction for a system call creates a
new resource for this set and passes it down to functions that call
\cc{log_write} to update the set in the proof.  \cc{end_op}, which
commits a transaction, requires that the set be no larger than
\cc{MAXOPBLOCKS}.

\paragraph{Group commit for \cc{namei}.}

Handling absorption within a transaction was not enough to reason
about the budget of log blocks in xv6.
Consider a process
that calls \cc{namei("/a/..")}.  The kernel runs \cc{namei} inside
a transaction because when \cc{namei} is done with an inode (e.g.,
the inode for ``/a''), it calls \cc{iput}, and \cc{iput} may discover that
\cc{namei} has the last reference to the inode because concurrently
another process unlinked ``/a'', and thus \cc{namei}'s \cc{iput} logs the effects
of the \cc{unlink}.

Now consider a process calling \cc{namei("/a/../b/../c/../d/..")} with
a concurrent process unlinking each directory.
If a racing \cc{unlink} happens for every path element (``/a'', ``/b'',
``/c'', and ``/d''),
\cc{namei}'s transaction runs the risk of growing to an arbitrary number of
blocks.  Each time it calls \cc{iput} for the inode
of a path element, \cc{iput} will observe a zero link count and log
the \cc{unlink}.

The reason that \cc{namei}'s transaction stays within
the \cc{MAXOPBLOCKS} budget
is that if a concurrent \cc{unlink} races with \cc{namei}, then the
\cc{unlink} must have also already written that exact inode into the log
(when it did the \cc{iupdate} to drop the inode's \cc{nlink}), so
the \cc{iupdate} in \cc{namei}'s \cc{iput} gets
absorbed with that other transaction.  So \cc{namei} stays within its
budget because of cross-transaction absorption.  We asked the agent to
extend the proof to do cross-transaction accounting of log blocks, and
that allowed the agent to complete the proof for \cc{namei}.

\paragraph{Buffer cache.}

In xv6, allocating a buffer in the cache to hold a disk block for a block number
is a distinct operation from loading the bytes of the block's content
from disk.  To read a disk block, the kernel calls \cc{bread}, which
calls \cc{bget} to check the buffer cache.  If a block is not present
in the buffer cache, \cc{bget} allocates an entry in the buffer cache
under the \cc{bcache.lock} lock, sets \cc{b->blockno} to the block
number of the block, increases \cc{b->refcnt} from 0 to 1, sets
\cc{b->valid} to false, and releases the \cc{bcache.lock}.  Then,
\cc{bget} acquires the sleeplock \cc{b->lock} and returns \cc{b} to
\cc{bread}, which reads the data from disk and marks the buffer
valid.  When the kernel is done with the buffer \cc{b}, it calls
\cc{brelse}, which decreases \cc{b->refcnt}.

A complication for the proof is that between releasing
\cc{bcache.lock}
and acquiring \cc{b->lock} in \cc{bget},
the kernel holds no lock, and the proof must argue that the
buffer was not reused for a different block number, based on the fact
that the kernel does not reuse a buffer with \cc{b->refcnt} $\ge 1$, even if
\cc{b->valid} is false.
Another
complication is that the proof must argue that a later \cc{bget} for a
block number in the buffer cache returns the current content of the
block (which may differ from the on-disk contents while the block
sits in the write-ahead log).

The proof addresses both complications by giving each buffer its own
invariant that owns the buffer's contents (\cc{b->valid}, \cc{b->data},
and a share of \cc{b->dev} and \cc{b->blockno}) together with a ghost
resource describing the block's current logical content, rather than
associating the contents with either lock.  The sleeplock \cc{b->lock}
protects only a checkout token: \cc{bget} exchanges this token for the
buffer's contents when it acquires the sleeplock, and \cc{brelse} puts
the contents back into the invariant in exchange for the token before it
calls \cc{releasesleep}, since a waiting \cc{bget} may return as soon as
the sleeplock is free, before \cc{brelse} has decremented
\cc{b->refcnt}.

The proof tracks reference counts using ghost state under
\cc{bcache.lock} and fractional ownership: every outstanding reference
owns a fragment of a per-buffer counter, and the invariant ties the
counter to \cc{b->refcnt}.  The recycling path of \cc{bget} needs the
counter to be zero in order to rewrite \cc{b->blockno}, so any thread
holding a reference fragment---including one that has released
\cc{bcache.lock} but not yet acquired \cc{b->lock}---can conclude that
its buffer still holds its block number, regardless of \cc{b->valid}.

For a buffer with \cc{b->valid} false, the invariant holds the ghost
resource for the block's content, which the thread that performs the
disk read (whichever thread first acquires the sleeplock) uses to
establish that the bytes it read match the block's logical content;
for a valid buffer, the invariant states that \cc{b->data} agrees with
that content, which is what a later \cc{bget} relies on.  The proof
of the buffer cache never interprets the content resource itself: it is a parameter
supplied by the logging layer, which instantiates it with its own
ghost state tracking whether the block's current content is on disk or
in the write-ahead log.

\paragraph{Link consistency without global statements.}

The consistency of the file system, both on disk and in memory, requires
that the directory structure be consistent with the link counts.  At some
level, this is a global property: the link count of each inode depends on
the directory entries present in the rest of the file system.  However,
for performance, the xv6 implementation uses fine-grained per-inode locks
when it runs.  To avoid reasoning about the global file system state
each time the on-disk or in-memory state is modified, we introduced a
ghost resource to keep track of inode references in a modular way.

Specifically, each inode owns a ghost variable tracking its outstanding
link count, and the inode's own invariant says that its \cc{ip->nlink}
field is equal to the number of outstanding links according to
the authoritative ghost state.  Every directory that has a link to
an inode maintains ownership of a unit fragment of this ghost state,
representing the fact that it incremented that inode's \cc{nlink} by 1.
This resource allows \cc{sys_unlink} to justify decrementing an inode's
\cc{nlink} when it is removed from a parent directory, because the parent
directory can return this unit fragment of ownership, proving that the
\cc{nlink} will not underflow when decremented.

A further invariant that the file system must maintain is that the
\cc{..} entry in every directory correctly refers to its parent inode.
The reason is that, when a subdirectory is deleted, the implementation
of \cc{sys_unlink} in xv6 (which subsumes \cc{rmdir}) decrements the
parent directory's link count.  The proof must establish that the child
directory indeed held a reference on the parent directory.  To do this
without explicit global consistency properties, the xv6 proof includes
the parent directory's inode number in the link-counting ghost state,
for inodes that happen to be directories.  (The parent directory's
inode for files is not well-defined, because xv6 supports hard links.)
This part of the ghost state allows each directory's invariant to say
that (1) the inode number of the \cc{..} entry is the parent directory's
inode number from the ghost state, and (2) any child directory refers
back to the correct parent directory's inode number.

The key benefit of this design is that it allows the xv6 proof to state
file system consistency properties, such as link-count consistency and
\cc{..} consistency, in a local manner, specified in the independent lock
invariant of each inode.  This reduces the amount of global reasoning
needed when performing local operations.

\paragraph{\cc{iget}.}

The \cc{iget} function takes an inode number, \cc{inum}, as argument
and returns the inode associated with it.  The na\"ive spec for
\cc{iget} states that it returns an allocated inode.  That spec is not
strong enough, however, for concurrent system calls: another
thread may have freed the inode and reused the \cc{inum} for another
inode.  The na\"ive spec does not forbid it: the inode returned is
allocated but is not the inode the caller intended to look up, which
makes it difficult for callers of \cc{iget} to reason about their
correctness.  Instead, \cc{iget} has a stronger premise: it
requires that the caller provide a resource that pins down the
identity of the inode named by \cc{inum} and rules out its being
freed.

Callers of \cc{iget} fulfill the premise in different ways.  For
example, \cc{dirlookup} in \cc{namex} fulfills it with evidence that
it holds the parent directory locked and that the directory is still
linked (i.e., \cc{nlink} $> 0$).  \cc{ireclaim}, which scans the disk at
boot for orphaned inodes and frees them, fulfills the premise with an
exclusive \emph{boot token}, which exists only during boot and
demonstrates that \cc{ireclaim} runs while it owns all
inodes and xv6 runs sequentially.  \cc{ialloc} fulfills it with
exclusive ownership of the freshly allocated inode, which, as we
discuss below, is the hardest instance.

The premise relies on an invariant that entries exist only in linked
directories (an unlinked directory is empty except for \cc{.} and
\cc{..}); to preserve it, system calls that create names (e.g.,
\cc{create} and \cc{sys_link}) must check that the directory they
write to is still linked.

The xv6 kernel had none of the necessary \cc{nlink} checks: \cc{namex}
would follow \cc{..} in an unlinked directory, and \cc{create} and
\cc{sys_link} would write names into one.  The
proof got stuck trying to fulfill \cc{iget}'s premise in \cc{namex};
the stuck proof turned out to reflect a real bug with several
manifestations, including a kernel panic (\autoref{sec:bugs}).

\paragraph{\cc{ialloc}.}

When allocating an inode, the xv6 file system
calls \cc{ialloc} to get an inode number that is not in
use.  \cc{ialloc} scans the inode blocks to find an unused inode number, marks the
inode on disk as allocated by calling \cc{log_write}, and calls
\cc{brelse} to release the block holding the inode.  Then, it calls
\cc{iget} on the inode number to get a \cc{struct inode} representing
that inode.  \cc{ialloc} does not hold any locks between calling
\cc{brelse} and \cc{iget}, and thus the proof must establish that no other thread
could have modified the inode (e.g., freed it);
that is,
\cc{ialloc} must discharge \cc{iget}'s premise for an \cc{inum} that is in
no directory at all.

Initially, the agent could not prove this claim, and on one attempt changed the
xv6 source to not do the \cc{brelse} to make the proof go through, which we
caught because the agent could not push to the official xv6 repo.  On
another attempt the agent introduced a new axiom \cc{create_fresh_ty}.
Our CI report of the axioms that the top-level theorem depends on caught this
attempt (\autoref{sec:agent}).  In the end, it turned out that \cc{ialloc} was not provable
because \cc{iput} had a bug (\autoref{sec:bugs}).

Proving \cc{ialloc} after the \cc{iput} fix turned out to still be
difficult: the
proof must make the argument that the allocated \cc{inum} is not in the file
system tree and thus no thread could have found the \cc{inum} and modified
the corresponding inode.
This argument is further complicated because
an inode might have been unlinked (and not be in the file system tree)
while a file descriptor is still
pointing to that inode; the agent had to argue that \cc{iput} could not have freed
such an inode yet because the file descriptor still holds an outstanding reference to it.
Finally, \cc{ialloc} must also rule out \cc{ireclaim} as a concurrent
writer, which it does using the same boot token that \cc{ireclaim}'s own premise
rests on: once boot finishes, the proof irreversibly exchanges the boot token for
a persistent witness that boot is over, and \cc{ialloc} requires this witness.

This proof relies on the link-count ghost state, together with a
companion ghost state recording each inode's type (zero for a free
inode), which lets the proof conclude that any inode with an
outstanding link is allocated.

\paragraph{\cc{iput}.}

The \cc{iput} function puts an inode \cc{ip} back into xv6's inode cache: it
decreases \cc{ip->ref}, so that the \cc{struct inode} can be recycled for
another inode once the count reaches zero.  If \cc{ip->nlink} is also 0, it
frees the inode too.  To read \cc{ip->ref}, \cc{iput} holds
\cc{itable.lock}, and if \cc{ip->ref} is 1, it acquires the \cc{ip->lock}
sleeplock to free the inode.  The implementation of \cc{acquiresleep},
in turn, acquires the process's \cc{p->lock} if another thread is
holding the sleeplock.  However, \cc{kfork} holds \cc{p->lock} across
\cc{idup}, which takes \cc{itable.lock}.  In this scenario there is no
global ordering of locks.

The proof of \cc{iput} must establish that this scenario cannot
happen, which is true because, if \cc{ip->ref} is 1, no other
thread could be holding \cc{ip->lock}, and \cc{acquiresleep} will not
go to sleep.  To handle this case, we have two specifications for
\cc{acquiresleep}: the
general one, which may sleep and therefore requires an empty lock-held
set, and a non-blocking one, which requires the caller to prove that no
thread can currently hold the sleeplock and therefore allows any
lock-held set.

To be able to use the second specification, the proof associates with
each inode's sleeplock a ghost counter of potential holders.  The
counter has two parts: an authoritative total, which is stored in the
lock invariant of \cc{itable.lock} next to the inode's reference
count, and fractions that add up to this total, which the proof hands out
to the holders of references to the in-memory inode (a fraction per
reference, since many \cc{struct file}s may share one inode reference
and race for its sleeplock).  Acquiring the sleeplock deposits the
caller's fraction in the sleeplock's invariant, and releasing it takes
the fraction back; a thread can thus hold or wait for the sleeplock
only if it owns a fraction.

When \cc{iput} observes \cc{ip->ref} equal to 1 while holding
\cc{itable.lock}, the ghost state for the reference count tells the
proof that the reference \cc{iput} owns is the only one, and therefore
that \cc{iput}'s fraction is the entire total recorded in the lock
invariant of \cc{itable.lock}.  \cc{iput} returns its fraction to the
lock invariant, which brings the total to zero.  A total of zero means
that no thread owns a fraction, so no thread holds or is waiting for
the sleeplock, which is exactly the precondition of the non-blocking
specification of \cc{acquiresleep}.  Acquiring the sleeplock then
hands \cc{iput} a fresh fraction of the same size, which restores the
total in the lock invariant of \cc{itable.lock} to what it was, so
that \cc{iput} can release \cc{itable.lock} afterwards.

\subsection{Process exit}

When a process exits in xv6, the kernel must reclaim ownership of the
process's kernel stack.  However, the code path of the process exiting
involves a trap vector at the top level, \cc{uservec}, invoking the
C-level trap handler \cc{usertrap}, which calls \cc{syscall}, which
in turn dispatches \cc{sys_exit}, and ultimately this sets the process
state to \cc{ZOMBIE} and switches to the scheduler.  The challenge is that,
at each level of this call stack, the callers (such as \cc{usertrap} and
\cc{syscall}) still have CSL-level ownership of resources representing
their parts of the stack, corresponding to their stack variables, saved
registers, etc.  At the same time, in order to reuse the kernel stack
of this exited process, the kernel needs to regain ownership of the entire
4~KB page of memory.

The xv6 proof addresses this challenge by extending the specification
of the calling convention with explicit support for function calls
that might never return.  In particular, if a function chooses not to
return to its caller, its caller should give up ownership of its stack
memory locations that might be implicitly captured inside the caller's
continuation $\wpcycle$.  We encode this using
separation logic's $\land$ operator (AND).  Unlike the $\ast$ separating
conjunction, where $A\ast B$ means that $A$ and $B$ exist separately,
and ownership of $A\ast B$ allows using both of them, $A\land B$ means
that $A$ and $B$ are both true but they might overlap, so they are not
facts about disjoint parts of the state.  The rule for using $A\land
B$, then, is that a proof that owns $A\land B$ must choose which one to
get: either get ownership of $A$, but that makes $B$ disappear, or get
ownership of $B$, but that makes $A$ disappear.

\autoref{fig:syscall-stack-reclaim-spec} shows this pattern in a
simplified version of the spec for the \cc{syscall} function.  With this
specification, the proof of \cc{syscall} can choose to either use the
left side of the $\land$, and return to the caller, or use the right
side of the $\land$, whereby the $\textlog{stack_reclaim}~m[\cc{sp}]$
resource allows it to reclaim ownership of all stack memory from the
current stack pointer up towards the top of the stack ($\textlog{sie_cap_gpr}$
holds ownership of the other half of the stack, namely, all of the free
stack locations below the current stack pointer).  The proof of
the caller of \cc{syscall}, namely \cc{usertrap}, has to supply both of
these resources to \cc{syscall}'s proof.  A function's proof may need
to selectively invoke its own caller's WP or $\textlog{stack_reclaim}$
resources to satisfy these two obligations when calling a function that
might not return.

\begin{figure}[ht]
\begin{specblock}
\specline{\ktext \wand} \\
\specline{\pc{\cc{syscall}} \wand} \\
\specline{\textlog{sie_cap_gpr}~m~n~\ldots \wand} \\
\specline{\ldots \wand} \\
\specline{(~( \forall~m'.~~\pc{m[\cc{ra}]} \wand} \\
\specline{\phantom{(~( \forall~m'.~~}\textlog{sie_cap_gpr}~m'~n~\ldots \wand} \\
\specline{\phantom{(~( \forall~m'.~~}\textlog{callee_saved}~m = \textlog{callee_saved}~m' \wand} \\
\specline{\phantom{(~( \forall~m'.~~}\ldots \wand} \\
\specline{\phantom{(~( \forall~m'.~~}\wpcycle ) ~ \land} \\
\specline{\phantom{(~} \textlog{stack_reclaim}~m[\cc{sp}]~) \wand} \\
\specline{\wpcycle}
\end{specblock}
\caption{Simplified spec for the \cc{syscall} function, demonstrating
  how the specification captures the possibility of \cc{syscall} not
  returning, and in that case, being able to reclaim ownership of the
  caller's stack frame.}
\label{fig:syscall-stack-reclaim-spec}
\end{figure}

For functions that never return, such as \cc{sys_exit}, the specification
does not require any $\wpcycle$ continuation; the caller must always
supply just the $\textlog{stack_reclaim}$ resource.  At the bottom level,
the specification of \cc{swtch} requires the caller to supply either a
$\wpcycle$ for returning, which is placed into a $\textlog{valid_context}$
for eventually resuming the kernel thread, or a $\textlog{stack_reclaim}$
resource that allows the scheduler to reclaim the entire kernel stack.
In particular, the xv6 scheduler specification requires the caller to
supply the $\wpcycle$ continuation if the process is \cc{RUNNABLE} or
\cc{SLEEPING}, and supply the $\textlog{stack_reclaim}$ resource if the
process is \cc{ZOMBIE}.

\subsection{Compiler optimizations}

One advantage of verifying the kernel at the RISC-V level in \sys is
that the proof subsumes any compiler optimizations.  One surprising
optimization that the xv6 proof ran into is GCC's computation of a pointer
difference within a struct array.  In particular, the \cc{procinit} function initializes the
kernel stack for all \cc{struct proc}s, at kernel boot time, as shown in
\autoref{fig:procinit-c}.  The \cc{p - proc} expression, in C, computes the
index of pointer \cc{p} in the \cc{proc} array, which requires dividing the
difference between the two pointer addresses by \cc{sizeof(struct proc)},
which happens to be 360.

\begin{figure}[ht]
\input{code/procinit.c}
\caption{Example of code from the \cc{procinit} function in xv6 that triggers
  a surprising optimization when compiled with GCC.}
\label{fig:procinit-c}
\end{figure}

When GCC compiles this code, it avoids emitting what would be the
direct translation using a divide-by-360 instruction.  Instead, GCC's
optimization first observes that $360=8\times 45$, so to divide by 360,
it suffices to divide by 8 (by doing a right-shift by 3), and then divide
by 45.  Then, to divide by 45, GCC computes the modular inverse of 45,
$M = 45^{-1} \bmod 2^{64} = \mathtt{0x4fa4fa4fa4fa4fa5}$.  Finally,
since GCC knows that the difference between the pointer addresses \cc{p}
and \cc{proc}, which we will call $\delta$, is divisible by 360, and
thus that $\frac{\delta}{8}$ is divisible by 45 and less than $2^{63}$,
dividing it by 45 is equivalent to multiplying by $M$.  Thus,
GCC computes $\cc{p}-\cc{proc}=\frac{\delta}{360}$ as $(\delta \gg
3)\times M$.  The proof of \cc{procinit} in xv6 proves the correctness
of this modular-inverse optimization.

\subsection{UART output}

The specification for \cc{uartwrite}, used in xv6 to print
output to the console, is stated in terms of a ghost variable
tracking an append-only list of bytes printed to the console,
$\textlog{uart_sent}~\gamma~\textit{bs}$.  The specification for
\cc{printk}, which we will discuss next, says that, whatever the console
output was before, the result will be that the console now contains
the output of \cc{printk} as well.  Since the console output is append-only,
\cc{printk} can return this fact in its postcondition as a fact about a
\emph{prefix} of the console output; more bytes might have been written
by other processes, or by \cc{printk} calls on other cores, but
\cc{printk}'s output is guaranteed to be present.

\subsection{Specifying \cc{printk}}

xv6 implements an in-kernel variant of \cc{printf}, which it calls
\cc{printk}.  Specifying \cc{printk} is challenging because it takes a
variable number of arguments, and because the facts that the caller's
proof has to establish about each of those arguments depend on the
contents of the format string passed to \cc{printk} in the first argument.

\autoref{fig:printk-spec} shows a simplified version of the specification
for \cc{printk} in xv6.
The pure function $\textlog{kinds}$ walks the format string $f$ and returns,
in order, the kind of every argument that \cc{printk} will consume: a
$\textlog{Num}$ for the integer conversions (\cc{\%d}, \cc{\%u},
\cc{\%x}, \cc{\%p}, \cc{\%c}, and their \cc{l} and \cc{ll} variants,
which the figure elides), and a $\textlog{Str}$ for \cc{\%s}.  Escapes such as \cc{\%\%}
and unrecognized conversions consume no argument, exactly as in the C
code.  Since $\textlog{kinds}$ is a structural fixpoint, Rocq proofs can
compute its result---for example, for the format string used by \cc{syscall} to report
an unknown system call, $\textlog{kinds}~\texttt{"\%d \%s: unknown sys call \%d\textbackslash{}n"}
= [\textlog{Num}; \textlog{Str}; \textlog{Num}]$ by reflexivity.

The resource that the caller must supply for each kind, $\textlog{kind_res}$,
is where the argument types differ.  An integer, character, or pointer printed
with \cc{\%p} costs nothing: in RISC-V, every variadic argument is passed in a 64-bit
register, \cc{printk} only reads the value of a $\textlog{Num}$, and its
integer-printing routine can handle any value.  The C-level
distinctions between \cc{int}, \cc{long}, and \cc{char} are
invisible to the logic because they are invisible at the instruction level.
For strings (\cc{\%s}), the caller's proof must either prove that the
pointer is null (in which case \cc{printk} prints \cc{(null)}), or supply
a string points-to resource $\strptsto{v}{s}$ for the argument's value
$v$.

$\textlog{args}$ combines all of the required resources, given the format
string $f$, and also encodes the calling convention: argument $j$ lives
in register \cc{a}$(j{+}1)$.  The bound $|\textlog{kinds}~f| \le 7$
encodes the assumption that \cc{printk} gets no more than 7 arguments;
relaxing this assumption would require extending $\textlog{args}$ to know
about how the calling convention passes further arguments on the stack.
The postcondition returns the resources back to the caller, and also says
that some byte list $\textit{cs}$ has been appended to the bytes that
this caller has provably sent to the UART.  The spec does not precisely
pin down the resulting byte sequence sent to the UART, since the exact
format of kernel \cc{printk} debug messages was not important for our
verification goal.

\begin{figure}[ht]
\centering
\begin{minipage}[t]{0.44\columnwidth}
\begin{specblockt}
\specline{\textlog{kind} \triangleq \textlog{Num} \ALT \textlog{Str}} \\
\specline{\textlog{kinds} : \textdom{string} \ra \textdom{list}~\textlog{kind}} \\
\quad
  & \textlog{kinds}~(\texttt{"\%d"} \dplus r) = \textlog{Num} \cons \textlog{kinds}~r \\
  & \textlog{kinds}~(\texttt{"\%s"} \dplus r) = \textlog{Str} \cons \textlog{kinds}~r \\
  & \textlog{kinds}~(\texttt{"\%\%"} \dplus r) = \textlog{kinds}~r \\
  & \textlog{kinds}~(c \cons r) = \textlog{kinds}~r \quad\text{(any other $c$)} \\
  & \textlog{kinds}~\texttt{""} = [] \\
& \\

\specline{\textlog{kind_res}~v~\textlog{Num} \triangleq \textlog{True}} \\
\specline{\textlog{kind_res}~v~\textlog{Str} \triangleq
          v = 0 \lor \exists~s.~~\strptsto{v}{s}} \\
& \\

\specline{\textlog{args}~m~f \triangleq} \\
  & \mathop{\scalebox{1.4}{$\ast$}}_{j \mapsto k \in \textlog{kinds}~f}
    \textlog{kind_res}~m[\cc{a}(j{+}1)]~k
\end{specblockt}
\end{minipage}%
\hfill
\begin{minipage}[t]{0.52\columnwidth}
\begin{specblockt}
\specline{|\textlog{kinds}~f| \le 7 \ra} \\
\specline{\ktext \wand} \\
\specline{\pc{\cc{printk}} \wand} \\
\specline{\textlog{sie_cap_gpr}~m~\ldots \wand} \\
\specline{\strptsto{m[\cc{a0}]}{f} \wand} \\
\specline{\textlog{args}~m~f \wand} \\
\specline{\textlog{uart_sent}~\gamma~\textit{bs} \wand \ldots \wand} \\
(~\forall~m'~\textit{cs}.~~
   & \pc{m[\cc{ra}]} \wand \\
   & \textlog{sie_cap_gpr}~m'~\ldots \wand \\
   & \textlog{callee_saved}~m = \textlog{callee_saved}~m' \land \\
   & \quad m'[\cc{a0}] = 0 \wand \\
   & \strptsto{m[\cc{a0}]}{f} \wand \\
   & \textlog{args}~m~f \wand \\
   & \textlog{uart_sent}~\gamma~(\textit{bs} \dplus \textit{cs}) \wand \\
   & \wpcycle ~) \wand \\
\specline{\wpcycle}
\end{specblockt}
\end{minipage}

\caption{Simplified spec for \cc{printk}.}
\label{fig:printk-spec}
\end{figure}

\subsection{Bounded physical memory}

The implementation of \cc{exec} in xv6, the \cc{kexec} function,
initializes the page table of the fresh process based on the supplied ELF
image of the user binary that is being \cc{exec}'ed.  This is tricky code
because it populates the page table based on an untrusted ELF image.  In particular, consider
the simplified snippet of this code shown in \autoref{fig:uvmalloc-c}.
\cc{kexec} passes arbitrary values from the ELF program header \cc{ph} as
the \cc{newsz} argument of \cc{uvmalloc}, and \cc{uvmalloc}'s call to
\cc{mappages} will panic if it turns out this address is already mapped.
Every user page table has a trampoline mapping at the very top of the
virtual address space, used to switch between user and kernel page tables,
and thus the proof had to consider whether \cc{kexec} could possibly
call \cc{panic} on some adversarial ELF binary.

\begin{figure}[ht]
\input{code/uvmalloc.c}
\caption{Simplified code from \cc{kexec}, \cc{uvmalloc}, and
  \cc{mappages} in xv6.}
\label{fig:uvmalloc-c}
\end{figure}

The reason this panic turns out to be unreachable is that, in order
for \cc{uvmalloc}'s for loop to reach the trampoline virtual address
(which would have triggered the panic), it has to first allocate physical
memory pages all the way up to that high virtual address.  This is not
possible; the system simply does not have that much physical memory.
Thus, the kernel proof is able to conclude that the panic is unreachable,
by contradiction: if the execution reached that panic, it would mean that the
current proof holds separation-logic resources for more physical memory
than is possible on the machine.

\subsection{Safety of arbitrary user-mode execution}
\label{sec:proof-user-safety}

An OS kernel must ensure that, when it starts executing processes in
user mode, those processes cannot corrupt the execution of
the kernel.  The xv6 kernel proof handles this by proving a WP theorem
about execution in user mode.  \autoref{fig:user-spec} shows a simplified
version of this theorem.  Unlike every other $\wpcycle$ specification,
this spec does not talk about execution at a specific PC address.  Instead,
it covers execution at any PC, as long as the processor remains at the
user privilege level, for an arbitrary number of cycles.

\begin{figure}[ht]
\begin{specblock}
\specline{\textlog{user_cfg}~C \triangleq} \\
\quad
  & \regptsto{\cc{stvec}}{C.\cc{stvec}} \ast
    \regptsto{\cc{medeleg}}{C.\cc{medeleg}} \ast \ldots \ast {} \\
  & \textlog{direct_mode}~C.\cc{stvec} \ast {} \\
  & \textlog{user_exceptions_delegated}~C.\cc{medeleg} \ast \ldots \\
& \\
\specline{\textlog{user_pt}~\textit{pt} \triangleq} \\
\quad
  & \regptsto{\cc{satp}}{\textit{pt}.\textit{root}} \ast
    \textlog{pt_pages}~\textit{pt} \ast
    \textlog{pt_wf}~\textit{pt} \ast {} \\
  & \exists~M.~~\textlog{dom}~M = \textlog{mapped}~\textit{pt} \ast {} \\
  & \phantom{\exists~M.~~}
    \mathop{\scalebox{1.4}{$\ast$}}_{\textit{va} \mapsto b \in M}
    \memptsto{(\textlog{translate}~\textit{pt}~\textit{va})}{b} \\
& \\
\specline{\textlog{user_inv}~C~\textit{pt} \triangleq
  \exists~\textit{hs}~\textit{ms}~\textit{pc}~m~c~t~e.} \\
\quad
  & \regptsto{\cc{cur\_privilege}}{\cc{User}} \ast {} \\
  & \regptsto{\cc{hart\_state}}{\textit{hs}} \ast
    \textit{hs} \in \set{\cc{Active}, \cc{Waiting}} \ast {} \\
  & \regptsto{\cc{mstatus}}{\textit{ms}} \ast
    \textlog{user_mstatus_ok}~\textit{ms} \ast {} \\
  & \pc{\textit{pc}} \ast \textlog{gpr_file}~m \ast {} \\
  & \regptsto{\cc{scause}}{c} \ast
    \regptsto{\cc{stval}}{t} \ast \regptsto{\cc{sepc}}{e} \ast {} \\
  & \textlog{user_pt}~\textit{pt} \ast \textlog{user_cfg}~C \\
& \\
\specline{\textlog{user_trap_frame}~C~\textit{pt} \triangleq
  \exists~\textit{ms}~m~c~t~e.} \\
\quad
  & \regptsto{\cc{cur\_privilege}}{\cc{Supervisor}} \ast {} \\
  & \regptsto{\cc{hart\_state}}{\cc{Active}} \ast {} \\
  & \regptsto{\cc{mstatus}}{\textit{ms}} \ast
    \textlog{trap_mstatus_ok}~\textit{ms} \ast {} \\
  & \pc{C.\cc{stvec}} \ast \textlog{gpr_file}~m \ast {} \\
  & \regptsto{\cc{scause}}{c} \ast
    \regptsto{\cc{stval}}{t} \ast \regptsto{\cc{sepc}}{e} \ast {} \\
  & \textlog{user_pt}~\textit{pt} \ast \textlog{user_cfg}~C \\
& \\
\specline{\textlog{user_inv}~C~\textit{pt} \wand} \\
\specline{(\textlog{user_trap_frame}~C~\textit{pt} \wand \wpcycle) \wand} \\
\specline{\wpcycle}
\end{specblock}
\caption{Simplified spec for arbitrary user-mode execution.  The
  user-mode execution invariant $\textlog{user_inv}$ owns the entire
  machine state a user instruction can touch, with all mutable parts
  existentially quantified; the only way out is a trap that delivers
  $\textlog{user_trap_frame}$ to the kernel's handler at \cc{stvec}.}
\label{fig:user-spec}
\end{figure}

The user-mode execution invariant, $\textlog{user_inv}$, describes what
it means for the system to be correctly executing user-mode code: the
current privilege is \cc{User}; the HART is either active or waiting
(sleeping);
the program counter and the general-purpose registers have arbitrary
contents; and \cc{mstatus} is arbitrary up to a few pinned bits that user
execution cannot change (for instance, \cc{MPRV} and \cc{MXR} are clear,
so the user program's loads and stores are translated as user accesses).
The supervisor trap CSRs (\cc{scause}, \cc{stval}, \cc{sepc}) hold
arbitrary values; they are updated when the processor takes a trap.

The invariant also owns the user page table,
$\textlog{user_pt}~\textit{pt}$: the \cc{satp} register, the
page-table pages that make up the tree rooted at $\textit{pt}.\textit{root}$, and
$\memptsto{\textit{pa}}{b}$ for every byte of every physical page mapped
with the user bit set, with arbitrary contents.  The domain
of this memory map is pinned to exactly the mapped virtual addresses,
so $\textlog{user_pt}$ says ``this is \emph{all} of the user's memory'',
and no other part of the kernel proof can own any of it.  The pure side
conditions, abbreviated $\textlog{pt_wf}$, say that the translation is
injective (no two user virtual addresses map to the same physical byte,
so a store through one address cannot invalidate the points-to for another)
and that the page-table entries are well-formed.  CSL's exclusive
ownership rules mean that the physical pages that are in the user's
page table are disjoint from any other state owned by the kernel.

Finally, the invariant holds the configuration parameters in
$\textlog{user_cfg}~C$: ownership of \cc{stvec} and of the delegation
registers, together with the facts that \cc{stvec} is in direct mode
(so every trap lands exactly at its base address), that every exception
that user code can raise is delegated to supervisor mode, and that no
interrupt is enabled that would be handled in machine mode.  These last
facts are what guarantee that a trap out of user mode lands at the
kernel's handler rather than somewhere the proof knows nothing about.

The specification then says that, given $\textlog{user_inv}$ and
a $\wpcycle$ for the trap handler, the machine may run forever.
The trap-handler $\wpcycle$ receives the $\textlog{user_trap_frame}$
resource, which is what a trap from user mode hands the kernel: privilege
\cc{Supervisor}, the PC at $C.\cc{stvec}$, the trap CSRs freshly written,
and the same page table and configuration as before the trap.  Note that
\cc{scause}, \cc{stval}, and \cc{sepc} are existential in the trap frame,
since the handler must handle every kind of trap; the per-cause lemmas
that produce this frame know their exact values.

The proof itself is an induction over any execution that starts at the user
privilege level.  At each cycle, it considers the possibility that the user
code might trap, jumping to the trap handler (which is proven by using
the WP of the trap handler address), or it might execute arbitrary code.
The theorem considers every possible outcome of executing the RISC-V
fetch phase, followed by an arbitrary decode of the fetched instruction
bytes, followed by an arbitrary execution of the decoded instruction.
For every single one of these cases, the proof concludes that either
$\textlog{user_inv}$ is re-established, and the proof continues by
induction, or the machine has trapped and $\textlog{user_trap_frame}$
holds, and the proof concludes with the trap handler's WP.

One surprise we discovered in the process of doing this user-mode inductive
proof is that user-mode code in RISC-V can put the core to sleep.
At first, the user-mode execution invariant included the fact that the
RISC-V core running the user code is active (i.e., not sleeping).  As it
turns out, there is a RISC-V instruction, \cc{WRS.NTO}, that can put the
CPU to sleep while executing in user mode.  This is safe because
interrupts are still enabled, so the kernel will safely regain execution
at the next timer interrupt.  However, this required us to
extend the user-mode invariant to allow being at user privilege level
while the core is actually sleeping.

\subsection{Visibility of writes under TSO}

A key challenge that TSO introduces is that different cores can see
different results when reading the same memory location, because of
store buffering.  There are several places where this shows up in an
interesting way in the xv6 proof, as we now describe.

One example is shared state protected by a spinlock.  When one core
releases a lock, it passes ownership of the shared state to the
spinlock's lock invariant.  When another core acquires the lock, it
should get ownership of the same lock invariant, but in the proof, the
lock invariant was originally true on the first core, where it might have
referred to writes in that core's store buffer.  The proof of the second
core, however, must now establish that those same writes are visible on
the second core as well.

A second example is the xv6 spinlock implementation: it attempts to guard
against deadlock by checking whether the current CPU already holds the
spinlock before trying to acquire it.  This \cc{holding} check does a
non-atomic read of the spinlock's \cc{lk->cpu} field, which stores a
pointer to the \cc{struct cpu} of the CPU holding this lock, if any.
However, under TSO, the read is not guaranteed to see the latest value
of this field, which makes it difficult for the proof to establish that
this \cc{holding} check will never panic---after all, it might be that,
at some point in the past, this CPU did hold the lock.

A third example is context switching of processes between cores: when
an interrupt comes in, a running kernel thread gets suspended, and
another CPU's scheduler may choose to resume running it on another core.
However, under TSO weak memory, different cores might have different
views of shared memory.  This makes it non-trivial to prove that it is
safe to resume the suspended kernel thread on a different core.

To address this, the xv6 proof introduces a notion of a \emph{view}, which
logically captures the view of memory from the perspective of some core.
Any CSL resources in the xv6 proof that are intended to be shared across
cores are stated as being view-dependent (that is, the resource is a
function of the view).  The specification of a spinlock associates a
logical view with the spinlock itself---this captures the notion that
the lock invariant is true with respect to some kind of logical spinlock
view, rather than the view of any particular core.  The lock invariant
pins down exactly what this view means: it is the view corresponding
to the timestamp of the latest \cc{release} write to the lock word.
This allows the \cc{acquire} proof to establish that, if \cc{acquire}
saw the lock in an unlocked state, its local CPU timestamp must be high
enough to see all of the writes in the lock's view.  This in turn allows
\cc{acquire}'s proof to hand back the lock invariant to the caller's
proof relative to the caller's own view on the local CPU.

Using views, the xv6 proof also introduces a resource to
represent non-sequentially-consistent shared memory locations,
$\ctxvalues{a}{$\xi$}{S}$.  This resource says that, in view $\xi$,
reading address $a$ can result in any value from set $S$.  This is
effectively a statement about a suffix of the history of address $a$
starting from the timestamp associated with view $\xi$.  This allows
us to capture the invariant for the lock's \cc{lk->cpu} field: namely,
$\ctxvalues{\cc{lk->cpu}}{$\xi$}{S}$ where $S$ cannot contain this
CPU.  The view $\xi$ is important for dynamically allocated spinlocks.
In particular, xv6 dynamically allocates pages of memory for pipes, which
contain a spinlock, and the proof must establish that, when acquiring
a pipe spinlock, the \cc{holding} check does not inadvertently see an
old value that happened to be at \cc{lk->cpu} from before this memory
location was reused for a spinlock.

Finally, to reason about context switching of kernel threads, the xv6 proof
introduces the notion of suspending one view under another view.  xv6 has
a kernel thread for every process, and also a kernel thread for every
CPU's scheduler.  The per-process kernel threads migrate between cores,
but the scheduler kernel threads are pinned to their respective cores.
When a process kernel thread context-switches into the scheduler, the
scheduler's proof suspends the process thread's view under its own view.
This means that the set of writes visible to the process thread is a
subset of the writes visible to the scheduler thread.  The scheduler
then releases the lock associated with the process, \cc{p->lock}, which
moves the suspended view from being relative to the scheduler thread's
view to being suspended under the spinlock's view.  When another core
acquires \cc{p->lock} to run that process, the lock invariant gives it
ownership of the suspended process view, now under the new scheduler
thread's view.  Finally, when the new scheduler jumps to the resumed
kernel thread, it un-suspends the process view.

View-relative predicates also show up in other places where resources
move across cores.  For example, in xv6, the first CPU is responsible
for initializing many kernel subsystems, including the memory allocator,
the page table, etc.  The invariants that the first CPU establishes
about these subsystems are themselves view-relative, which means they
might not be visible to other cores yet.  xv6 uses a barrier variable,
\cc{started}, that other cores poll to see if the first core has
finished initializing yet.  The invariant associated with \cc{started}
is $\ctxvalues{\cc{started}}{0}{0,1}$, which means that it can only have
the values 0 and 1, starting from the system boot time, along with a
predicate that says that, if the value is 1, the reader can learn the
view-relative resources initialized by the first CPU.

\subsection{Interrupt exclusivity}
\label{sec:proof:plic}

The UART driver in xv6 relies on the PLIC (platform-level interrupt
controller) to ensure that only one core is handling UART receive
interrupts at any given time, to avoid race conditions in accessing
the UART's receive FIFO.  The PLIC on RISC-V provides a way for a
core to ``claim'' an interrupt, which both returns the IRQ of the
pending interrupt and prevents other cores from claiming it until
released.  The simplified specifications for these PLIC functions in
xv6, \cc{plic_claim} and \cc{plic_complete}, are shown in
\autoref{fig:plic-spec}.  When \cc{plic_claim} returns an interrupt,
it also returns exclusive resources intended for that interrupt's
handler.  In the case of the UART, it is the $\textlog{uart_rx_tok}$
token that allows the proof of the UART receive interrupt handler
to prove that no other
cores are accessing the receive FIFO at the same time.

\begin{figure}[ht]
\centering
\begin{minipage}[t]{0.50\columnwidth}
\begin{specblockt}
\specline{\ktext \wand} \\
\specline{\pc{\cc{plic_claim}} \wand} \\
\specline{\textlog{sie_cap_gpr}~m~n~\textlog{false}~\ldots \wand} \\
\specline{\textlog{plic_inv}~\gamma \wand} \\
(~\forall~m'.~~
   & \pc{m[\cc{ra}]} \wand \\
   & \textlog{sie_cap_gpr}~m'~n~\textlog{false}~\ldots \wand \\
   & \textlog{callee_saved}~m = \textlog{callee_saved}~m' \land \\
   & \quad m'[\cc{a0}] \in \{0, \textlog{UART_IRQ}, \textlog{VIRTIO_IRQ}\} \wand \\
   & (m'[\cc{a0}] = \textlog{UART_IRQ} \wand \exists~k.~\textlog{uart_rx_tok}~\gamma~k) \wand \\
   & \wpcycle ~) \wand \\
\specline{\wpcycle}
\end{specblockt}
\end{minipage}%
\hfill
\begin{minipage}[t]{0.46\columnwidth}
\begin{specblockt}
\specline{\ktext \wand} \\
\specline{\pc{\cc{plic_complete}} \wand} \\
\specline{\textlog{sie_cap_gpr}~m~n~\textlog{false}~\ldots \wand} \\
\specline{\textlog{plic_inv}~\gamma \wand} \\
\specline{m[\cc{a0}] \in \{0, \textlog{UART_IRQ}, \textlog{VIRTIO_IRQ}\} \wand} \\
\specline{(m[\cc{a0}] = \textlog{UART_IRQ} \wand \exists~k.~\textlog{uart_rx_tok}~\gamma~k) \wand} \\
(~\forall~m'.~~
   & \pc{m[\cc{ra}]} \wand \\
   & \textlog{sie_cap_gpr}~m'~n~\textlog{false}~\ldots \wand \\
   & \textlog{callee_saved}~m = \textlog{callee_saved}~m' \wand \\
   & \wpcycle ~) \wand \\
\specline{\wpcycle}
\end{specblockt}
\end{minipage}

\caption{Simplified PLIC specifications for \cc{plic_claim} and \cc{plic_complete}.}
\label{fig:plic-spec}
\end{figure}

The specification allows multiple cores to concurrently invoke
\cc{plic_claim}.  By promising to return the exclusive resources of
the UART for the UART IRQ, the specification promises that it will not
return the UART IRQ to a second core until the first core releases it
using \cc{plic_complete}.  The proof of \cc{plic_claim} works by opening
an invariant at the instant it performs the MMIO access to the PLIC's
claim register, similar to the proof of \cc{acquire}.

\section{Proving user-level code}
\label{sec:user}

The previous section describes the specification and proof of the xv6
kernel itself, including the safety of executing arbitrary user-level code.
This suffices to conclude that user-level programs cannot corrupt the
kernel, and all kernel state will remain intact (e.g., the file system
remains consistent and no invariants are violated).

This section describes how we extend specifications and proofs to allow
proving the correctness of user-level programs running on top of the
xv6 kernel.  For example, we would like to prove our application theorem
from \autoref{sec:theorem}, which implies that if the user types
\cc{echo hello world} on the UART console, then the only response the user
should see is \cc{hello world} printed back to them (or one of the allowed error messages).  Proving such a
theorem requires reasoning about the correctness of multiple user-level
processes, including \cc{init} (which is responsible for setting up the
stdin, stdout, and stderr file descriptors and launching the shell),
\cc{sh} (which is responsible for taking user input from the console,
parsing it, and running commands), and \cc{echo} (which prints its
string arguments from \cc{argv}), as well as precisely reasoning about
the interaction between those processes and the kernel.

This theorem is succinct---it is simply a statement about the sequence
of bytes sent on the UART console---yet it implies that a wide range of
problems cannot occur.  For example, it implies that the file system must
contain the correct binaries for \cc{/init}, \cc{/sh}, \cc{/echo}, etc.,
across any possible sequence of reboots.  It also implies that the kernel
correctly implements all of the system calls, including \cc{fork},
\cc{exec}, \cc{read}, \cc{write}, \cc{wait}, and \cc{open}.  It also
implies that the user-level programs themselves are correct---e.g.,
\cc{init} does not spawn multiple shells that might race for reading
input from the console, and \cc{sh} correctly runs \cc{echo}.  It also
means there are no memory safety violations---the stack of each user-level
process never overflows, taking into account the initial \cc{argv} placed
on the user-level stack by \cc{exec}.  Finally, this also implies that
the kernel correctly executes user-level processes, such as ensuring
that the instruction cache is flushed when a process is migrated from
one core to another.

An important aspect of our approach to proving user-level programs is
that the kernel's specification and proof remain generic.  In order
to prove the correctness of the \cc{echo hello world} application in
our example above, it suffices to prove theorems about the user-level
programs involved in orchestrating this execution (\cc{init}, \cc{sh}, and
\cc{echo}), but the kernel's specifications and proofs can be used as-is.

\subsection{User-mode execution environment}
\label{sec:user-env}

To enable proofs that reason about the execution of user-level code, the
xv6 kernel's specification defines the complete state of a user-level
process, in terms of what the user-level code can observe.  This is
captured by the $\textlog{uvis}$ record as shown in \autoref{fig:uvis}.
All user-level execution, including instructions, memory accesses,
and system calls, is defined in terms of how it accesses and updates
this $\textlog{uvis}$ state.

\begin{figure}[ht]
\centering
\begin{specblockc}
\speclinec{\textlog{uvis} \triangleq \{} \\
\quad & \textit{tf} : \textdom{list}~\textdom{word}; & \text{trapframe: registers and PC to resume at} \\
      & M : \textdom{vaddr} \rightharpoonup \textdom{byte}; & \text{memory image} \\
      & \pi : \textdom{page} \rightharpoonup \textdom{perm}; & \text{per-page R/W/X permissions} \\
      & \textit{sz} : \textdom{Z}; & \text{program break} \\
      & \textit{fds} : \textdom{list}~\textlog{fdstate}; & \text{one state per file descriptor} \\
      & \textit{cwd} : \textdom{inum}; & \text{working directory} \\
      & \textit{pid} : \textdom{Z}; & \text{process ID} \\
      & \textit{gen} : \textdom{gname}; & \text{this process's generation} \\
      & \textit{ch} : \textdom{set}~\textdom{gname} & \text{generations of live children} \\
\} & & \\
\end{specblockc}
\caption{A simplified version of the user-visible state of a process, $\textlog{uvis}$.}
\label{fig:uvis}
\end{figure}

$\textlog{uvis}$ contains the trapframe words, representing the register
contents when the process enters the kernel; the process's memory image
$M$; the per-page permission view $\pi$; the program break $\textit{sz}$;
the state of each file descriptor (closed, or open at a particular
file); the current working directory; etc.  The user-mode process can
observe and modify this state both through explicit system calls (e.g.,
\cc{getpid} returns the process's own PID) and through executing
RISC-V instructions (e.g., executing code at a particular program
counter \cc{pc} depends on whether that page is mapped executable or
not according to the $\pi$ permission table).  The user-visible state
does not explicitly list the page table, or the current CPU on which
the process is executing, because those are transparently managed by the
kernel, and the execution of the user process can be described just in
terms of the $\textlog{uvis}$ state.

\paragraph{Separation logic for user-mode memory.}
Reasoning directly about execution on top of $\textlog{uvis}$ can
be cumbersome because that would require global specs: for example,
every instruction that touches memory would have to say what happens to
every location in memory.  To get the modularity advantages of CSL for
user-level code, we developed a user-level proof library that builds up
separation-logic abstractions on top of the raw $\textlog{uvis}$ state.

In particular, the user-level proof library creates two separation-logic
heaps that represent the raw memory image $M$: one corresponding to the
executable text of the process, and one corresponding to the rest of
the process memory.  Text bytes, $\utext{\gamma}{a}{b}$, are persistent,
because xv6 processes never modify their code; holding one is the right
to fetch an instruction from $a$.  The fact that text pages are read-only
ensures that the HART's icache is consistent with memory, and allows the
proof to conclude that fetching from that address will obtain the correct
instructions (as well as conclude that no page fault will occur due to
a missing page-execute bit).  Data bytes, $\ubyte{\gamma}{a}{b}$, are
exclusive, and holding one is the right to write $a$.  Read-only views
of data bytes, $\ubyteq{\gamma}{a}{b}$, are used for read-only state,
such as static strings.

All user-mode proofs keep around a $\textlog{urun}$ resource, as shown
in \autoref{fig:urun}, that maintains a correspondence between the
underlying $\textlog{uvis}$ state and the separation-logic view of the
state used in the user-level proofs.  The $\textlog{uheap}$ predicate
ties the two separation-logic heaps, $\gamma.t$ and $\gamma.d$, to the
memory image $M$.  The part of the last page of memory above the program
break is owned by $\textlog{urun}$ and not exposed in the data heap.

\begin{figure}[ht]
\centering
\begin{specblockt}
\specline{\textlog{uheap}~\gamma~M~\pi~\textit{sz} \triangleq
  \exists~M_t~M_d~M_{\textit{slack}}.} \\
\quad
  & M_t \subseteq M \ast M_d \subseteq M \ast M_t \mathrel{\#} M_d \ast {} \\
  & \textlog{dom}~M_t \subseteq \textlog{exec_only}~\pi \ast
    \textlog{dom}~M_d \subseteq \textlog{writable}~\pi \ast {} \\
  & \textlog{auth}_{\gamma.\textit{t}}~M_t \ast
    \textlog{auth}_{\gamma.\textit{d}}~M_d \ast
    \textlog{usz}_{\gamma.\textit{s}}~\textit{sz} \ast {} \\
  & M_{\textit{slack}} = \{ a \mapsto b \in M_d \mid a \ge \textit{sz} \} \ast
    \mathop{\scalebox{1.4}{$\ast$}}_{a \mapsto b \in M_{\textit{slack}}} \ubyte{\gamma}{a}{b} \\
& \\
\specline{\textlog{urun}~\gamma~m~\textit{pc}~n \triangleq
  \exists~M~\pi~\textit{sz}~\textit{fds}~\textit{cwd}~\textit{pid}~\textit{gen}~\textit{ch}~\textit{pt}~C.} \\
\quad
  & \textlog{uheap}~\gamma~M~\pi~\textit{sz} \ast
    \ustack{\gamma}{m[\cc{sp}]}{n} \ast {} \\
  & \textlog{auth}_{\gamma.\textit{fd}}~\textit{fds} \ast
    \textlog{auth}_{\gamma.\textit{cwd}}~\textit{cwd} \ast
    \textlog{auth}_{\gamma.\textit{ch}}~\textit{ch} \ast {} \\
  & \textlog{ukernel}~\textit{pt}~C~
    (\textlog{uvis_of}~m~\textit{pc}~M~\pi~\textit{sz}~\textit{fds}~\textit{cwd}~\textit{pid}~\textit{gen}~\textit{ch})
\end{specblockt}
\caption{Simplified definitions of the per-process user heap and
  running-process predicate $\textlog{urun}$.  $\gamma$ is a record of
  the process's ghost names.}
\label{fig:urun}
\end{figure}

The user-level proof library defines specifications for each
RISC-V instruction that user-level code can execute in terms of the
separation-logic state exposed by $\textlog{urun}$.  For example,
the specification for \cc{ADDI} needs only the instruction's text
bytes, $\textlog{uinstr}~\gamma~\textit{pc}~i$ (a persistent fact
derived from $\utext{\gamma}{}{}$), and $\textlog{urun}$, as shown in
\autoref{fig:urun-rules}.  Similarly, the specification for \cc{SD}
additionally consumes and returns the heap points-to resource for the
word being overwritten.

\begin{figure}[ht]
\centering
\begin{minipage}[t]{0.46\columnwidth}
\begin{specblockt}
\specline{\textit{rd} \notin \{\cc{x0}, \cc{sp}\} \ra} \\
\specline{\textlog{uinstr}~\gamma~\textit{pc}~(\cc{ADDI}~\textit{rd},\textit{rs1},\textit{imm}) \wand} \\
\specline{\textlog{urun}~\gamma~m~\textit{pc}~n \wand} \\
(~ & \textlog{urun}~\gamma~m[\textit{rd} := m[\textit{rs1}] + \textit{imm}] \\
   & \phantom{\textlog{urun}~\gamma~} (\textit{pc}+4)~n \wand \\
   & \wpcycle ~) \wand \\
\specline{\wpcycle}
\end{specblockt}
\end{minipage}%
\hfill
\begin{minipage}[t]{0.50\columnwidth}
\begin{specblockt}
\specline{a = m[\textit{rs1}] + \textit{imm} \ra a \bmod 8 = 0 \ra} \\
\specline{\textlog{uinstr}~\gamma~\textit{pc}~(\cc{SD}~\textit{rs2},\textit{imm}(\textit{rs1})) \wand} \\
\specline{\uword{\gamma}{a}{v} \wand} \\
\specline{\textlog{urun}~\gamma~m~\textit{pc}~n \wand} \\
\specline{(~\uword{\gamma}{a}{m[\textit{rs2}]} \wand} \\
\quad & \textlog{urun}~\gamma~m~(\textit{pc}+4)~n \wand \\
      & \wpcycle ~) \wand \\
\specline{\wpcycle}
\end{specblockt}
\end{minipage}
\caption{User-mode instruction rules for \cc{ADDI} (left) and \cc{SD}
  (right), stated on $\textlog{urun}$.  $m$ is the register file of $\textlog{urun}$.}
\label{fig:urun-rules}
\end{figure}

One detail that the \cc{SD} specification abstracts over is the fact
that the xv6 kernel implements lazy allocation of zeroed pages.  The
$\textlog{uvis}$ state represents such pages as containing all zeroes,
even though there might be no physical page allocated for that range
of memory.  When an \cc{SD} instruction stores into a page of memory
(or when an \cc{LD} instruction loads from memory), the process might
take a page fault if the page has not been allocated yet.  The kernel
allocates a fresh page, zero-fills it, installs it into the page table,
and resumes the execution of the process, with the same exact visible
state according to $\textlog{uvis}$.  After the page has been installed,
the \cc{SD} or \cc{LD} instruction can resume execution.

\subsection{System calls}
\label{sec:user-sys}

xv6 processes use the \cc{ECALL} instruction to invoke a system call.
The kernel provides a specification for a trap caused by \cc{ECALL},
in terms of how it affects $\textlog{uvis}$, and the user-level
proof library builds on that specification to provide an equivalent
separation-logic-based spec for \cc{ECALL}.  Instead of specifying the
complete behavior of \cc{ECALL} for all possible system calls that might
be invoked, the user-level proof library provides separate specifications
for each system call number.

\paragraph{File descriptor state.}
The state of each file descriptor is part of $\textlog{uvis}$ as a
list of $\textlog{fdstate}$s, each of which is either $\textlog{Closed}$ or
$\textlog{Open}~r~w~t$, where $r$ and $w$ are the readable and writable
flags and the type $t$ says whether the descriptor is an inode (with
its inode number, and the ghost name $\gamma_o$ of its offset), a pipe end, or a device (such as the UART console).

The user-level proof library introduces separation-logic resources to track
the state of each file descriptor.  There are two kinds of such resources,
to represent the Unix convention that file descriptors 0, 1, and 2 are
stdin, stdout, and stderr respectively.  These three lowest-numbered
file descriptors are represented by $\textlog{ustd}~\gamma~l$, and higher-numbered
file descriptors are represented by $\textlog{ufd}~\gamma~\textit{fd}~\textit{st}$.
A wrapper $\textlog{ufd_own}$ defines a valid file descriptor that is
either low- or high-numbered.

\paragraph{Specifying \cc{dup}.}

\autoref{fig:sys-dup} shows the specification for the \cc{dup}
system call, which clones an existing file descriptor, in terms of
these separation-logic resources.  The rule for \cc{dup} takes the
resource of the file descriptor being cloned, and returns either a
new descriptor allocated at the lowest closed slot with a copy of the
source's \emph{state} (so the program knows the new descriptor has the
same type and mode as the source), or $-1$ with everything unchanged.
The special treatment of low-numbered file descriptors allows proving
\cc{init}'s code that sets up stdin, stdout, and stderr, by calling
\cc{open(...)} to open file descriptor 0, and then calling \cc{dup(0)}
twice to clone it as file descriptors 1 and 2.

\begin{figure}[ht]
\centering
\begin{specblockt}
\specline{\textlog{fdstate} \triangleq \textlog{Closed} \ALT \textlog{Open}~(r~w : \textdom{bool})~(t : \textlog{fdtype})} \\
\specline{\textlog{fdtype} \triangleq \textlog{Inode}~\textit{inum}~\gamma_o \ALT \textlog{Pipe} \ALT \textlog{Device}~\textit{major}} \\
\specline{\textlog{ufd}~\gamma~\textit{fd}~\textit{st} \triangleq
  \textit{fd} \hookrightarrow_{\gamma.\textit{fd}} \textit{st} \ast \textit{st} \ne \textlog{Closed} \ast 3 \le \textit{fd}} \\
\specline{\textlog{ustd}~\gamma~l \triangleq |l| = 3 \ast \mathop{\scalebox{1.4}{$\ast$}}_{k < 3} k \hookrightarrow_{\gamma.\textit{fd}} l[k]} \\
\specline{\textlog{ufd_own}~\gamma~l~\textit{fd}~\textit{st} \triangleq
  (\textit{fd} < 3 \land l[\textit{fd}] = \textit{st}) \lor \textlog{ufd}~\gamma~\textit{fd}~\textit{st}} \\
\specline{\textlog{ualloc}~\gamma~l~\textit{fd}~\textit{st} \triangleq
  \textlog{ustd}~\gamma~(\textlog{fill}~l~\textit{st}) \ast {}} \\
\quad & \textrm{if}~\textlog{lowest_closed}~l = \textlog{Some}~k~\textrm{then}~\textit{fd} = k~\textrm{else}~
  3 \le \textit{fd} \ast \textlog{ufd}~\gamma~\textit{fd}~\textit{st} \\
& \\
\specline{m[\cc{a7}] = \cc{SYS\_dup} \ra m[\cc{a0}] = \textit{fd}_0 \ra \textit{st} \ne \textlog{Closed} \ra} \\
\specline{\textlog{uinstr}~\gamma~\textit{pc}~\cc{ECALL} \wand
  \textlog{urun}~\gamma~m~\textit{pc}~n \wand} \\
\specline{\textlog{ustd}~\gamma~l \wand
  \textlog{ufd_own}~\gamma~l~\textit{fd}_0~\textit{st} \wand} \\
\specline{(~\forall~r.} \\
\quad & \big( (\exists~\textit{fd}_1.~~r = \textit{fd}_1 \ast
      \textlog{ualloc}~\gamma~l~\textit{fd}_1~\textit{st} \ast
      \textlog{ufd_own}~\gamma~(\textlog{fill}~l~\textit{st})~\textit{fd}_0~\textit{st}) \\
      & \;\lor\; (r = -1 \ast \textlog{ustd}~\gamma~l \ast
      \textlog{ufd_own}~\gamma~l~\textit{fd}_0~\textit{st}) \big) \wand \\
      & \textlog{urun}~\gamma~m[\cc{a0} := r]~(\textit{pc}+4)~n \wand \wpcycle ~) \wand \\
\specline{\wpcycle}
\end{specblockt}
\caption{Descriptor resources and the rule for \cc{dup}.  The new
  descriptor's state is a copy of the source's.  $\textlog{fill}~l~\textit{st}$
  writes $\textit{st}$ into the lowest closed slot of $l$, if
  there is one.}
\label{fig:sys-dup}
\end{figure}

\subsection{File system state}
\label{sec:user-fs}

To enable user-level proofs to reason about file system state, the xv6
kernel specification exposes an abstract view of the file system state,
as shown in \autoref{fig:fs-abs}.  This abstract view is used to specify
the state of the file system in memory, as well as the on-disk durable
state of the file system that will be present if the system crashes
and reboots.  This abstract view maps each live inode number to an
$\textlog{anode}$: a file with its byte contents, a directory with its
entries (including \cc{.} and \cc{..}), or a device with its major and
minor numbers, together with the link count.  Unlinked inodes are not
present in this abstract state, since their state can be lost on crash;
the current specification does not support reasoning about programs that
read/write unlinked-but-still-open inodes.  This abstract state is not
a tree because Unix allows hard links: specifying the file system as an
abstract tree would require duplicating the state of each file at each
of the file's linked names, and writing to a file would require finding
and updating all of the links to that file in the tree.

\begin{figure}[ht]
\centering
\begin{specblockt}
\specline{\textlog{absnode} \triangleq
  \textlog{File}~(\textit{bs} : \textdom{list}~\textdom{byte}) \ALT
  \textlog{Dir}~(\textit{ents} : \textdom{fname} \rightharpoonup \textdom{inum}) \ALT
  \textlog{Dev}~(\textit{major}~\textit{minor} : \textdom{Z})} \\
\specline{\textlog{anode} \triangleq \{ \textit{node} : \textlog{absnode};\; \textit{nlink} : \textdom{nat} \}} \\
\specline{\textlog{aview} \triangleq \textdom{inum} \rightharpoonup \textlog{anode}} \\
\end{specblockt}
\caption{The abstract file system state.}
\label{fig:fs-abs}
\end{figure}

Since the file system is shared between all of the processes running on
top of the kernel, concurrent system calls might access or modify the
same file system state.  For example, one process might write data to
the end of a log file, while another process reads the same log file.

To capture a concurrent specification for system calls that access the
file system, the xv6 kernel provides atomic-update-style specifications.
The atomic-update specification exposes the linearization point (or
multiple linearization points, if necessary) of each system call, which
defines the instant at which the effect of the system call can be thought
of as taking place.

\paragraph{Specifying \cc{read}.}

\autoref{fig:sys-read} shows the specification for \cc{read} on a file
descriptor that refers to an inode, in the same style as the \cc{dup}
rule of \autoref{fig:sys-dup}.  The caller owns its descriptor resource
(which says that \textit{fd} is open for reading on inode $i$) and its
destination buffer, and supplies a proof callback,
$\textlog{read_commit}$, written using the $\vs$ notation that
\autoref{sec:csl-intro:inv} used for \cc{AMOSWAP}.  The kernel proof
invokes
the callback at the linearization point of the read, lending it the
current abstract state of the file system \textit{av}, the descriptor's
current offset \textit{off}, and the number of bytes $r$ that the read
is about to deliver.  The callback must hand the file system state back
unchanged, but it gets to learn facts about that state, captured by a
predicate $\Phi$ of the application's choosing.  For example, an
application reading a record out of a database file can choose a $\Phi$
that snapshots the file's contents \textit{bs} at the linearization
point; the postcondition then ties the returned bytes to a slice of
that same \textit{bs}, so the application concludes that it read its
record, and not an interleaving of two concurrent writers.

\begin{figure}[ht]
\centering
\begin{specblockt}
\specline{\textlog{read_commit}~i~\gamma_o~\Phi \triangleq
  \forall~\textit{av}~\textit{off}~r.~~
  \textlog{fs_auth}^{1/2}~\textit{av} \wand
  \textlog{off}_{\gamma_o}^{1/2}~\textit{off} \vs
  \textlog{fs_auth}^{1/2}~\textit{av} \ast
  \textlog{off_ret}~\gamma_o~\textit{off}~r \ast
  \Phi~\textit{av}~\textit{off}~r} \\
\specline{\textlog{off_ret}~\gamma_o~\textit{off}~r \triangleq
  \textlog{off}_{\gamma_o}^{1/2}~\textit{off} \lor
  \textlog{off}_{\gamma_o}^{1/2}~(\textit{off} + r)} \\
& \\
\specline{m[\cc{a7}] = \cc{SYS\_read} \ra m[\cc{a0}] = \textit{fd} \ra
  m[\cc{a1}] = \textit{dst} \ra m[\cc{a2}] = n \ra 0 \le n \ra} \\
\specline{\textlog{uinstr}~\gamma~\textit{pc}~\cc{ECALL} \wand
  \textlog{urun}~\gamma~m~\textit{pc}~k \wand} \\
\specline{\textlog{ufd}~\gamma~\textit{fd}~(\textlog{Open}~\textlog{true}~\_~(\textlog{Inode}~i~\gamma_o)) \wand
  \mathop{\scalebox{1.4}{$\ast$}}_{j < n} \ubyte{\gamma}{\textit{dst}+j}{f~j} \wand
  \textlog{read_commit}~i~\gamma_o~\Phi \wand} \\
\specline{(~\forall~r~g.} \\
\quad & \textlog{ufd}~\gamma~\textit{fd}~(\textlog{Open}~\textlog{true}~\_~(\textlog{Inode}~i~\gamma_o)) \wand
        \mathop{\scalebox{1.4}{$\ast$}}_{j < n} \ubyte{\gamma}{\textit{dst}+j}{g~j} \wand \\
      & \big(~r = -1 ~\lor~ \exists~\textit{av}~\textit{off}~\textit{bs}.~~
        \textit{av}[i].\textit{node} = \textlog{File}~\textit{bs} \land
        r = \min(n, |\textit{bs}| - \textit{off}) \land {} \\
      & \quad\quad (\forall~j < r.~~g~j = \textit{bs}[\textit{off}+j]) \land
        (\forall~j \ge r.~~g~j = f~j)
        \;\ast\; \Phi~\textit{av}~\textit{off}~r ~\big) \wand \\
      & \textlog{urun}~\gamma~m[\cc{a0} := r]~(\textit{pc}+4)~k \wand \wpcycle ~) \wand \\
\specline{\wpcycle}
\end{specblockt}
\caption{Atomic-update specification for \cc{read} on an open, readable
  inode file descriptor.  $\textlog{fs_auth}$ is the abstract file system state of
  \autoref{fig:fs-abs}, and $\textlog{off}_{\gamma_o}$ is the offset of
  the file descriptor; the kernel lends its half of each to the callback.}
\label{fig:sys-read}
\end{figure}

The offset is part of the callback because a file descriptor, and thus
its offset, can be shared by multiple processes that read from it
concurrently.  A process that shares its descriptor hands the offset
back unchanged, and the kernel advances it.  A process that is the only
user of a descriptor can instead own the other half of the offset
ghost state, in which case its callback advances the offset by $r$;
this allows the process to track its position in the file from one
\cc{read} to the next.

When \cc{read} returns, the caller gets its buffer back, with some
prefix modified.  On success, the return value is exactly
$\min(n, |\textit{bs}| - \textit{off})$, the rewritten prefix is the
corresponding slice of the file, and the caller receives its $\Phi$.
On failure (e.g., the kernel could not copy the data into the user's
buffer), the callback is either returned unused or exchanged for a
zero-length receipt, which the figure elides.

The \cc{read} rule has two more \emph{arms}, not shown in \autoref{fig:sys-read},
that can be selected by the caller's
knowledge of its descriptor's state: the console and a pipe end.  They
replace only the callback and the content clause.
For example, for a console file descriptor, the callback is an atomic update against the
console's \emph{input history} (the kernel's read event, written as
a $\vs$ callback) and the content clause says the delivered bytes are
the next segment of that history.

\paragraph{Specifying \cc{open}.}
The specification of \cc{open} must deal with the fact that resolving a
path name is not atomic: the kernel walks the path one directory at a
time, locking each directory only while it looks up the next path
element, so other processes can concurrently create, link, and unlink
entries along the path.  As a result, \cc{open} has not one linearization point
but one per path element, plus a final one at the inode being opened.
\autoref{fig:sys-open} shows the resulting specification, for \cc{open} without the
\cc{O_CREATE} and \cc{O_TRUNC} flags.

\begin{figure}[ht]
\centering
\begin{specblockt}
\specline{\textlog{hop}~P~P_\textit{miss}~k~s \triangleq
  \forall~d~\textit{ents}.~~
  P~k~d \wand \textlog{dir}~d~\textit{ents} \vs
  \textlog{dir}~d~\textit{ents} \ast
  (\textrm{if}~\textit{ents}[s] = \textlog{Some}~d'~\textrm{then}~P~(k+1)~d'~\textrm{else}~P_\textit{miss}~k~d)} \\
\specline{\textlog{walk}~c~P~P_\textit{miss}~\textit{path} \triangleq
  (\textlog{True} \vs P~0~(\textrm{if}~\textit{path}[0] = \cc{'/'}~\textrm{then}~\textlog{ROOT}~\textrm{else}~c)) \ast
  \mathop{\scalebox{1.4}{$\ast$}}_{k < |\textlog{elems}~\textit{path}|}
    \textlog{hop}~P~P_\textit{miss}~k~(\textlog{elems}~\textit{path})[k]} \\
\specline{\textlog{open_commit}~\Phi \triangleq
  \forall~\textit{av}~i.~~
  \textlog{fs_auth}^{1/2}~\textit{av} \vs
  \textlog{fs_auth}^{1/2}~\textit{av} \ast \Phi~\textit{av}~i} \\
& \\
\specline{m[\cc{a7}] = \cc{SYS\_open} \ra m[\cc{a0}] = p \ra
  m[\cc{a1}] = \textit{omode} \ra} \\
\specline{\textlog{uinstr}~\gamma~\textit{pc}~\cc{ECALL} \wand
  \textlog{urun}~\gamma~m~\textit{pc}~n \wand
  \textlog{ustr}~\gamma~p~\textit{path} \wand
  \textlog{ucwd}~\gamma~c \wand
  \textlog{ustd}~\gamma~l \wand} \\
\specline{\textlog{walk}~c~P~P_\textit{miss}~\textit{path} \wand
  \textlog{open_commit}~\Phi \wand} \\
\specline{(~\forall~r.} \\
\quad & \big(~(r = -1 \ast \textlog{ustd}~\gamma~l \ast \ldots) ~\lor~
        \exists~\textit{fd}~\textit{av}~i.~~r = \textit{fd} \ast
        P~|\textlog{elems}~\textit{path}|~i \ast \Phi~\textit{av}~i \ast {} \\
      & \quad\quad \textlog{ualloc}~\gamma~l~\textit{fd}~
        (\textlog{Open}~(\textlog{readable}~\textit{omode})~(\textlog{writable}~\textit{omode})~
         (\textlog{type_of}~i~\textit{av}[i].\textit{node}))~\big) \wand \\
      & \textlog{ustr}~\gamma~p~\textit{path} \wand
        \textlog{ucwd}~\gamma~c \wand
        \textlog{urun}~\gamma~m[\cc{a0} := r]~(\textit{pc}+4)~n \wand \wpcycle ~) \wand \\
\specline{\wpcycle}
\end{specblockt}
\caption{Specification for \cc{open} of an existing file.  $\textlog{dir}~d~\textit{ents}$ is the
  part of the abstract file system state (\autoref{fig:fs-abs}) saying
  that inode $d$ is a directory with entries \textit{ents};
  $\textlog{elems}$ splits a path into its elements;
  $\textlog{ustr}$ and $\textlog{ucwd}$ are the caller's path string and
  current directory; and $\textlog{ualloc}$ is from
  \autoref{fig:sys-dup}.}
\label{fig:sys-open}
\end{figure}

The caller supplies a proof callback for every \emph{hop} of the walk.
The callbacks are tied together by a \emph{cursor} predicate $P$ chosen
by the caller, where $P~k~d$ means that, after $k$ hops, the walk has
reached the inode $d$.  At hop $k$, the kernel
lends the callback the entries \textit{ents} of the directory that the
walk is currently in; the callback hands them back unchanged, and
exchanges its cursor $P~k~d$ for $P~(k+1)~d'$ if the directory maps the
path element $s$ to inode $d'$, or for $P_\textit{miss}~k~d$ if there is
no such entry, in which case \cc{open} will fail.  The walk starts at
the root directory for an absolute path, and at the caller's current
directory otherwise.  Once the walk reaches the final inode, the kernel
proof invokes $\textlog{open_commit}$, which works like the callback for
\cc{read}: it lets the caller learn facts $\Phi$ about the file being
opened, at the instant it is opened.  The callback is quantified over
the inode $i$; the postcondition delivers $\Phi~\textit{av}~i$ at the same
$i$ as the final cursor, which is how the caller relates the two.

On success, the caller gets back the final cursor, its $\Phi$, and a new
file descriptor at the lowest closed slot, as with \cc{dup}; the
descriptor's type is determined by the inode that the walk reached, so
that, for example, a program that opens \cc{/console} learns that its
descriptor refers to the console device.  On failure, the caller gets
back whatever cursor the walk ended with, along with its unused
callbacks, which the figure elides.

The caller's choice of $P$ determines what it learns from the walk.
A caller that knows nothing about the file system can choose a trivial
$P$, and learns only that \cc{open} returned some descriptor or failed.
A caller that holds ghost state pinning down the directories along its
path---as our top-level theorem application does for the root directory,
which contains \cc{/init}, \cc{/sh}, \cc{/echo}, etc.---can choose a $P$
that tracks the exact inode the walk has reached, and that makes $P_\textit{miss}$
contradictory; such a caller learns precisely which inode it opened.
The same path-walk callbacks appear in the specification of every system call that
takes a path name, including \cc{exec}.

\subsection{Processes}

The main system calls in xv6 used to manage processes are \cc{fork},
\cc{exit}, \cc{wait}, and \cc{exec}.  The specifications for
\cc{fork}, \cc{exit}, and \cc{wait}, as provided by the user-level proof
library, are shown in \autoref{fig:sys-fork}.

\begin{figure}[ht]
\centering
\begin{specblockt}
\specline{\textlog{Forkable}~P \ra m[\cc{a7}] = \cc{SYS\_fork} \ra} \\
\specline{\textlog{uinstr}~\gamma~\textit{pc}~\cc{ECALL} \wand
  \textlog{urun}~\gamma~m~\textit{pc}~n \wand
  P~\gamma \wand
  \textlog{ufds}~\gamma~D \wand
  \textlog{uch}~\gamma~S \wand
  R \wand
  \Box (T \wand Q~(-1)) \wand} \\
\specline{(~\forall~r.~~r \ne 0 \ra
  \hfill \textrm{\emph{(parent)}}} \\
\quad & \big( (r = -1 \ast \textlog{uch}~\gamma~S \ast R)
      ~\lor~ \exists~\gamma_c.~~
        \textlog{child_tok}~\gamma_c~r~Q \ast
        \textlog{uch}~\gamma~(S \cup \{\gamma_c\}) \big) \wand \\
      & P~\gamma \wand \textlog{ufds}~\gamma~D \wand
        \textlog{urun}~\gamma~m[\cc{a0} := r]~(\textit{pc}+4)~n \wand \wpcycle ~) \ast {} \\
\specline{(~\forall~\gamma'.~~\gamma'.\textit{exit} = Q \ra
  \hfill \textrm{\emph{(child)}}} \\
\quad & P~\gamma' \wand \textlog{ufds}~\gamma'~D \wand
        \textlog{uch}~\gamma'~\emptyset \wand R \wand
        \textlog{urun}~\gamma'~m[\cc{a0} := 0]~(\textit{pc}+4)~n \wand \wpcycle ~) \wand \\
\specline{\wpcycle} \\
& \\
\specline{m[\cc{a7}] = \cc{SYS\_exit} \ra
  \textlog{uinstr}~\gamma~\textit{pc}~\cc{ECALL} \wand
  \textlog{urun}~\gamma~m~\textit{pc}~n \wand
  \gamma.\textit{exit}~m[\cc{a0}] \wand \wpcycle} \\
& \\
\specline{m[\cc{a7}] = \cc{SYS\_wait} \ra
  \textlog{uinstr}~\gamma~\textit{pc}~\cc{ECALL} \wand
  \textlog{urun}~\gamma~m~\textit{pc}~n \wand
  \textlog{uch}~\gamma~S \wand} \\
\specline{(~\forall~r~S'.~~
  \big( (r = -1 \land S' = S) ~\lor~
  \exists~\gamma_c \in S.~\exists~\textit{xs}.~~S' = S \setminus \{\gamma_c\} \ast
  \textlog{exit_tok}~\gamma_c~r~\textit{xs} \big) \wand} \\
\quad & \textlog{uch}~\gamma~S' \wand
  \textlog{urun}~\gamma~m[\cc{a0} := r]~(\textit{pc}+4)~n \wand \wpcycle ~) \wand \wpcycle \\
& \\
\specline{\textlog{child_tok}~\gamma_c~\textit{pid}~Q \wand
  \textlog{exit_tok}~\gamma_c~\textit{pid}~\textit{xs} \wand \later Q~\textit{xs}}
\end{specblockt}
\caption{Specifications for \cc{fork}, \cc{exit}, and \cc{wait}, and the
  rule for redeeming a child's exit resources.
  $\textlog{ufds}~\gamma~D$ stands for the caller's file descriptor
  resources (\autoref{fig:sys-dup}) and its current directory;
  $\textlog{uch}~\gamma~S$ says that the process's un-waited-for
  children are $S$; and $T$ is the application's taint
  (\autoref{fig:echo-app}).}
\label{fig:sys-fork}
\end{figure}

\paragraph{Specifying \cc{fork}.}

\cc{fork}'s specification captures the fact that, when \cc{fork} succeeds,
it returns twice.  The specification represents this fact by requiring
the caller to supply two continuations, in case \cc{fork} succeeds.
One continuation, representing the parent's execution, continues running
with the same $\gamma$ that invoked \cc{fork} in the first place, and
the other continuation, representing the child's execution, runs with
a new $\gamma'$ representing the child's copy of the parent's process,
including the memory address space, the register set, the file descriptors,
and the program break.  Both continuations receive the caller's
file descriptor resources $\textlog{ufds}$, each at its own $\gamma$,
because the child's descriptor table is a copy of the parent's.
The parent additionally chooses a resource $R$ to hand over to the child
outright (e.g., \cc{init} hands the shell the right to read console
input); if \cc{fork} fails, the parent gets $R$ back.

One challenge with \cc{fork} lies in representing its effects in separation
logic.  For example, consider the shell, which uses \cc{fork} followed
by \cc{exec} to start a new process.  The parent process has a memory
buffer storing the command to be executed.  After \cc{fork} returns to
the parent, the parent process may want to recycle that buffer.  On the
other hand, the proof of the child process needs to know the contents
of that buffer to reason about what happens when it calls \cc{exec}.
Separation logic does not allow the same resource to be owned concurrently
by both the parent and the child, and indeed the parent and child have
separate copies of their memory after \cc{fork} succeeds.

To allow the proof to learn that the initial state of the child process
matches the state of the parent process at the time it called \cc{fork},
the user-level proof library uses a notion of a $\textlog{Forkable}$
predicate.  A predicate $P$ that depends on the $\gamma$ of the
current process is forkable if it can be represented in terms of a
set of memory locations whose contents are preserved across \cc{fork}.
This means that, if $P~\gamma$ holds before calling \cc{fork} in the
parent process, then $P~\gamma$ still holds in the parent after \cc{fork}
returns, and $P~\gamma'$ holds in the child after \cc{fork} returns there.
The definition of $\textlog{Forkable}$ is shown in \autoref{fig:forkable}.
For example, when the proof of the shell invokes the specification of
\cc{fork}, it might choose a predicate $P$ that covers the \cc{argv}
buffers that are about to be passed to \cc{exec}.

\begin{figure}[ht]
\centering
\begin{specblockt}
\specline{\textlog{Forkable}~P \triangleq
  \forall~\gamma.~~P~\gamma \wand
  \exists~F_t~F_p~F.} \\
\quad
  & \mathop{\scalebox{1.4}{$\ast$}}_{a \mapsto b \in F_t} \utext{\gamma}{a}{b} \ast
    \mathop{\scalebox{1.4}{$\ast$}}_{a \mapsto b \in F_p} \ubyteq{\gamma}{a}{b} \ast
    \mathop{\scalebox{1.4}{$\ast$}}_{a \mapsto b \in F} \ubyte{\gamma}{a}{b} \ast {} \\
  & \big( \mathop{\scalebox{1.4}{$\ast$}}_{a \mapsto b \in F} \ubyte{\gamma}{a}{b} \wand P~\gamma \big) \ast {} \\
  & \Box\, \big( \forall~\gamma'.~~
    \mathop{\scalebox{1.4}{$\ast$}}_{a \mapsto b \in F_t} \utext{\gamma'}{a}{b} \wand
    \mathop{\scalebox{1.4}{$\ast$}}_{a \mapsto b \in F_p} \ubyteq{\gamma'}{a}{b} \wand
    \mathop{\scalebox{1.4}{$\ast$}}_{a \mapsto b \in F} \ubyte{\gamma'}{a}{b} \vs P~\gamma' \big)
\end{specblockt}
\caption{The $\textlog{Forkable}$ class: a payload $P$ over the heap's
  ghost names factors through a footprint of text, read-only data, and
  exclusive data bytes, and can be rebuilt at any fresh names $\gamma'$.}
\label{fig:forkable}
\end{figure}

\paragraph{Specifying \cc{exit} and \cc{wait}.}

In separation logic, a running user-level process might own some exclusive
resources.  For example, the user-level shell spawned by \cc{init} might
own the exclusive resource for reading from the UART console input;
this proves that there are no concurrent processes also reading from
the console, and therefore the user's input is not misinterpreted by
the shell.  This resource is initially passed to the shell by \cc{init}'s
proof, when it invokes \cc{fork}; the proof of \cc{init} gets to choose
which of its resources it keeps for itself in the parent process, and which
of its resources it passes to the child.

However, when a child exits, these resources may need to be returned
to the parent.  In the case of xv6, when \cc{init} observes that the
shell exits, it will spawn another shell; to do this, the proof needs
to obtain ownership of the UART console input back from the old shell,
in order to give it to the new shell.

To represent this resource transfer, the specification of processes in xv6
associates an exit predicate, $\gamma.\textit{exit}$, with each process.
This predicate defines what resources must be returned by the process when
it exits.  This predicate is chosen in the parent's proof at the time it
invokes \cc{fork}, and the child process is responsible for ensuring that
it returns these resources whenever it exits, as shown in the precondition
of \cc{exit} in \autoref{fig:sys-fork}.  Since \cc{exit} terminates the
process and never returns, its specification has no continuation.

A process may also be terminated by the kernel, with exit status $-1$,
without ever calling \cc{exit}: because of an exception (such as
an illegal memory access), or because another process killed it using
\cc{kill}.  The child cannot be expected to pay its exit resources in
this case, since it can be terminated at any instruction.  Instead, the
specification makes $\gamma.\textit{exit}$ a function of the exit
status, and \cc{fork} requires the parent to show that the resources for
status $-1$ follow from the application's \emph{taint} $T$, which we
describe in \autoref{sec:user-app} (\autoref{fig:echo-app}).
Informally, $T$ holds only once the application's top-level claim has
been abandoned, at which point nothing more needs to be proven about
the exit resources.  This corresponds to the ``unspecified output''
case of \autoref{fig:traces}, which captures the case of a user
typing in \cc{echo xx > /sh}.  Verified programs, in turn, prove that
they trigger no exceptions, and that they do not kill processes that
do not expect it.

Finally, when the parent calls \cc{wait}, it obtains the resources
associated with the exit predicate of the now-dead child process.  When the
process calls \cc{wait}, it does not know yet which PID will be reported
as exited.  The specification deals with this by having \cc{wait}
return an exit token, $\textlog{exit_tok}$, for a particular child
generation and PID.
The proof of the caller (e.g., \cc{init}) can combine this exit token
together with its knowledge of what the exit predicate of the child
process is, $\textlog{child_tok}$, to obtain ownership of the child's
exit resources $Q$, using the last rule in \autoref{fig:sys-fork}.

One complication is that Unix allows PIDs to be reused over time, so
a PID alone does not identify a child.  The specification instead
identifies each incarnation of a process by a ghost name, its
\emph{generation}: \cc{fork} returns the child's generation $\gamma_c$
in $\textlog{child_tok}$ and adds it to the parent's set of children
$S$, and \cc{wait} reports which generation it reaped, which is what lets the
parent match the exit token against the right $\textlog{child_tok}$.

\paragraph{Specifying \cc{exec}.}

The \cc{exec} system call introduces a new complication: if the system
call succeeds, then it returns to a whole new process image, as defined
by the ELF binary that was executed, and starts executing at the entry
point of that ELF binary.  \autoref{fig:sys-exec} shows how the xv6
specification for \cc{exec} captures this fact.  In particular, there are
two continuations needed by the \cc{exec} specification.  If \cc{exec}
errors out, it returns $-1$ to the caller as usual, and the caller must
supply a continuation for resuming at the next instruction with the
return value being $-1$; this is the explicit continuation in the
specification.  The continuation for the success case is part of the
\emph{exec deposit}, $\textlog{exec_deposit}$, which bundles the one-shot
resources that the caller hands to the kernel: the path-walk callbacks
for resolving the path name of the binary (the same $\textlog{walk}$ as
for \cc{open} in \autoref{fig:sys-open}); the same
$\textlog{open_commit}$ callback, for the kernel's single observation of the file's
contents, which it makes while holding the inode's lock; and the
\emph{slot} premise $\textlog{exec_slot_pre}$, which holds the success
continuation.

\begin{figure}[ht]
\centering
\begin{specblockt}
\specline{m[\cc{a7}] = \cc{SYS\_exec} \ra
  \textlog{uinstr}~\gamma~\textit{pc}~\cc{ECALL} \wand
  \textlog{urun}~\gamma~m~\textit{pc}~n \wand
  \textlog{exec_deposit}~\gamma~m~\textit{pc} \wand} \\
\specline{(~\textlog{urun}~\gamma~m[\cc{a0} := -1]~(\textit{pc}+4)~n \wand \wpcycle ~) \wand \wpcycle} \\
& \\
\specline{\textlog{exec_deposit}~\gamma~m~\textit{pc} \triangleq
  \textlog{walk}~c~P~P_\textit{miss}~\textit{path} \ast
  \textlog{open_commit}~\Phi \ast
  \textlog{exec_slot_pre}~Q~\textit{na}~\textit{argv}~\textit{fds} \ast \ldots} \\
& \\
\specline{\textlog{exec_slot_pre}~Q~\textit{na}~\textit{argv}~\textit{fds} \triangleq
  \forall~f~W'.~~\textlog{observed}~f \wand
  \textlog{loadable}~f \ra {}} \\
\quad & \textlog{image_ok}~f~\textit{na}~\textit{argv}~\textit{fds}~W' \ra
  \textlog{my_pay}~W'.\textit{gen}~Q \wand \textlog{uslot}~W' \\
& \\
\specline{\textlog{uslot}~W \triangleq \forall~\textit{pt}~C.~~\textlog{ukernel}~\textit{pt}~C~W \wand \wpcycle} \\
& \\
\specline{\textlog{image_ok}~f~\textit{na}~\textit{argv}~\textit{fds}~W' \triangleq} \\
\quad
  & W'.\textit{tf}.\cc{pc} = \textlog{entry}~f \land
    W'.\textit{sz} = \textlog{top}~f \land
    W'.\textit{tf}.\cc{sp} = W'.\textit{tf}.\cc{a1} = \textlog{sp_final}~f~\textit{argv} \land
    W'.\textit{tf}.\cc{a0} = \textit{na} \land {} \\
  & \textlog{elf_image}~f \subseteq W'.M \land
    \textlog{args_at}~(\textlog{top}~f)~\textit{na}~\textit{argv}~W'.M \land
    \textlog{stack_zero}~(\textlog{top}~f)~\textit{argv}~W'.M \land {} \\
  & \textlog{perm_ok}~f~W'.\pi \land
    W'.\textit{fds} = \textit{fds}
\end{specblockt}
\caption{Simplified specification for \cc{exec}.  The success case is
  the slot $\textlog{uslot}~W'$ inside $\textlog{exec_slot_pre}$, which is in turn part of
  $\textlog{exec_deposit}$; $\textlog{ukernel}$ is from \autoref{fig:urun}; $\textlog{walk}$ and $\textlog{open_commit}$
  are from \autoref{fig:sys-open}, and $\textlog{observed}~f$ is what they
  deliver when the path resolves to a file with contents $f$.}
\label{fig:sys-exec}
\end{figure}

On the other hand, if \cc{exec} succeeds, the caller is responsible
for supplying a continuation for running in a whole new process state,
as defined by $\textlog{uvis}$: for every file $f$ that the kernel's
observation may deliver, and every process state $W'$ that the kernel may
build from it, the slot premise requires the caller to supply $\textlog{uslot}~W'$:
a $\wpcycle$ for running the new program, given the kernel's resources
for a user process whose visible state is $W'$ (the same
$\textlog{ukernel}$ that underlies $\textlog{urun}$ in \autoref{fig:urun}).
The $\textlog{image_ok}$ predicate
defines the correspondence between the contents of the ELF binary that
was passed to \cc{exec} and the new state of the process (the initial
memory contents, register values, etc.).  The file descriptors in the
new process image are inherited from the previous process image.  To reason about the contents
of the ELF file, the \cc{exec} specification uses an atomic-update-style
pattern much like other file system syscalls that we described earlier.

\section{Specifying user-level programs and shell commands}
\label{sec:user-app}

The previous section describes the specifications of system calls for
xv6, which allow proving theorems about user-level applications such as
\cc{echo}, \cc{cat}, \cc{grep}, etc.  However, we would like to reason
about console-level behavior of a user interacting with a shell, where the
user can invoke combinations of these commands, such as \cc{cat README |
grep xv6 > out.txt}.

\paragraph{Abstract specs for shell programs.}

To avoid having to prove every possible shell pipeline separately,
we write an abstract specification capturing each program's behavior in
terms of what it does to its file descriptors.  In particular, we write an
abstract model of the program's behavior in terms of an interaction tree
(ITree)~\cite{xia:itrees}, and prove that the implementation matches
this specification.  This allows us to abstract over the different
choices for what stdin and stdout might be at runtime, such as the input
or output of a given program being the console, a file, or a pipe.

\paragraph{Sandboxing with \cc{seccomp}.}

To demonstrate that xv6 can provably isolate arbitrary user-level code,
we implemented a simple sandboxing mechanism in xv6, \cc{seccomp},
mirroring the Linux mechanism with the same name.  This \cc{seccomp}
mechanism allows specifying a set of syscalls that are blocked, and
this blocked set is inherited by all child processes.  We implement a
simple wrapper command that allows a user to run arbitrary binaries under
this sandbox; the wrapper blocks the \cc{open}, \cc{link}, \cc{unlink},
\cc{mkdir}, \cc{mknod}, and \cc{kill} system calls.  It allows other
syscalls, including \cc{read} and \cc{write} (so that the program can
interact with open file descriptors inherited from its parent), memory
allocation, \cc{fork} and \cc{exec}, etc.  The blocked system calls,
however, ensure that the sandboxed program cannot modify the contents
of the file system (except through any inherited file descriptors),
or affect other processes running on the kernel.

\paragraph{Application is an invariant.}

To prove a theorem about the behavior of shell commands, we need to give
ownership of different parts of the system (such as exclusive ownership
of console input and output, or knowledge of the contents of the file
system) to different processes over time.
For instance, when \cc{init} calls \cc{exec} on \cc{/sh}, its proof
needs to know that the file \cc{/sh} contains the shell's (verified)
binary, even though any process could modify that file; and when
\cc{echo} writes its arguments to the console, its proof needs to know
that nothing else is writing to the console at the same time.

\sys captures these facts as an \emph{application invariant}: a CSL
invariant, chosen by the application, that holds in addition to the
kernel's own invariants.  \autoref{fig:echo-app} sketches the invariant
for the top-level theorem's application.  It says that the binaries of \cc{/init},
\cc{/sh}, \cc{/echo}, etc., in the file system are the ones from the
initial disk image ($\textlog{pinned}$), and that the bytes sent to the
console so far form a legal transcript for the input received so far
($\textlog{legal}$), in the sense of \autoref{fig:traces}.

The application invariant is tied to the kernel's state by splitting
ownership of that state: the application invariant holds one half of the
authoritative ghost state for the abstract file system and for the
console's history, and the kernel's invariants hold the other half.
The two halves always agree, and updating the state requires both.
As a result, the kernel cannot change the file system or send a byte to
the console without the application invariant being re-established for
the new state, even though nothing in the kernel's own invariants
mentions the application.

\paragraph{Processes maintain the invariant at their system calls.}

The kernel does not know how to re-establish an application's
invariant, so this is the job of the process that issued the system
call.  This is the role of the proof callbacks ($\vs$) in the system
call specifications, such as $\textlog{read_commit}$ in
\autoref{fig:sys-read} and the hops of $\textlog{walk}$ in
\autoref{fig:sys-open}: at the system call's linearization point, the
kernel proof invokes the caller's callback, and the callback opens the
application invariant, combines its half of the state with the half
lent by the kernel's proof, updates the state, and re-establishes the invariant.
For example, the callback that \cc{echo}'s proof passes to \cc{write}
must show that appending \cc{echo}'s output to the console history keeps
the transcript legal.

To prove this, \cc{echo} needs to know that it is its turn to write,
and what it is supposed to print.  This knowledge takes the form of
resources that flow between the process proofs along with control:
\cc{init} passes them to the shell through \cc{fork}, the shell passes
them to \cc{echo}, and \cc{wait} returns them through the exit predicate
(\autoref{fig:sys-fork}).  The same callbacks let a process \emph{use} the
invariant: the callbacks that \cc{init} passes to \cc{exec} for \cc{/sh}
open the invariant to learn that the path resolves to the shell's binary,
which allows \cc{init}'s proof to continue with the proof of the shell as
its continuation.  Each process proof is thus a proof about one binary,
and the application invariant is what connects them.

\paragraph{The invariant implies the trace property.}

The kernel's top-level theorem is parameterized by the application.
To instantiate it, the application supplies three things: a proof that
its invariant holds for the initial disk image (a
computation over the bytes of \cc{fs.img}) and is re-established at
every boot; a proof of the first user process, \cc{init}, against the
system call specifications, which in turn needs proofs of the
programs that it forks and execs; and a proof that the invariant implies
the desired property of the externally visible trace.  The last step is
simple for our top-level theorem's application, because the invariant
directly states that the console output is a
legal transcript for the console input.  The resulting theorem is the
one from \autoref{sec:theorem}: every execution of the machine is safe,
and its trace is one of the traces allowed by \autoref{fig:traces}.

\paragraph{Taint.}

The application invariant cannot hold unconditionally, because
the user at the console may type anything.  If the user types
\cc{echo xx > /sh}, as in \autoref{fig:traces}(d), the shell will
overwrite its own binary, and $\textlog{pinned}$ no longer holds.
The kernel must remain correct when this happens (i.e., the user
should not be able to corrupt the kernel), and indeed the theorem
still promises that the kernel maintains all of its own invariants;
what is lost is only the application's promise about the console output.

\sys represents this using a \emph{taint} $T$: a persistent fact saying
that the console input has departed from the discipline that the
application expects.  The application states every one of its claims as
a \emph{lease}, which is either the claim or the taint.  The invariant
tracks whether the input so far is disciplined using a monotone
counter, which goes from 0 to 1 when the input first leaves the
discipline, and can never go back; $T$ is the knowledge that the counter has
reached 1.  Process proofs case-split on each lease they hold.  In the
untainted case, they have the claim, and proceed as described above.  In
the tainted case, they hold $T$, which trivially re-establishes every
lease in the invariant, so a tainted process has no further application-level
obligations; its proof falls back to the kernel's generic guarantee for
arbitrary user code (\autoref{sec:proof-user-safety}), which is also what covers
any unverified binary that a tainted system might run.  The taint also
covers the kernel terminating a process, which is why it appears in the
specification of \cc{fork} (\autoref{fig:sys-fork}).
Finally, since $T$ contradicts the counter being 0, the invariant implies
that a disciplined input has a legal transcript, which is exactly the
shape of the theorem: the trace property is promised only for inputs
that follow the discipline, and kernel safety is promised always.

\sys also maintains a per-era taint resource that represents the fact
that the user ran \cc{seccomp xx} for some untrusted binary \cc{xx}.
The reason that this taint is necessary is that, once a user runs \cc{xx}
under the seccomp sandbox, that binary could keep a copy of the console
file descriptor until the system reboots.  As a result, any input or
output of the system is now suspect; e.g., even if the user sees what
appears to be a legitimate shell prompt, it might actually be a fake shell
prompt printed by the \cc{xx} program.  Thus, the top-level theorem says
that, after a user runs \cc{seccomp xx}, the output could be anything,
until the machine reboots, at which point it becomes well-defined again.

\begin{figure}[ht]
\centering
\begin{specblockt}
\specline{T \triangleq \textlog{mono_lb}~\gamma_T~1
  \qquad\qquad
  \textlog{lease}~R \triangleq R \lor T} \\
& \\
\specline{\textlog{app_inv} \triangleq
  \knowInv{}{\begin{array}{l}
    \exists~\textit{av}~h.~~
    \textlog{fs_auth}^{1/2}~\textit{av} \ast
    \textlog{cons_auth}^{1/2}~h \ast
    \textlog{mono_auth}~\gamma_T~(\textrm{if}~\textlog{disc}~h~\textrm{then}~0~\textrm{else}~1) \ast {} \\
    \quad \textlog{lease}~\lceil \textlog{pinned}_{\cc{/init},\cc{/sh},\cc{/echo}}~\textit{av} \rceil \ast
    \textlog{lease}~\lceil \textlog{legal}~h \rceil
  \end{array}}} \\
& \\
\specline{\textlog{app_inv} \ast \textlog{cons_auth}^{1/2}~h \vdash
  \lceil \textlog{disc}~h \ra \textlog{legal}~h \rceil}
\end{specblockt}
\caption{A sketch of the top-level theorem's application invariant.
  \textit{av} is the abstract state of the file system
  (\autoref{fig:fs-abs}) and $h$ is the history of console input and
  output; the kernel's invariants hold the other halves of
  $\textlog{fs_auth}$ and $\textlog{cons_auth}$.  $\textlog{disc}$ and
  $\textlog{legal}$ are the allowed inputs and the legal transcripts
  of \autoref{sec:theorem} (\autoref{fig:traces}).  The last line is how the invariant implies
  the trace property.}
\label{fig:echo-app}
\end{figure}

\section{Validating the model}
\label{sec:validate}

To gain confidence in the trusted RISC-V model that underlies \sys,
we developed a set of conformance tests that cross-check the model
against a reference implementation.  In our case, we use three reference
implementations: the QEMU rv64 whole-system emulator, whose \cc{virt}
machine is the hardware platform targeted by xv6; the StarFive
VisionFive~2 RISC-V board (JH7110 SoC), manipulated via JTAG; and
the CVA6 implementation of a RISC-V system, simulated in Verilator.
The goal of the conformance checker is to check that \sys's formal model
captures the possible behaviors of the reference implementations.

Formally stated, the conformance checker's job is to find executions of
the reference implementation and prove that the observed executions are a subset of the
executions allowed by the model as defined in Rocq.
This gives us confidence that our theorem,
which applies to every execution of the xv6 kernel on top of the trusted
model, covers the executions on top of the reference implementation
as well.

One subtle aspect of the correspondence between the reference
implementation and the model is that, in some cases, the model gets
``stuck'', meaning that the semantics of a certain operation is undefined.
For example, the \sys model of the disk does not support certain request
types, which are not used by xv6; the semantics is undefined for those
operations.  The \sys theorems prove that the xv6 kernel never triggers
this undefined behavior.  Thus, for the purpose of the conformance checker,
a test case on which the model gets stuck (i.e., has no transition at all)
counts as passing: it is treated as a valid
correspondence between the reference implementation and the model.

Conformance checking is purely a property of the model itself, and does
not involve any of the CSL machinery used to prove correctness of
the concurrently executing xv6 kernel.  This makes the conformance checker's job easier: in many
ways, it suffices to simply let the model execute for many cycles, compute
the result of that execution, and check whether it matches the reference output.

There are two key challenges in conformance checking.  The first is to
generate enough test cases to cover all of the interesting aspects of
the system's execution.  The second is to deal with non-determinism,
which has two facets.  On the reference implementation side,
if some test case produces multiple different outputs, due to some
non-determinism, it is the job of the conformance checker to generate
as many of those outputs as possible.  On the model (Rocq) side,
proving that a certain output can be observed in the model may require
a specific choice of non-determinism, such as uninitialized initial
register contents or interleaving between concurrent CPUs and devices;
it is then the conformance proof's job to synthesize this schedule of
non-deterministic events to demonstrate that the reference output can
be reproduced by the model.

We implement the test cases as short snippets of RISC-V assembly
instructions.  The output of a test case is the contents of a 4~KB page of
memory; the particular test case chooses what relevant data it wants
to include in that 4~KB page.  The conformance checker must prove that
every 4~KB page snapshot observed in the reference implementation
(QEMU, JH7110, or CVA6) can also be reproduced on top of the model.

Our conformance tests include tests for basic CPU operation: simple
instructions, contents of the registers at system boot time, and
page-table and interrupt machinery.  These tests are largely deterministic.

The conformance tests also include tests for shared-memory behavior.
In order to exhibit memory access races, the
test first runs a setup phase, which waits for all of the CPUs to boot,
followed by a single barrier that releases all of the CPUs into the
second phase, which tries to execute a racing shared-memory access.
This generates multiple possible results (4~KB page snapshots) for a
single test, reflecting multiple possible race outcomes.  On the model
side, for each observed result, the proof constructs an explicit schedule
of steps that leads to it.

The tests also include test cases for devices, including the UART,
the PLIC, and the virtio disk, as well as the disk's DMA machinery
(the UART and disk tests run only on QEMU, since the JH7110 and CVA6
have a different UART and no virtio disk).  The RISC-V
assembly instructions directly access the device MMIO registers, set up
shared DMA ring buffers, etc.

\section{Agent-based verification}
\label{sec:agent}

At a high level, using agents for this verification project looks just
like writing code with an agent.  We have many Claude Code agents running
in parallel, each with its own directory that contains a checkout of
the project's git repository.  The agent changes files, runs \cc{make}
to check whether the proofs are correct, modifies specs, proofs,
and any other files in the directory, commits, pushes, resolves merge
conflicts, etc.  The repository contains Rocq files;
\cc{make} runs Rocq to check that the proofs are all valid, and we have
a CI job that continuously makes sure that the proofs check out (and
reports the axioms that the top-level theorem depends on).

As with other agent-based software development efforts, we find it is
useful to keep design notes, optimization techniques, ongoing and
completed projects, failed approaches that should not be tried again,
etc., as documentation (in Markdown format) in the git repository itself.
This keeps the agents' memory in the git repository, shared between all
of the agents working on the project, as opposed to in per-agent local
files on the agent's machine.

The project's proof development is large (\nTotalMLines million lines of
Rocq), but most of it consists of proofs, which are, at some level,
irrelevant as long as they are validated by Rocq and prove the desired
top-level theorem.  The one caveat is performance: we continuously
keep an eye on proof-checking times because proofs that take a long
time to check directly impact the iteration cycle needed to make
further progress, as each rebuild to check proof validity gets more
expensive.

Even the specifications are, for the most part, not fully trusted.  They are
important to get right, in order to ensure that different parts of the
proof will compose correctly with one another.  However, the top-level
adequacy theorem cancels out all of the intermediate specs and proves
a much simpler theorem.  This ensures that, if the agents do not get the
individual function specs quite right, in a way that creates conflicts
between different kernel subsystems, this will be caught because, when \cc{make}
runs Rocq to check the top-level adequacy theorem, Rocq will reject
the proof.  This also allows us to be less diligent about auditing the
intermediate specs, as long as the top-level theorem can still be proven.

\subsection{Abstractions}

Agents, such as Claude Code, are effective at automating proofs,
but one thing that we found agents are not at all good at
is introducing appropriate abstractions.  For example, when we were
verifying the file system, the agent's specifications and proofs about
all file system functions involved direct reasoning about the contents
of on-disk blocks.  As a result, the specifications were extremely
verbose and fragile.  We forced the agent to introduce abstract state
capturing the notion of an inode, a directory, and a file system tree.
This greatly simplified the specifications, made them more robust to
low-level changes, and allowed the agent to finish the proofs,
whereas before this abstraction, it could not scale to the level of
complexity generated by the explicit statements about all on-disk bits.
We had many other examples of this failure mode, including abstractions
for reasoning about the kernel's instructions living in physical memory,
abstractions for page-table translation and TLB consistency, abstractions
for a kernel thread running with its own kernel stack, abstractions for
ABI calling conventions, abstractions for interrupt handling, etc.

Another weakness of current agents is that they are not good at
reasoning about concurrency.  In particular, setting up ghost state,
invariants, and ownership disciplines in terms of which functions own
which separation-logic facts requires manual effort.  On the other hand,
once a function has a clear separation-logic spec, and clear
invariants owning separation-logic facts about shared resources, agents
are effective at doing the actual proofs.

\subsection{Iterating on the design}

Several times, we asked an agent to make a substantial change to the specs
and proofs, and the agent was ultimately unable to complete the task.
The failure mode, in each case, was that the agent entered an infinite
spiral, continuing to introduce more lemmas, theorems, specs, case
analyses, etc., without really making progress on the overall problem.
For example, when we asked the agent to prove the safety of executing
user-mode code, it started formulating abstractions and theorems about
all the possible instructions that a user process might execute, all the
possible memory contents that the process might have, etc., all without
a clear end in sight.  When we asked the agent to
add the notion of a current HART ID to every WP, to allow talking about
process migration across cores, the agent devolved into rewriting every
specification and proof, without converging.
When we asked the agent to prove \cc{ialloc}, it went as far as modifying the xv6 code and adding a new
axiom to paper over a bug (that it did not spot), rather than stepping back.
When we asked an agent to introduce a layer of abstraction for virtual
memory translation backed by the kernel's real page table (as opposed
to the identity mapping), the agent started extending every single
specification to talk about the bit-level contents of page-table entries
corresponding to every address mentioned in that spec.

A final example of this kind of failure showed up when we were proving
file system consistency across reboots.  Agents initially came up with
internal invariants that were sufficient to prove kernel safety (bounds
on link counts, etc.), but not sufficient to connect on-disk block-level
state to abstract state like files and directories.  As a result, all
of the specs leaked the low-level byte details, and then committing
a transaction required reasoning about the byte-level effects of the
transaction's writes, and whether the file system invariant would still
hold if log recovery ran after a crash with those byte-level writes.
Every syscall had to prove that, if log recovery ran on its changes,
the effect would still satisfy the global FS consistency invariant.
We had to intervene and force two changes to the specs: (1) a hard
layer boundary at the write-ahead log, proving an atomic update from old
logical disk state to new logical disk state, and avoiding any reasoning
about individual writes or log recovery, and (2) invariants that connect
block-level state to abstract file/directory state, rather than invariants
that define inequalities to avoid panics.

The common theme in all of these failure modes is that we did not fully
understand the problem that we were asking the agent to solve.  In each
case, after observing the agent entering this spiral for several days,
we better understood the problem space, including the challenges that
we had to address in formulating our goal for the agent.  The solution
in each case turned out to be stopping the agent, discarding its work,
and starting over with a more precise goal statement for the agent,
informed by the previous agent's failures.  In three cases, it took
us three iterations to finally reach a workable solution (the kernel
page-table support, the user-space execution theorem, and file system
crash consistency), because these cases turned out to require a lot of
attention to specific details that were not obvious to us from the outset.

We also found it helpful to use agents to help us fine-tune specifications.
When we had a rough specification that we were not sure about, we asked
the agent to go ahead and try to prove it, knowing full well that it was
likely to fail.  However, since the cost of synthesizing the proof using an agent
is relatively low, this was a highly effective way for us to learn where
the remaining tricky cases were in our spec; having a fully worked-out proof
that fails at exactly those tricky cases was helpful feedback for us to
improve the spec.

\subsection{Multi-agent design review}
\label{sec:agent:review}

In several cases, we found it effective to ask the agent to write a design
document describing its proposed plan for a complex change, and have
other agents explicitly review that design, cross-reference it with the
actual code, and sketch out the core proof argument in Rocq to get the
syntax correct and type-checked (without fully proving it).

This came up when an agent was porting particularly tricky proofs (the
buffer cache and inode cache) from a sequentially consistent memory model
to the TSO weak-memory model.  The agent was in an endless spiral for
about two days, continuously finding problems with the design, revising
the design, finding more problems, etc.  Several times, we asked the
agent to take a step back and re-think the design, but that did not help.

What turned out to work well was to ask a second agent to take a look
at all of the design documents written by the first agent, analyze its
multiple attempts at solving the problem, and come up with a plan to avoid
the pitfalls that led to the first agent's spiral.  This quickly surfaced
a number of deep issues that made the first agent's design unworkable.
A third agent, acting as a second reviewer, spotted further issues that
the first reviewer missed on its own.
We handed this new design document back and forth
between the implementation agent and the two reviewers about ten times,
in the span of a few hours; each iteration spotted fresh issues that none
of the agents found by themselves.  Along with the design document, we
asked the agents to develop (and iterate on) a prototype of the overall
proof plan, in Rocq, containing the relevant definitions, specifications,
and invariants, with the actual proof part assumed for the time being.
Afterwards, the implementation agent completed the task (finishing the
buffer-cache proofs) in a single pass of about two hours.  A similar
approach worked for completing the inode-cache proofs, albeit with more
review round-trips, owing to the higher complexity of the inode cache.

\subsection{Iterative discovery of specifications is slow}
\label{sec:agent:iter}

The part of the overall proof development process that takes the longest
is making changes to existing specifications---e.g., strengthening an
existing syscall specification to promise more facts, changing a kernel
invariant to track more information about the state of a process, etc.
These changes take a long time because they often require modifying many
existing files that touch the specs or invariants that are being updated.

As a result, one style of proof development that is particularly slow
is iterative specification discovery, where the complete specification
or invariant is not precisely known ahead of time.  In this setting,
the proof developer, together with the agent, repeatedly tries to prove
some theorem, fails because the existing invariants or specifications
are not strong enough, revises the specs and invariants, and tries again.
This takes a long time because the issues are discovered one at a time,
and each iteration through the cycle takes a long time.

One example of this showed up when we were trying to prove the user-level
application property for the \cc{echo hello world} application.  When we
started doing the proof, our syscall specifications had many limitations,
where their statements were not quite strong enough to allow our proof to
conclude that \cc{echo hello world} typed at the UART console input will
indeed produce the correct output.  We made many iterations, together
with the agent, in revising the specs, with each iteration taking on
the order of an hour.

Multi-agent design review, described in \autoref{sec:agent:review}, is
helpful from this perspective, because it cuts down this slow iterative
cycle, by finding potential specification issues faster (without having
to go through a complete hour-long iteration), and finding many potential
issues in one go (allowing them to be addressed all in a single iteration,
instead of one at a time).

\subsection{Agent memory is sticky and implicit}

We have found that, while agents run, they make implicit design decisions
and assumptions about what the goal is, what the proof should be like,
what constraints exist on the specs and proofs they are working on, what
the implementation intends to do, etc.  It is infeasible
to fully review the agent's internal assumptions, especially when using
a coordinator agent to orchestrate many sub-agents.  One failure mode that
results from this is that, once an agent (or an agent with its
sub-agents) makes some bad implicit assumption, it is hard to get the
agents to complete the task at hand, and it is often unclear why the
agents are unable to finish it.

To address this problem, we find it is useful to start agents from scratch.
When an agent reaches a reasonable checkpoint---either finishing its
work, or reaching some consistent unfinished state where it seems to
be unable to make progress---we ask it to checkpoint its state into a
project status document, and start the next agent fresh.  The next agent
reads the project status, along with general design documents and notes in
the git repository, and tries to solve the problem with a fresh context.
We review the project status to understand what implicit assumptions the
agent might be making, and we also ask the fresh agent to explain
to us what the status of the project is, what the issues are, and what
its plan of attack is.  This often helps uncover implicit assumptions
that are preventing the agents from finishing certain proofs.

One place where we have found it hard to ``clear'' the agent's thinking is
in the code that is already committed in the git repository.  In particular,
we found it is much easier to get agents to write new specs and proofs
than to refactor or revise existing ones, and when agents are working
with existing specs and proofs, it requires effort to get them to adopt
a new design.  Writing new specs and proofs tends to work well without
much supervision; minimal prompting suffices.  On the other hand,
performing some refactor, even if relatively simple, requires careful
prompting and supervision to ensure the agent does the right refactor,
does not ``reinvent the wheel'' along the way, does not fall back to
the original design, etc.

Finally, we have sometimes discovered that agents already ``knew'' of a
substantial issue but failed to tell us about it until we explicitly asked.
For example, towards the end of finishing up the entire kernel proof, we
asked the agent whether all of the in-flight projects had been completed
(expecting the answer to be ``yes''), and to move those completed project
documents to an archived directory.  However, the agent came back telling
us that, no, one of the projects was clearly documented as not being
completed---file system crash safety.  Indeed, it turned out that the
project document for the file system crash safety effort said, as one
of its many bullet points, that the crash safety theorem was vacuous
(the theorem assumed that every time the system rebooted, it started
with a fresh copy of the \cc{fs.img} image from \cc{mkfs}, rather than
the state of the disk at the moment of the crash).  This was written in
the design document, but the design document had a long list of irrelevant
low-level details, so we did not read it in full before the agent surfaced
this issue when forced to decide whether the project was done.

\subsection{File organization discipline}

We find that it is highly effective to have a strict organizational structure
for the files in our project's git repository.  In particular, for every kernel
function \cc{F}, we have:

\begin{itemize}
\item a \cc{SpecF.v} file that states the function's specification.

\item a \cc{CodeF.v} file that captures low-level facts about the
RISC-V instructions that comprise the function itself (this is largely
auto-generated by a Python script, written with the help of an agent, but
the resulting \cc{CodeF.v} file is still checked by Rocq just like every other
file).

\item a \cc{ProofF.v} file that proves that the implementation of the
function, as captured by \cc{CodeF.v}, meets the precise spec as stated
in \cc{SpecF.v}.

\item a \cc{LinkF.v} file that links the proof of \cc{F}, from \cc{ProofF.v},
with proofs of any other functions that \cc{F} calls; this ensures that all
of the functions agree on the specs for all other functions.

\end{itemize}

The reason why we have link files is that our specification of a function
\cc{F} can be dependent on other specifications---namely, the specs of
all of the other functions that \cc{F} invokes.  Formally, this is stated
as a premise of the theorem about \cc{F}'s execution---to conclude that
\cc{F} executes correctly, the caller of the theorem has to first prove
the correctness of all the functions that \cc{F} might call, with the
expected specifications for those functions.  A link file instantiates
the theorem about \cc{F}'s correctness by filling in these premises, using
proven theorems about those called functions.  The linking ultimately
converges to a proof for the system initialization code that lives at
address \cc{0x80000000}, where the CPU starts executing at power-on.

In addition to having this family of files for every function, we also
have well-defined files for each level of abstraction in the kernel.
For example, we have a file defining the abstractions for disk blocks,
for inodes, etc., and also files defining the invariants that hold on
top of those abstractions.

The most substantial benefit of this structure is that it makes it
clear to agents which files they are expected to modify, and which
files they are expected to keep unchanged.  While Rocq is ultimately
responsible for checking that all of the proofs are correct, we find
that this code structure helps guide agents in the correct direction.
As a counterexample, early on in this project, we had a single file that
would define the spec for a function, the abstractions it provides,
the proof of its correctness, its invariants, and its dependencies on
other functions.  Agents would often get confused as to the rules we were
expecting them to follow, and would change the spec of a function to make
the proof easier, re-define the abstractions to make postconditions less
useful (or even vacuous), etc.

A second benefit of this file structure is that it enables substantial
parallelism in terms of build times.  By having each function proof
be independent of any other function proof, \cc{make} can run Rocq in
parallel on all function proofs at the same time; these proofs comprise
the bulk of the CPU time required to do a build.

A third benefit of this approach is that it allows for cleaner separation
of changes between concurrent agents.  When agents are working on different
files, they can first reach agreement on what the specs are going to be at
the boundary between them, and then independently work on the proof files
for their functions without conflicting with any changes from other agents.

\subsection{Performance optimization}

Fast proof checking enables agents to iterate more quickly when developing
proofs.  To ensure that our proofs have reasonably fast proof-checking
times, we continuously prompt an agent to improve performance, profile the
proofs to find expensive statements, optimize, etc.  We find this to be
effective: this approach finds many performance pitfalls, and fixes them.

One class of optimizations made possible by the fact that we use agents is
writing highly specialized proofs that do not rely on proof automation.
This trades away proof automation that would be important for human
proof engineers (e.g., a single \cc{iFrame} tactic that does a lot
of search to do a big proof step, or a single \cc{set_solver} tactic
that solves a large family of set-theoretic goals) in favor of
special-case proofs at each point (spelling out exactly how to prove
resource ownership instead of \cc{iFrame}, or spelling out exactly the
proof argument for a set-theoretic goal without using \cc{set_solver}), which
improves performance of proof builds.

Occasionally, when we asked the agent to do performance optimization,
it uncovered new avenues for optimizing proof performance---in other
words, the agent's behavior was non-deterministic in terms of what the
agent would find each time, and it was worthwhile to keep asking it to
do the same task multiple times.  For example, at one point it decided
to profile the OCaml implementation of Rocq itself, found some sources
of quadratic performance blow-up, and figured out a way to avoid them by
being careful about what tactics it used in its proofs.

Some Rocq proof files took several minutes to compile, and iteratively
fixing up those proof files through an edit-recompile cycle took a long
time, especially when the broken part of the proof was towards the end
of the file.  The agents were not familiar with the interactive style
of using Rocq through \cc{rocq repl}, so instead we developed a tool
that cached the state of \cc{rocq repl} for a large proof file, and
incrementally re-checked a modified file by rewinding the \cc{rocq repl}
process to the common prefix of the old and new files, and replaying
new proof commands from that point~\cite{rocq-warm}.

\subsection{Orchestration for large projects}

When undertaking a large change that requires significant design effort,
we use highly capable models (such as Fable in the case of Claude Code)
to drive the top-level agent's loop, and we ask the agent to start
sub-agents to do specific well-scoped tasks (e.g., using Opus or Sonnet).
This is analogous to how agent-based software development workflows use agent
orchestration, and works well for proofs too.  We find that highly capable
top-level agents do a good job of ensuring that the specs are meaningful,
and that the overall project will converge, whereas agents running less
capable models like Opus and Sonnet are adequate for executing individual proof
tasks when they do not have to design the overall proof plan.

On the other hand, for smaller-scale tasks that do not require high-level
design decisions, we use Opus or even Sonnet agents directly.  For
example, most of the performance profiling, as described above, was done
by Opus and Sonnet agents, especially once the optimization techniques
had been discovered and described in the shared notes in the repo, so
that the agents simply needed to match the performance problem observed in
some proof against the well-known optimization techniques.

\subsection{Kernel changes}

One challenge of verifying the xv6 kernel at the level of its ELF binary
is that any change to the xv6 source code changes the entire ELF binary.
The compiler might choose different instructions, different relative
offsets, the symbol addresses might move around, etc.  This is one
aspect of the proof where agents make the approach viable: it is unlikely
that human proof engineers would be interested in fixing up all of the
proofs after every recompile.  On the other hand, we have found that,
each time we made a change to the xv6 source code and recompiled
it, an agent was able to fix up all of the proofs within several hours.
Throughout these iterations, we asked the agent to develop tools to make
future iterations cheap, such as: making as many address references as
possible symbolic, referring to symbols from the ELF symbol table;
developing scripts to automate generating the \cc{CodeF.v} files from
the ELF dump as opposed to writing them by hand; developing scripts to
automate replacing PC-relative offsets in instructions throughout the
proofs, instead of having an agent reason about each broken proof step
iteratively; etc.

\section{Bugs found}
\label{sec:bugs}

In the process of specifying and proving the xv6 kernel, we found 13 bugs,
summarized in \autoref{fig:bugs}: one in the Sail RISC-V model that we use
as our specification of the hardware, one in Sail's Lean extraction of
that model, 10 in the xv6 kernel itself, and one in the \cc{grep} user
program.
The rest of this section describes each of them in turn.

\begin{figure*}[t]
\centering
\small
\setlength{\tabcolsep}{4pt}
\newcommand{\bugcol}{\raggedright\arraybackslash}
\begin{tabular}{>{\bugcol}p{0.44\textwidth}>{\bugcol}p{0.11\textwidth}>{\bugcol}p{0.4\textwidth}}
\toprule
\textbf{Bug} & \textbf{Component} & \textbf{Consequence} \\
\midrule
A/D writeback was a blind write & Sail model & One CPU could resuscitate a page-table entry that another CPU cleared \\
\addlinespace[4pt]
Incorrect short-circuiting boolean semantics & Sail extraction & Wrong semantics of boolean operators in the Lean version of the CPU model \\
\addlinespace[4pt]
\cc{iput} freed inode before dropping \cc{struct inode} reference & fs & Racing \cc{iput} and \cc{ialloc} could orphan an on-disk inode until the next reboot \\
\addlinespace[4pt]
Creating files in disconnected directories & fs & \cc{..} could name a free inode, and the kernel panicked because its type was not \cc{T_DIR} \\
\addlinespace[4pt]
Overflowing \cc{nlink} & fs & Too many links to one inode overflowed \cc{nlink}; unlinking it panicked \\
\addlinespace[4pt]
Scheduler did not reset \cc{push_off}'s \cc{intena} flag & scheduler & Scheduler re-enabled interrupts, and could miss a wakeup between scanning the process table and \cc{WFI} \\
\addlinespace[4pt]
Partial \cc{writei} was not logged & fs & Partially written block was never logged, so reads changed once the buffer cache recycled it \\
\addlinespace[4pt]
Missing icache fence when running user code & exec, trampoline & Stale instruction cache entries from an earlier process could be executed after \cc{exec} \\
\addlinespace[4pt]
\cc{freeproc} updated \cc{p->parent} without \cc{wait_lock} & proc & \cc{p->parent} was written without the lock that protects it \\
\addlinespace[4pt]
Inconsistent UART TX FIFO handling & uart & \cc{printk} racing with console writes could overflow the TX FIFO \\
\addlinespace[4pt]
Boot code did not enable \cc{ADUE} & boot & On hardware that follows the spec, the kernel did not handle the A/D traps it then received \\
\addlinespace[4pt]
\cc{kill(0)} marked an unused \cc{struct proc} as killed & proc & A future process was killed immediately after \cc{fork} \\
\addlinespace[4pt]
\cc{grep} stopped at a line longer than its buffer & user (\cc{grep}) & A line of 1023 or more bytes silently ended the search \\
\bottomrule
\end{tabular}
\caption{Bugs found during verification.}
\label{fig:bugs}
\end{figure*}

\paragraph{Sail model bug: A/D writeback.}

We discovered a bug in the Sail RISC-V model itself, when we could not
fully prove the correctness of our page-table and TLB invariant.  The bug
comes down to the hardware model's implementation of the writeback of A/D
(accessed/dirty) bits into the page-table entry.  When the Sail model
decided to write the A/D bits back into the actual page table, it
failed to check whether the page-table entry was still valid, because
it did a blind write rather than an atomic read-modify-write operation.
This meant that one CPU could have cleared a page-table entry (e.g., in
order to free a page of memory), and then another CPU would inadvertently
``resuscitate'' this mapping by doing a blind write to the PTE as part
of the A/D writeback.  We submitted a fix for this issue to the Sail
RISC-V model developers, who agreed that this is a real issue.  The fix
involves doing an atomic read-modify-write to perform the A/D writeback.

\paragraph{Sail Lean extraction bug: Short-circuiting semantics.}

When we ported our proofs from Rocq to Lean, the agent discovered that
the Lean extraction of the Sail RISC-V model incorrectly implements the
semantics of boolean operators.  In particular, Sail defines the semantics
of $\textlog{and}$ and $\textlog{or}$ operators as short-circuiting:
e.g., when the model says $A \mathbin{\textlog{and}} B$, if $A$ is false, then
it does not run $B$.  However, the Lean extraction of the Sail model
implemented eager evaluation of both sides of the boolean operators:
the Lean representation first evaluated both $A$ and $B$, together
with their side-effects, and then chose the resulting boolean value.
As a consequence, if $B$ had side-effects when $A$ returned false, $A
\mathbin{\textlog{and}} B$ would be executed incorrectly in the Lean version of
the Sail model.  We fixed this by changing the Sail tool to generate
correct Lean code, so that the expected semantics of boolean operators
are faithfully represented.

\paragraph{Kernel bug: \cc{iput} could lose an inode.}

The kernel had a bug in reclaiming inodes once the inode's refcount
reached zero; \autoref{fig:iput} shows the code.  When dropping the last refcount (from 1 to 0) on an
unlinked inode, the \cc{iput} function first marked the on-disk inode
as free while holding the lock on the \cc{struct inode} (\cc{ip->lock}),
then released
that lock, and only then acquired another lock (\cc{itable.lock})
to actually drop the
refcount on the \cc{struct inode}.  The problem is that,
between releasing \cc{ip->lock} and acquiring \cc{itable.lock},
the inode number was already available for allocation, while the
\cc{struct inode}, representing an inode cache entry, still carried
\cc{iput}'s reference.  Thus, a concurrent file creation could
allocate the same inode number and obtain a second reference to the
same \cc{struct inode}.

\begin{figure}[ht]
\newcommand{\iputrace}[1]{\textcolor{red}{\textup{\textbf{#1}}}}
\begin{minipage}[t]{0.49\textwidth}
\input{code/iput-bug.c}
\end{minipage}%
\hfill
\begin{minipage}[t]{0.49\textwidth}
\input{code/iput-fix.c}
\end{minipage}
\caption{Simplified code of \cc{iput} in xv6 with the bug (left),
  showing where a concurrent file create and unlink by another process
  P2 can interleave, and after the fix (right).  In the fixed code,
  \cc{ifree} marks the on-disk inode \cc{inum} as free by setting
  its type to 0.}
\label{fig:iput}
\end{figure}

This bug manifested itself as a race.  Consider a process P1 unlinking a
file: its \cc{iput} sets the inode's type to 0 (marking the inode as free),
but P1's \cc{iput} still holds its reference to \cc{ip}, and has yet to decrement \cc{ip->ref}.  Before
the decrement, there is a short period of time during which P1 holds
no locks (between releasing \cc{ip->lock} and acquiring
\cc{itable.lock}).
Thus, some other process P2 creating a new file may allocate the inode (since its type is
0), setting the type to some non-zero value (e.g., \cc{T_FILE} for a regular
file).  P2's \cc{iget} in \cc{ialloc} finds the same \cc{struct inode} in the inode cache, since P1's \cc{iput} has not
yet decremented \cc{ip->ref}, and \cc{ip->ref} becomes 2.  Now P2 may \cc{unlink} the file;
its \cc{iput} decrements \cc{ip->ref} from 2 to 1, but it will not free the on-disk
inode because it is not dropping the last reference.  Then, P1's \cc{iput} finally decrements \cc{ip->ref}
to 0.  At this point, the on-disk inode has \cc{type} $\ne 0$ and \cc{nlink} $= 0$.
The inode is orphaned without a crash.  A reboot will fix it through
\cc{ireclaim}, but before then the inode is lost.

The fix, shown on the right of \autoref{fig:iput}, was to free the inode by its inode number, after decrementing
\cc{ip->ref}, rather than by its
\cc{struct inode} pointer.  We were initially not confident in the fix,
because it required the kernel to operate on inode numbers directly,
which is not the case for any other file system code, and the fix had
the potential to affect some other part of file system correctness that
we might not have thought about.  However, after we proved the kernel
correct with the fix, we had sufficient confidence to commit this fix
in the upstream xv6 repository.

\paragraph{Kernel bug: Creating files in disconnected directories.}
If a process's cwd was a directory that had been deleted, and the
parent directory was also deleted, then \cc{..} referred to a free inode,
and trying to access \cc{..} caused the kernel to panic because the
inode type was no longer \cc{T_DIR}.  The fix was to make \cc{namei}
return an error if it is traversing a directory whose own link count
is zero.  Similarly, we fixed \cc{create} and \cc{link} to check that
they are not creating a new filename in an unlinked directory.

\paragraph{Kernel bug: Overflowing \cc{nlink}.}
xv6 did not check for overflow on the inode's link count.  As a result,
creating many links to a single inode could overflow that inode's \cc{nlink}
and wrap it around to a negative value.  The implementation of \cc{unlink}
panics if it finds \cc{nlink} $< 1$.  The fix was to check for overflow,
and return an error if creating a new link would cause the inode's
\cc{nlink} to overflow.

\paragraph{Kernel bug: Scheduler did not reset \cc{push_off}'s interrupts-enabled flag.}
When a kernel thread context-switches to the scheduler using \cc{swtch},
the scheduler inherits the per-CPU flag tracking whether the top-level
\cc{push_off} had interrupts enabled or disabled (this flag determines
whether the last \cc{pop_off} will re-enable interrupts on the CPU).
The scheduler needs to have interrupts remain disabled: it checks the
entire process table to see if there are any more runnable processes,
and if it makes a complete loop finding no runnable processes, it goes
to sleep using the \cc{WFI} instruction.  If an interrupt were to come in
between the scan and the \cc{WFI}, and the interrupt woke up some process,
the scheduler would not notice.  However, the scheduler forgot to reset
this per-CPU flag when it resumed execution after context-switching from
a kernel process thread, which inadvertently caused it to re-enable
interrupts when the scheduler released the spinlock protecting the
currently running \cc{struct proc}.  The fix involved explicitly clearing
this bit in the scheduler before releasing any locks.

\paragraph{Kernel bug: Failing to log a partial file write.}
In xv6, \cc{sys_write} calls \cc{writei}, which needs to copy data
from the user-supplied pointer into the file's data block.  To do so,
\cc{writei} gets the data block, and then calls \cc{copyin} to safely
copy data from a user-space pointer into the data block.  The user buffer
could span a page boundary; in this case, \cc{copyin} validates the
mapping of the first page, copies the bytes from there, and then
proceeds to validate and copy data from the second page.  If the second
page turns out not to be mapped, then \cc{copyin} returns an error after
having already copied data from the first page.  On receiving an error
from \cc{copyin}, \cc{writei} returned an error back to \cc{sys_write},
without putting the partially updated data block into the write-ahead log.
This caused the file's data block to enter an unstable state: as long as
it remained in the buffer cache, reads would return the partially written
data, but once the buffer was recycled, reads would start returning
the old data.  We found this bug while proving that every disk buffer
write is logged to the write-ahead log.  The fix was to have \cc{writei}
log the data block even when \cc{copyin} returns an error.

\paragraph{Kernel bug: Missing icache fence when switching to a user process.}

RISC-V does not require the instruction cache to be coherent with
writes to memory.  This is not a
concern for the kernel's code, because the kernel code is loaded at boot
time and is never modified afterwards.  However, user code \emph{does}
get loaded dynamically during \cc{exec}, and gets stored in dynamically
allocated pages of memory.  As a result, after the kernel runs \cc{exec}
and starts executing the process in user space, the instruction cache could still
contain old cached entries back from when this same physical page of
memory contained the code of a different user process.  To fix this
bug, we added \cc{fence.i} to the xv6 kernel's \cc{userret} trampoline,
before it returns to user space.

\paragraph{Kernel bug: \cc{freeproc} failed to hold \cc{wait_lock} while updating \cc{p->parent}.}

xv6 uses a separate lock, \cc{wait_lock}, to protect the parent field
of a \cc{struct proc}, \cc{p->parent}, instead of using the spinlock
associated with the rest of the process, \cc{p->lock}.  The reason is
that operations that deal with \cc{p->parent} need to operate on multiple
processes at a time, such as re-parenting an orphan process to \cc{init}
(PID 1).  However, the \cc{freeproc} function modified \cc{p->parent}
without holding \cc{wait_lock}.  We found this while proving \cc{freeproc};
the proof did not have the ownership of \cc{p->parent} needed to write to
that field, despite holding the \cc{p->lock} spinlock, since ownership
of the \cc{p->parent} fields lived in the lock invariant of \cc{wait_lock}.

\paragraph{Kernel bug: Inconsistent UART TX FIFO handling.}

The xv6 kernel was not consistent about its locking discipline when
writing data to the UART.  There were two ways that the kernel would write
to the UART: one coming from user processes writing to a console file
descriptor, and the other coming from the kernel's calls to \cc{printk}.
Both paths checked whether the UART's TX FIFO was full by checking a status
register in the UART, and if it was not full, wrote bytes to the UART's TX
FIFO register.  However, the two were not synchronized with one another,
which meant that it was possible for the TX FIFO to overflow when the
two raced.  This showed up in the proof when we could not assign consistent
ownership of the UART's TX FIFO state to either of the two lock invariants.
Each was independently proven to be correct, but when we tried to compose
\cc{printk} together with user-space processes writing to the console,
the adequacy theorem could not assign the same device ownership to both
invariants.  We fixed this by using a single spinlock to protect the UART
TX FIFO state (i.e., both \cc{printk} and user console writes grab this
spinlock, check if the TX FIFO is available, and if so, write data to it).

One reason the xv6 kernel had this bug is that xv6 is intended to be a
teaching operating system, where students modify its source code through a
series of lab assignments.  Students inevitably make mistakes, including
bugs that lead to deadlock, corruption, etc.  Thus, it was important for
xv6 to ensure \cc{printk} can run and print out debugging information
to help the student, even if, say, the console lock is already held by
some deadlocked process.  On the other hand, for the verified version
of xv6, the \cc{acquire} specification enforces the lock-ordering
discipline and the proof rules out corruption, so this aspect
of \cc{printk}'s design was no longer important.

\paragraph{Kernel bug: Boot code did not enable \cc{ADUE}.}

RISC-V provides two ways to manage page-table accessed and dirty bits:
either hardware-managed (``Svadu''), where the hardware writes back A
and D bits into page-table entries when pages are accessed or modified
through the TLB, or software-managed (``Svade''), where the hardware
traps into the kernel in order to update A and D bits.  The RISC-V
specification, and the Sail model, state that by default the system
starts in software-managed mode, and switching to hardware-managed
mode requires setting the \cc{ADUE} bit in the \cc{menvcfg} register.
However, QEMU starts in hardware-managed mode by default, for
backward-compatibility reasons.  xv6 was tested only on QEMU, and failed
to set \cc{ADUE} in \cc{menvcfg}.  We discovered this when the proof
required reasoning about traps on accesses to every virtual memory address;
the xv6 kernel did not handle these traps.  The fix was to set the
\cc{ADUE} bit in \cc{menvcfg} in xv6's machine-mode boot code.

\paragraph{Kernel bug: \cc{kill(0)}.}

As part of specifying the system calls, we discovered that the
implementation of \cc{kill(0)} set the \cc{killed} flag on the first
unused \cc{struct proc}, which happened to have a \cc{p->pid} of zero.
This did not violate any internal kernel safety invariants, but it did
make for a surprising specification of \cc{kill}, because, with this
buggy implementation, some unspecified future process would
get killed immediately after \cc{fork}.  We fixed the implementation of
\cc{kill} to do nothing when invoked with a zero PID.

\paragraph{User-program bug: \cc{grep} silently dropped input.}

xv6's \cc{grep} reads its input into a fixed 1024-byte buffer,
repeatedly calling \cc{read} to fill up the buffer until it sees a
newline.  Each repeated call to \cc{read} asks the kernel to read the
number of bytes corresponding to how much space remains in the buffer.
When the buffer reaches 1023 bytes of data with no newline, the loop
calls \cc{read} with a zero length, which causes the kernel to return 0.
The code interprets that as an end-of-file indicator, and \cc{grep} exits.

To fix this bug, we changed \cc{grep} to continue reading bytes until the
next newline.  Our fixed version of \cc{grep} does not match on lines that
are at least 1023 bytes long (since we did not want to introduce unbounded
memory allocation), but it does continue to match subsequent lines.

\section{Implementation}
\label{sec:impl}

The implementation effort took \nDevDays days and \nCommits commits.
\autoref{fig:loc} breaks the development down by component.  Of the
\nTotalLines lines of Rocq in the tree, \nGeneratedLines
lines are generated by tools and never edited: the Sail RISC-V semantics
compiled to Rocq, the linked kernel and user ELFs dumped to Rocq data,
the instruction-decode layer re-derived from that dump, and the
reference-implementation results that the conformance tests are checked
against.  The remaining \nWrittenLines lines are written by agents.

\nGenericLines lines are not about xv6 in particular.
\nLogicLines of those are the logic itself: the Iris language instance
over the Sail execution monad, the points-to assertions for registers,
memory, and devices, the per-node proof interface, the TSO memory model,
and adequacy.  \nInstrLines lines are instruction rules: the
per-instruction WP leaves and the step engines they compose, for
machine mode, supervisor mode under the kernel's translation regime, and
the trampoline page.  \nArchLines lines are RISC-V architecture theory
with no separation logic in it (Sv39 page-table trees and walks, the
\cc{mstatus} transforms, decode totality, bitvector arithmetic), and
\nDevmodelLines lines are the device models: the UART, the PLIC, and the
virtio disk, as part of the machine model beside the Sail core.  Two
more rows are about user-mode execution in general:
\nUsafeLines lines prove that arbitrary user code runs safely
under a user page table (\autoref{sec:proof-user-safety}),
and \nUlogicLines lines
are the verified-user tier that runs a program with a known image
instruction by instruction, together with the user-side heap and
descriptor resources (\autoref{sec:user}).

\nKernelSideLines lines are about the xv6 kernel.  \nKinvLines lines define
resources for kernel state: the process table and
scheduler protocol, the buffer cache, the inode cache and its escrow,
the log, pipes, spinlocks and sleeplocks, the kernel and user page
tables, the file system's state, durability, and crash predicates, and
the pure on-disk semantics those read.  \nDevprotoLines lines are the
driver-side proof plan for the devices: the DMA lease and the keyed queue
protocol the disk driver shares with the device thread, the kernel's plan
for the PLIC, and the three device threads with the invariants they run
under.

The function specifications are \nSpecLines lines (\cc{Spec*.v} plus the
statements they import: the abstract file system state, the per-syscall
definitions and log budgets, and the user/kernel trap contract).
\nProofLines lines are whole-function proofs, \nLinkLines lines link
each function proof to the proofs of the functions it calls, and
\nBootLines lines boot the system---the M-mode carve of the boot image,
the ghost allocation of every subsystem, the era-0 facts read off the
\cc{mkfs} image---and assemble the top-level theorem.

\nUserLines lines are the user side: \nUprogLines lines of proof about
\cc{init}, \cc{sh}, \cc{cat}, \cc{echo}, \cc{sync}, \cc{grep}, \cc{seccomp},
and the \cc{ulib}
routines they call, and \nAppLines lines of application claims---the
pure session models of what the user may type and what the console must
show, the Iris claims over the abstract file system and console, and the
application theorems.  The rest of the written
Rocq is \nAuxLines lines of tests, pilots, and superseded drafts kept as
records; \nVtestRocqLines lines of conformance-test proofs
(\autoref{sec:validate}); and \nModelWrittenLines lines of platform
hooks for the Sail model.

Outside Rocq, the development carries \nToolsLines lines of Python (the ELF
dumper, the decode-layer and code-catalog generators, the conformance-test
harness, and the reporting tools), \nVtestImageLines lines of agent-written
RISC-V assembly for the conformance-test images, and \nNotesLines lines
of Markdown design notes in \nNotesFiles files.

\begin{figure}[t]
\centering
\small
\setlength{\tabcolsep}{4pt}
\begin{tabular}{lrr}
\toprule
\textbf{Component} & \textbf{Files} & \textbf{Lines} \\
\midrule
\multicolumn{3}{l}{\emph{Written: not specific to xv6}} \\
\quad Program logic: Iris over Sail, memory model, adequacy & \nLogicFiles & \nLogicLines \\
\quad Instruction rules: WP leaves and step engines           & \nInstrFiles & \nInstrLines \\
\quad RISC-V architecture theory                              & \nArchFiles  & \nArchLines \\
\quad Device models (UART, PLIC, virtio disk)                 & \nDevmodelFiles & \nDevmodelLines \\
\quad Arbitrary user-mode execution                           & \nUsafeFiles & \nUsafeLines \\
\quad Verified user-execution tier                            & \nUlogicFiles & \nUlogicLines \\
\multicolumn{3}{l}{\emph{Written: the xv6 kernel}} \\
\quad Kernel abstractions and invariants                      & \nKinvFiles  & \nKinvLines \\
\quad Device driver protocols and device threads              & \nDevprotoFiles & \nDevprotoLines \\
\quad Specifications (\cc{Spec*.v} and their vocabulary)      & \nSpecFiles  & \nSpecLines \\
\quad Whole-function proofs (\cc{Proof*.v}, \ldots)           & \nProofFiles & \nProofLines \\
\quad Linking (\cc{Link*.v})                                  & \nLinkFiles  & \nLinkLines \\
\quad Boot, system assembly, top-level theorem                & \nBootFiles  & \nBootLines \\
\multicolumn{3}{l}{\emph{Written: user programs}} \\
\quad User-program proofs (\cc{Uk*.v})                        & \nUprogFiles & \nUprogLines \\
\quad Application claims and theorems                         & \nAppFiles   & \nAppLines \\
\multicolumn{3}{l}{\emph{Written: other}} \\
\quad Tests, pilots, superseded drafts                        & \nAuxFiles   & \nAuxLines \\
\quad Conformance tests                                       & \nVtestRocqFiles & \nVtestRocqLines \\
\quad Sail platform hooks                                     & \nModelWrittenFiles & \nModelWrittenLines \\
\midrule
\multicolumn{3}{l}{\emph{Generated}} \\
\quad Sail RISC-V model in Rocq             & \nModelFiles      & \nModelLines \\
\quad Kernel image in Rocq                  & \nKernelDumpFiles & \nKernelDumpLines \\
\quad User program images in Rocq           & \nUserDumpFiles   & \nUserDumpLines \\
\quad Instruction-decode layer              & \nDecodeFiles     & \nDecodeLines \\
\quad Captured QEMU, JH7110, and CVA6 results & \nVtestCaptureFiles & \nVtestCaptureLines \\
\midrule
\textbf{Total Rocq}                         &                   & \textbf{\nTotalLines} \\
\bottomrule
\end{tabular}
\caption{Size of the development, in lines of Rocq.  ``Generated'' files are
  produced by tools and are never edited by hand.}
\label{fig:loc}
\end{figure}

\paragraph{Trusted definitions.}

Rocq's kernel checks the proofs, so most of the lines shown in
\autoref{fig:loc} are not trusted, as long as the top-level theorem statement is
correct.  To measure how many lines are required to define the meaning
of the top-level theorems, we used Rocq's \cc{Print All Dependencies}
command to track down what definitions are transitively used in the
theorem statements (as opposed to their proofs).  \autoref{fig:tcb}
shows the results.

\begin{figure}[t]
\centering
\small
\setlength{\tabcolsep}{4pt}
\begin{tabular}{lrrr}
\toprule
\textbf{What the statement rests on} & \textbf{\cc{xv6\_power}} & \textbf{\cc{xv6\_fs}} & \textbf{\cc{union}} \\
\midrule
Sail RISC-V model, compiled to Rocq & 21{,}585 & 21{,}212 & 21{,}212 \\
Machine and shared-memory model & 1{,}535 & 1{,}207 & 1{,}202 \\
Device models (UART, PLIC, virtio disk) & 1{,}806 & 1{,}797 & 1{,}797 \\
Initial disk image (plus generated Rocq representation of a 2.0\,MB binary file) & 37 & 37 & 37 \\
Kernel image (generated Rocq representation of a 41\,KB binary file) & --- & --- & --- \\
Iris ghost state and CSL specifications & 1{,}788 & --- & --- \\
The conclusion statement (FS consistency, UART console trace, etc.) & 1{,}494 & 994 & 1{,}591 \\
\bottomrule
\end{tabular}
\caption{Lines of definition each adequacy statement unfolds to, by category.}
\label{fig:tcb}
\end{figure}

The generic adequacy theorem, \cc{xv6_power}, which
proves that any pure property that can be derived from the invariants
maintained by xv6 holds on every trace-level execution according to the
operational semantics, depends on \nTcbPowerDefs definitions spanning
\nTcbPowerLines lines in our proofs, plus external dependencies (the
Sail model, the dump of the ELF kernel, the stdpp and Iris libraries,
etc.).  The statement of this
theorem includes some of the CSL machinery, because it
talks about any pure invariant that can be derived from CSL invariants,
which inherently needs to talk about Iris CSL properties.

We also have a specialized instantiation of the above generic theorem,
\cc{xv6_fs}, where we prove it for a specific trace property: that the
file system
remains consistent (across all concurrent executions, all crashes, etc.).
This adequacy corollary unfolds to \nTcbFsDefs
definitions spanning \nTcbFsLines lines.  These definitions are notably
pure---they do not mention any CSL machinery at all---because the proof
of this corollary establishes that the xv6 CSL invariants imply the pure
file system consistency property, and therefore the theorem statement
itself does not need to refer to CSL in any way.  On the other hand, this
corollary does include additional dependencies that define what file
system consistency means.

Finally, the application-level theorem of \autoref{sec:theorem},
\cc{union}, instantiates the
same generic theorem with a different trace property: the console output
that a shell session may produce when the user types any sequence of
commands of the forms \cc{echo \ldots}, \cc{echo \ldots\ > f}, \cc{cat f},
pipelines such as \cc{echo \ldots\ | cat | grep w}, using \cc{sync} to flush
file system changes to disk, and using \cc{seccomp} to run unverified binaries
in a sandbox, across any number of power cycles.
This theorem unfolds to \nTcbUnionDefs definitions spanning \nTcbUnionLines lines.
Its statement is free of CSL machinery, since the conclusion is a property
of the bytes on the console UART alone; the additional definitions describe
what each command, and each stage of a pipeline, may print.

\paragraph{Invariants.}

The kernel's shared state is owned by Iris invariants, allocated in
\nInvariants distinct namespaces.  Thirteen of them belong to the file
system: the log's block bytes, an escrow through which a \cc{sync} that
finds a commit in flight collects its result from that commit, the block bitmap, the inode table, the
inode reference escrow, the per-inode escrow, the free-inode pool,
the inode region, the file-table top, the superblock, and three
families of per-object \emph{transit boxes} (invariants for reasoning
about shared memory locations that are accessed under different locks,
which is tricky to reason about under weak shared memory).  Four belong to the
device fabric (\autoref{sec:model}): the UART, the PLIC, the virtio disk, and
the external-interrupt pins the PLIC drives into each HART.  One is the kernel
page table; two belong to the process
layer (the per-lock invariant, instantiated once per spinlock, and the
process-slot availability ghost); and four are machine-level resources the
boot path configures and every later proof depends on: the PMP configuration,
the timer comparator, the boot handshake, and the power/crash state.  One
further invariant, which holds the trace property, belongs to the whole-system
statement rather than to any one part of the kernel.  The remaining five
support application-level proofs (\autoref{sec:user}): the application
invariant (\autoref{sec:user-app}), the user-level half of each file
descriptor's offset (\autoref{fig:sys-read}), an escrow for the right to
read console input, an invariant for concurrent user reads/writes of
a pipe, and one for a pipe shared with a \cc{seccomp}-sandboxed program,
whose reads and writes follow no protocol.

\paragraph{Build time.}

Building the whole development takes about \nBuildAuditVtestMin~minutes on a
24-core virtual machine with Intel Xeon E5-2699 2.2~GHz vCPUs and 96~GB of
memory: \nModelBuildMin~minutes for the Sail model and the ELF dumps,
\nProofBuildMin~minutes for the Iris proofs, \nAuditMin~minutes
to run \cc{Print Assumptions} on the top-level theorem to confirm
the set of assumed axioms, and \nVtestMin~minutes to check the
conformance-test proofs.

\paragraph{Porting to Lean.}

After we had finished doing all of this verification work in Rocq,
we asked an agent to port the whole verification framework and proofs
to Lean.  This required some supervision, but far less than the original
development.  The resulting Lean port was about half the size of the Rocq
development, in terms of lines of proof, and it compiled (checking all of
the proofs) about 5$\times$ faster, through a combination of cleanup,
simplifications, and better optimizations in Lean itself.
(This paper, including all of the reported numbers,
describes the original Rocq development.)

One surprising finding is that, when one of the authors of this paper
switched the git repo from Rocq to Lean, the other author barely noticed
(aside from the performance improvements), since the level of discussion
with the agent was far above the specific syntax differences between
Rocq and Lean; the agent was able to switch all of its work from Rocq
to Lean transparently.

\subsection{Changes to xv6}

We made almost no changes to xv6 itself, other than fixing the bugs
described in \autoref{sec:bugs}.  Even in tricky cases, we forced the proof
to work with the code, rather than changing the code.  One example is the
file reference counting, which does not overflow for a subtle reason.
Another example is \cc{kexec} not causing a panic when remapping the
trampoline page at the top of the address space.

We made some changes where the xv6 kernel was purposely sloppy to
support students doing lab assignments on top of xv6.  In particular,
it was important for xv6, as used for class assignments, to ensure that
\cc{printk} continues working even if the students modify the kernel, run
into a bug, and call \cc{printk} when the rest of the kernel is corrupted.
To this end, \cc{printk} tried to avoid acquiring locks, at the risk of
corrupting the UART TX FIFO (see \autoref{sec:bugs}).  In our verified xv6
kernel, the proof forces every caller of \cc{printk} to follow the
lock-ordering discipline,
so we changed the implementation to be correct by holding locks.

Along the same lines, we removed support for \cc{procdump}, which printed
a listing of currently running processes when the user entered \cc{Control-P} on
the UART console.  This, too, existed to support student debugging,
and to implement a cheap version of the \cc{ps} command.  However,
\cc{procdump} did not acquire appropriate locks, again in order to ensure
that some version of \cc{procdump} can be executed even if the student's
lab assignment has deadlocked the kernel.  We removed the support for
the \cc{Control-P} special input handler in the UART console.

We proved that many of the calls to \cc{panic} were unreachable because the
kernel maintains all of its invariants, and therefore many assertions about
kernel data structure consistency never fail.  However, the kernel still
has some calls to \cc{panic} that represent running out of resources.
For example, the kernel panics if it runs out of in-memory
\cc{struct inode} entries.  On the other hand, the kernel correctly
handles running out of memory without panicking.

We also deleted some lines of code where the proof helped us realize they
were superfluous.  One example is that \cc{kexec} did an explicit check
on the \cc{argc} value to ensure that it did not exceed the maximum
allowed number of arguments.  It turns out that \cc{sys_exec}, which is
the caller of \cc{kexec}, already performed this check, and moreover,
the check inside \cc{kexec} was not sufficient (it was off by one in an
unsafe direction).

Finally, we fixed PID wraparound in xv6; the implementation previously
assumed that a 32-bit integer PID would never overflow.  To prove that
PIDs uniquely identify processes, and that PIDs are never negative
or zero, we modified the implementation to check, at allocation time,
whether the PID has wrapped around or is already in use by an existing process.

\subsection{TCB}
\label{sec:impl-tcb}

The theorem about the xv6 kernel's correctness requires trusting several
components, namely:

\begin{itemize}

\item The Sail model of RISC-V.

\item Sail's Rocq backend that generates a Rocq definition based on the
  semantics of RISC-V defined in the Sail language.

\item The Rocq proof-checking kernel.

\item The ELF dumper, which is a Python script that translates the
  kernel's ELF binary into a Rocq file representing all of the bytes in
  the ELF image, along with the text and data sections of the kernel
  (since the semantics assumes the kernel is already loaded in memory
  based on the ELF program headers, rather than loading the literal ELF
  binary into DRAM).

\item The \cc{Print Assumptions} command that confirms the axioms assumed
  by the top-level adequacy theorem (which, in our case, are the standard
  logical functional-extensionality axiom, as well as two axioms required
  by the Sail model to represent its behavior for memory reservations, used
  by the LR and SC instructions).

\item The model of how the system executes, which is built on Sail's semantics,
  and includes a model of weak shared memory (TSO with load-load
  reordering) with
  handling of exclusive read-modify-write operations, introduced by us,
  as well as models of devices (UART, PLIC, and virtio disk) that we defined, and
  a model of power handling (power-off and power-on) to represent
  crashes.  We validate this model using conformance tests
  (\autoref{sec:validate}), but it is not a proof that considers all
  possible cases.

\end{itemize}

There are also a number of notable components that are not in the TCB:

\begin{itemize}

\item The kernel implementation, including both the C code and all of the
  assembly: the boot code, context switching, hardware
  configuration, page-table setup and management, switching to user space,
  taking interrupts from user space, taking interrupts while running
  the kernel, TLB A/D writeback, I/O memory fences when dealing with
  disk DMA regions, etc.

\item The compiler, linker, assembler, and all other parts of the kernel build
  process that lead to producing a single kernel ELF binary.

\item The correctness of the initial file system image, as produced by \cc{mkfs}.

\end{itemize}

One shortcoming of the Sail model is that it is over-specified in some
cases; for instance, it models the TLB as a 64-entry direct-mapped cache,
rather than allowing any associativity or different sizes.  We also found
(and fixed) the PTE A/D writeback bug in the Sail model.

The Sail RISC-V model comes with an extensive configuration file format
that requires any user of the Sail model to define the specific features
they want to include.  We chose a specific set of features that correspond
to the platform targeted by the xv6 kernel (e.g., not including B, V,
FP, and control-flow integrity features, among others).

\section{Evaluation}
\label{sec:eval}

This section answers the following questions:

\begin{itemize}

\item How much agent work was required to verify xv6?

\item How much effort is needed to adapt the proofs when the xv6
  source code changes and the ELF kernel binary is recompiled?

\item How much effort is required to adapt to changes in the
  Sail RISC-V model?

\item What issues were uncovered by model conformance tests?

\end{itemize}

\subsection{Agent effort}

To report on the agent effort required to verify xv6, we extracted data
from both authors' Claude Code transcripts, which record every message,
its timestamp, and its token usage; \autoref{fig:effort} summarizes
them.  Unfortunately, part of the record is lost, because Claude
Code deletes transcripts older than \nRetentionDays days by default.
Roughly \nLostShare\% of the sessions are missing, most of them from one
window of about two weeks early in the development process.  The rest of the
record is nearly complete, including the project's first two weeks.

Work was spread over \nSessions agent sessions in \nWorkDirs working
directories---separate checkouts of the same repository, so that
several agents could work at once (\autoref{sec:agent})---and those
sessions spawned a further \nSubagents sub-agent runs.  Across all
of them we typed \nPrompts prompts, totaling \nPromptWords words.
The typical prompt is short: the median is \nPromptMedianWords words and
the mean \nPromptMeanWords.  Most of them are of the form ``finish the
\cc{consoleintr} proof'' or ``git pull the latest changes; specify and
prove \cc{sys_close}''; the long ones are design rulings, where we tell
the agent what abstraction to use, or reject one it proposed.  A further
\nFollowups prompts were generated by the harness rather than by us,
continuing a session that had been told to keep going with \cc{/goal}.

Against those \nPrompts prompts, the agents produced \nAssistantMsgs
messages and made \nToolCalls tool calls, and Claude was actively
running for \nAgentHours~hours.  Because up to \nPeakSessions sessions
ran concurrently, that is \nWallHours~hours of wall-clock time.
(``Actively running'' excludes gaps longer than five minutes between
consecutive messages in a session, so a session left open overnight is
not counted as active.)  The token cost was \nOutputMTokens~million
output tokens, of which \nThinkingMTokens~million were thinking
tokens, against \nInputBTokens~billion input tokens---almost all of
them (\nCacheReadBTokens~billion) served out of the prompt cache.
Claude's running time includes the time it spent waiting for Rocq,
which is substantial.

This development cost less than \$3,000 in Claude subscriptions (for
multiple Max 20x plans), plus several multi-core development VMs on AWS
and GCP.

\begin{figure}[t]
\centering
\small
\setlength{\tabcolsep}{4pt}
\begin{tabular}{lr}
\toprule
Elapsed time                                  & \nDevDays days \\
Git commits                                   & \nCommits \\
Agent sessions / working directories          & \nSessions ~/ \nWorkDirs \\
Sub-agent runs                                & \nSubagents \\
Human prompts                                 & \nPrompts \\
\quad total words / median words per prompt   & \nPromptWords ~/ \nPromptMedianWords \\
Agent messages / tool calls                   & \nAssistantMsgs ~/ \nToolCalls \\
Claude running time (summed over sessions)    & \nAgentHours h \\
\quad wall-clock (union of sessions)          & \nWallHours h \\
Output tokens                                 & \nOutputMTokens M \\
Input tokens (\nCacheReadBTokens{}\,B cached)   & \nInputBTokens B \\
\bottomrule
\end{tabular}
\caption{Development effort, extracted from the Claude Code transcripts.}
\label{fig:effort}
\end{figure}

\subsection{Re-proving after kernel changes}

The proofs are about a specific kernel binary, so any change to the xv6
sources moves symbol addresses and re-encodes call targets, and every
proof that names an address is affected.  The development was updated
(``bumped'') to a new version of the xv6 source code \nXvBumps times.

Almost all of the changes are mechanical.  Across the \nXvBumps commits
that updated the development to a new version of xv6, \nXvBumpGenChurn lines
of generated code changed---the kernel dump, the instruction-decode layer,
the user-program dumps---along with \nXvBumpWrittenChurn lines of written
specification and proof, a median of \nXvBumpWrittenMedian lines per bump.

To measure what a bump costs, we located, for each bump, the session in
which the agent made that commit and measured from the prompt that asked
for the bump to the commit.  We could do this for \nXvBumpsMatched of the
\nXvBumps commits; the transcript of the remaining one was pruned by Claude Code.
The median bump took \nXvBumpMedianMin~minutes of agent time and
\nXvBumpMedianPrompts prompts; the cheapest took \nXvBumpMinMin~minutes; the
most expensive, \nXvBumpMaxHours~hours, was one where the C source gained a
check (so the specification moved, not just the addresses) on top of a
relayout that shifted most of the kernel's symbols.  In total the
\nXvBumpsMatched measured bumps account for \nXvBumpHours~hours,
\nXvBumpPrompts prompts, and \nXvBumpTimeShare\% of the project's
agent time.

\subsection{Re-proving after Sail model changes}

During the development of this project, we updated the Sail RISC-V model
\nSailBumps times: adding a fix for the PTE A/D writeback bug,
changing how Sail extracts axioms, and tagging instruction vs.\ data loads
to enable modeling a non-coherent icache.
The PTE A/D writeback fix required significant effort to adopt,
because it brought a number of other updates to the Sail model
along with it, and because it required revising the page-table proofs.
That bump required \nSailAdBumpHours~hours of agent time.  Updating
the axiom extraction required \nSailAxiomBumpHours~hours, and the
tagging change \nSailTagBumpHours~hours.

\subsection{Model discrepancies}

We generated a total of \nVtestCases test cases covering behaviors of the CPU, page
tables, interrupts, shared memory, CLINT, PLIC, UART, and the disk.
The JH7110 and CVA6 do not implement the virtio disk device, and their UART
devices are different from the one in QEMU and our model, so we did not
run those tests on the JH7110 or CVA6.  The JH7110 also implements an earlier
revision of the RISC-V specification, which means that it does not have
certain CSRs that xv6 uses (and that QEMU implements), and the JH7110
does not implement hardware writeback of PTE A/D bits.

Comparing the runs generated by these tests on our three reference
implementations against the model surfaced \nVtestFindings discrepancies;
we describe the main ones below.

The UART, PLIC, and disk tests uncovered several issues with our model
of those devices.  To give a few examples: the UART device exposes
several single-byte registers at consecutive addresses; the model failed to
consider the possibility of a multi-byte read accessing the address of one
of the UART's single-byte MMIO registers.  The PLIC model failed to correctly
implement priority thresholds.  The disk model failed to correctly return
the length of a used ring slot.  xv6 did not depend on any of these
features, but conformance tests found these differences between QEMU
and our model, and after confirming what the correct behavior should be,
we fixed these issues.  The PLIC of the JH7110 matched our model in some
cases, but differed in cases where the PLIC had more interrupt sources;
our model was initially intended to match QEMU.

The disk test also uncovered a more serious issue with our disk model:
our disk model executed all submitted disk requests in order, whereas
the QEMU device can execute requests out of order (and indeed our tests
observed such reordering).  We changed our model to allow executing
requests in any order, which required updating the invariant used by
the xv6 disk driver, as well as the associated proofs.  The file system
already assumed that writes could be reordered, because this was the
spec of the \cc{virtio_disk_rw} function.

The CPU tests uncovered several cases where the initial register contents
on QEMU differed from our model.  These were all ``benign'' differences:
as one example, QEMU reports in the \cc{misa} register that it supports
the hypervisor feature (H); our model does not support the hypervisor
feature, and does not report this bit in \cc{misa}.  Similar issues showed
up with the initial register contents on the JH7110.

The shared-memory tests uncovered the fact that QEMU can generate shared-memory
accesses that are not sequentially consistent (we ran QEMU on an x86
machine, which implements a TSO shared-memory model).  Our initial model
of shared memory was sequentially consistent, and could not reproduce
that memory behavior.  We have since extended our shared-memory model
to TSO with load-load reordering, which can represent this behavior.

The CPU tests observed that the instruction cache on the JH7110 is
non-coherent: self-modifying code keeps executing old instructions until
the machine executes \cc{fence.i}.  QEMU did not exhibit such behavior.
We proved that our model of a non-coherent instruction fetch captures
both of these behaviors.

The tests also uncovered a difference between the QEMU TLB behavior and
the behavior of the Sail model.  Sail's RISC-V model implements a 64-entry
direct-mapped (non-associative) TLB.  One of our page-table tests observed
a difference between QEMU and the model because the model forced eviction
of a certain TLB entry and re-walked the page table, whereas QEMU kept
that same PTE in the TLB and did not re-walk.  Ideally, the Sail model
would specify a non-deterministic TLB that allows a wider range of
observed behaviors, but fixing that is beyond the scope of this project.

The tests also uncovered a known shortcoming of the Sail model: the Sail
model allows access only to HART 0's CLINT MMIO registers.  When we ran
the tests on a non-0 HART of the JH7110, we discovered that we could
not match that behavior in our model.  According to comments in the
Sail source code, this is a known, deliberate limitation of the model.

\section{Related work}
\label{sec:relwk}

\paragraph{OS kernel verification.}

Formal verification of OS kernels goes back to the
security kernels of the 1970s: the MITRE PDP-11/45
kernel~\cite{schiller:pdp11-kernel,millen:kernel-validation}, UCLA
Secure Unix~\cite{walker:uclaunix}, PSOS~\cite{feiertag:psos}, and
KSOS~\cite{mccauley:ksos}.  These projects verified designs or partial
implementations by hand or with early tools; Kit~\cite{bevier:kit}
was the first machine-checked proof of a kernel implementation, at the
machine-code level of a small idealized processor without interrupts,
devices, or multiple processors.

seL4~\cite{klein:sel4,klein:sel4-tocs} established that full functional
correctness of a microkernel is feasible, by refinement from an abstract
specification down to C and, later, to the binary~\cite{sewell:sel4-tv}.
The verified kernel is single-core, is mostly non-preemptible
with interrupts handled by polling at a few points, treats address
translation and the TLB through a discipline assumed of the hardware
rather than a model of it, and rules DMA out of scope.

Verisoft~\cite{alkassar:verisoft} verified a microkernel on a simplified
processor, and Verisoft XT~\cite{leinenbach:hyperv} targeted
Hyper-V, with the Hyper-V proofs left incomplete.

CertiKOS~\cite{gu:certikos,gu:certikos-layers,gu:certikos-ccal} verified
a multi-core kernel with fine-grained locking by layered refinement,
with interrupts and device drivers~\cite{chen:certikos-drivers}
modeled over an abstract machine; assembly code is handled through
a verified compiler for a subset of x86, and the hardware page-table
walker and DMA are not modeled.

Push-button verifiers such as Hyperkernel~\cite{nelson:hyperkernel},
Serval~\cite{nelson:serval}, and Nickel~\cite{sigurbjarnarson:nickel} avoid
interactive proofs by restricting the kernel design (finite interfaces,
no interrupts during system calls) so that symbolic evaluation of LLVM
IR or RISC-V binaries is tractable.  One consequence is that they do
not support concurrency (or many other features present in xv6).

Other verified low-level systems include
Komodo~\cite{ferraiuolo:komodo}, verified in Dafny
at the level of Arm assembly; SeKVM~\cite{li:sekvm}
and its relaxed-memory successor~\cite{tao:vrm}, which
verify the core of KVM on Arm across cores; the Arm CCA
firmware~\cite{li:armv9cca}; {\"u}berSpark~\cite{vasudevan:uberspark};
Ironclad~\cite{hawblitzel:ironclad}; and recent systems written
in verified Rust such as VeriSMo~\cite{zhou:verismo} and
Atmosphere~\cite{chen:atmosphere}.

Nearly all of this prior work reasons above the hardware: at the level
of C, Rust, LLVM IR, or an idealized assembly language, over a machine
model that abstracts away the page-table walker, the TLB, the precise
semantics of traps and interrupts, and devices that access memory.
Most were designed or redesigned for verification.  \sys instead verifies
an existing Unix-like kernel, against the Sail RISC-V semantics itself,
with concurrent HARTs, hardware page-table walks, interrupts at any
instruction, DMA, and power failures all part of the model rather than
assumptions about it.  Prior verified kernels also do not include
a complete Unix-like environment: processes, fork/exec, pipes, file
descriptors, and a crash-safe file system running inside the kernel.

\paragraph{Application-level theorems.}

Several projects have proven end-to-end theorems about an application's
externally visible behavior down to the machine, as \sys does, but
without a general-purpose OS kernel in between.  The CLI
stack~\cite{bevier:cli-stack} and
Verisoft~\cite{alkassar:verisoft-approach,alkassar:verisoft} pursued
pervasive verification from a gate-level processor through a compiler
and a small kernel up to applications (in Verisoft's case, an
email client), over simplified uniprocessor hardware.
Ironclad Apps~\cite{hawblitzel:ironclad} verifies complete
applications down to assembly, on top of the minimal Verve
OS~\cite{yang:verve}, for a single-threaded stack.  Bedrock's web
application case study~\cite{chlipala:bedrock-web} verifies a multithreaded
application together with its cooperative thread library at the
assembly level.  CakeML's
verified stack~\cite{loow:cakeml} and the Bedrock2 lightbulb and garage-door
systems~\cite{erbsen:lightbulb,erbsen:garage-door} state their top-level
theorems, much as \sys does, as predicates on the I/O trace of
a verified processor running the application's machine code; in both cases
the application runs on bare metal, with system calls or MMIO provided
directly by a thin runtime.  In contrast, \sys's application theorem is
about user-level binaries (\cc{init}, \cc{sh}, \cc{echo},
\cc{cat}, \cc{grep}, \cc{sync}, and \cc{seccomp})
running as separate processes on a multi-core Unix-like kernel: the theorem covers
\cc{fork}, \cc{exec}, \cc{wait}, file descriptors, loading binaries from a
crash-safe file system across power cycles, preemptive scheduling across
HARTs, and the kernel's handling of allocation failures, all of which
show up in the legal console traces of \autoref{fig:traces}.
\sys's kernel specifications are also generic, so that the application
proof is layered on top of the kernel's proofs without modifying them
(\autoref{sec:user}).

Several other kernel verification projects provide theorems about
application execution but do not prove functional correctness of
particular applications.  For example, seL4's isolation proofs, as used on an
autonomous helicopter in the HACMS program~\cite{klein:sel4-cacm},
show that untrusted components cannot interfere with critical ones.
CertiKOS~\cite{gu:certikos-layers} proves contextual refinement,
which says that any user program behaves the same over the kernel's
implementation as over its abstract specification, but does not
itself verify any user program against that specification.  Mansky et
al.~\cite{mansky:vst-certikos} close this gap for a simple case,
connecting VST proofs of C programs that perform console I/O to CertiKOS's
specifications of the corresponding system calls, so that the program's
guarantee is stated in terms of the OS-level I/O events; the processes
there are sequential, and the connection is at the level of CompCert C and
the CertiKOS abstract machine.  Brun et al.~\cite{brun:beyond-isolation}
argue that enabling such application proofs should be a primary goal of
OS verification.

\paragraph{File system verification.}

Verified file systems have mostly been developed as standalone
artifacts.  FSCQ~\cite{chen:fscq,chajed:fscq-cacm} introduced crash
Hoare logic to specify and prove crash safety of a sequential file
system, and DFSCQ~\cite{chen:dfscq} extended it to deferred durability
and \cc{fsync} semantics; DiskSec~\cite{ileri:disksec} added
confidentiality.  Yggdrasil~\cite{sigurbjarnarson:yggdrasil} verified a
file system by push-button symbolic execution against crash-refinement
specifications, and Argosy~\cite{chajed:argosy} showed how to compose
layers with recovery.  Cogent~\cite{amani:cogent} and
Flashix~\cite{schellhorn:flashix} verified flash file systems, and
VeriBetrKV~\cite{hance:veribetrkv} a crash-safe key-value store.
Concurrent verified storage came with AtomFS~\cite{zou:atomfs}, which
proves linearizability of a concurrent (but not crash-safe) file
system, and with Perennial~\cite{chajed:perennial},
GoJournal~\cite{chajed:go-journal}, and DaisyNFS~\cite{chajed:daisynfs},
which extend Iris with crash reasoning to verify concurrent, crash-safe
journaling and an NFS server.  Ntzik et al.~\cite{ntzik:faults} give a
separation logic for fault-tolerant resources, and
SibylFS~\cite{ridge:sibylfs} a tested specification of POSIX file
system behavior.

The overall file system proof plan in \sys follows the approach taken
in Perennial~\cite{chajed:perennial} and PoWER~\cite{leblanc:power} in
terms of introducing abstract ghost state to keep track of the durable
on-disk state that survives reboots, as well as per-boot ghost state to
keep track of in-memory state of file system transactions that are in
progress but have not committed yet.

Prior work on verified file systems mostly assumes a disk with atomic sector
writes behind a synchronous or abstract interface; the disk driver, DMA,
and interrupts are out of scope, as are the process and file-descriptor
machinery that share state with the file system.

In \sys the file system is verified inside the kernel, together with
the disk driver and its DMA descriptors, the disk interrupt handler,
and the disk model's per-sector writeback behavior, so that the crash
invariant is established at the moment each sector reaches the disk
rather than at an abstract disk interface.  The proof also has to handle
the interaction between the file system and the rest of a Unix kernel:
files unlinked while open, file descriptors shared across \cc{fork},
and \cc{exec} reading a binary from the file system while replacing the
process's address space.

\paragraph{Verification at the instruction level.}

Realistic ISA models have become the ground truth for
machine-level reasoning.  Sail~\cite{armstrong:sail} and its symbolic
executor Isla~\cite{armstrong:isla} give executable, sub-instruction
semantics for Arm and RISC-V, and the CHERI project proved
architectural security properties (capability monotonicity) directly
over Sail models of the whole instruction set~\cite{nienhuis:cheri}.
Earlier ISA models in the same spirit include Fox and Myreen's HOL4
model of Armv7~\cite{fox:armv7} and the ACL2 x86isa
model~\cite{goel:x86isa}; notably, Goel verified a system-mode
program with paging enabled against x86isa~\cite{goel:thesis}, so
hardware page-table walks were part of the semantics, although the
reasoning was sequential and Hoare-style.  These efforts prove
properties \emph{of the architecture}, or of small programs on it;
\sys instead uses the Sail RISC-V model as the trusted semantics
underneath a proof of an entire concurrent kernel.

Program logics over machine code predate Iris: Myreen and Gordon's
Hoare logic for realistically modeled machine
code~\cite{myreen:machine-code}, Jensen et al.'s separation logic for
x86 in Coq~\cite{jensen:high-level-sep}, and
Bedrock~\cite{chlipala:bedrock2} all reason about assembly-level
programs, but sequentially and over simplified machines.  seL4's
binary verification~\cite{sewell:sel4-tv} and
CakeML~\cite{kumar:cakeml,tan:cakeml} reach the binary by
translation validation and verified compilation rather than by a
logic over the ISA.  The closest work to \sys in terms of connecting
software proofs to a machine model is the
Bedrock2 line~\cite{erbsen:lightbulb,erbsen:garage-door}, which verifies
software in separation logic against the riscv-coq ISA semantics,
models an MMIO device as externally observable events, and connects
down to a Kami-verified processor~\cite{choi:kami}.  That work is
single-HART and sequential, with no interrupts, page tables, or DMA;
\sys's theorem is about a multi-HART kernel under all of these.

\paragraph{Concurrent separation logic at the machine level.}

Islaris~\cite{sammler:islaris} is the most direct precedent for \sys: an
Iris program logic over Sail-derived instruction semantics for Arm and
RISC-V, using Isla symbolic traces as the specification of each
instruction.  Islaris is sequential and its case studies are small
(exception handlers and a few hundred instructions), with no page
tables or devices; \sys can be seen as scaling this approach to an
entire OS kernel with concurrency, interrupts, and devices.  Kuru and
Gordon's logic for x86-64 assembly~\cite{kuru:modal}, discussed below,
is likewise an Iris logic at the instruction level, sequential, and
focused on address translation.

Cerise~\cite{georges:cerise} is an Iris logic at
the assembly level of an idealized capability machine, with logical
relations for reasoning about untrusted code, building on earlier
capability-machine work~\cite{skorstengaard:stktokens}; later work adds MMIO
devices and interrupts in a simplified form~\cite{vanstrydonck:full-system}.
Cerise's treatment of unknown code running arbitrary instructions is
analogous to \sys's invariant for user-mode execution
(\autoref{sec:proof}), although \sys relies on the page table and
privilege mechanism of a real ISA rather than on capabilities.

Feng et al.~\cite{feng:interrupts} used CSL-style ownership
transfer at interrupt enable and disable points at the assembly level, on
an idealized machine with an idealized interrupt model, building on their
earlier CSL for low-level code~\cite{feng:csl-ag}.  At the C level,
Certi$\mu$C/OS~\cite{xu:certiucos} verifies a preemptive real-time kernel with
CSL and interrupts, and CertiKOS~\cite{chen:certikos-drivers} models
devices as concurrent threads in a refinement framework.  \sys differs
from all of these in reasoning about the actual RISC-V interrupt, trap,
and privilege machinery as specified by Sail, including the interaction
of interrupts with page-table switches and context switches.

\paragraph{Hardware page tables and TLBs.}

Several program logics have made address translation first-class, but
all of them are sequential.  In the seL4 line, Kolanski and Klein's
mapped separation
logic~\cite{kolanski:mapped-sep-logic,kolanski:mapping-sep,kolanski:phd}
interprets assertions relative to a page-table root: a virtual
points-to fact existentially quantifies the physical address that the
virtual address translates to, and owns both the page-table entries
traversed by the walk and the physical location itself, so that the
frame rule survives page-table updates.  Syeda and Klein~\cite{syeda:tlb}
add an explicit TLB to the machine model and a logic that tracks
addresses whose cached translations may be stale, and verify the Arm
instruction sequence that installs a new page-table root and flushes
the TLB.  Kuru and Gordon~\cite{kuru:modal,kuru:thesis} build an Iris
logic over a RISC-like fragment of x86-64 assembly in which a virtual
points-to fact is a ghost translation token,
$\textit{va} \hookrightarrow \textit{pa}$, together with ownership of
the physical location, while a separate per-address-space invariant
owns the page-table memory that justifies the ghost map.  This is the
same decomposition that \sys uses for $\vmemptsto{\textit{va}}{v}$
(\autoref{sec:csl-riscv}); in both cases it allows the page tables to be
modified without collecting the virtual points-to facts that depend on
them.  On top of it, Kuru and Gordon introduce a hybrid-logic modality
$[r]P$, stating that $P$ holds in the address space rooted at $r$, and
the proof rule for writing the page-table root register moves
assertions into and out of this modality.  With it they verify a
context switch that changes address spaces, and a software page-table
walk and page-mapping routine from an xv6-derived kernel, including the
kernel's use of constant-offset (identity) mappings to reach
page-table pages from their physical addresses.  Their machine model
has a single core and no TLB, and page tables are modified only by
software.

Kernel-scale proofs have modeled address translation more coarsely.
Verisoft verified a paging mechanism against a simple
hardware MMU~\cite{alkassar:paging,alkassar:baby-hypervisor}, and
Verisoft XT modeled the TLB walk of a hypervisor's shadow page tables
as a separate concurrent thread whose memory accesses are governed by
VCC's ownership discipline~\cite{alkassar:shadow-page-tables,cohen:multicore-hypervisor};
this is the closest prior treatment of a hardware page-table walk
racing with kernel code, though on a simplified x86 model rather than a
machine-checked ISA semantics.  Verified hypervisors
reason about page tables and TLB invalidation across cores by
refinement: SeKVM~\cite{li:sekvm} and its extension to Arm relaxed
memory~\cite{tao:vrm}, Komodo~\cite{ferraiuolo:komodo}, and the PROSPER
line, which verifies MMU virtualization for Linux on Arm against an
Armv7 ISA model in HOL4~\cite{guanciale:memory-isolation,nemati:armv7-mmu}.
Simner et al.~\cite{simner:relaxed-vm} give a relaxed virtual-memory
model for Armv8-A in which concurrent hardware walks are first-class,
but as a model of the architecture rather than a proof of software
over it.  \sys treats the RISC-V page-table walker and its A/D-bit
writeback as concurrent hardware activity whose accesses to page-table
memory are governed by CSL invariants, on the same footing as accesses
by other HARTs, and the TLB is part of the model rather than a
discipline assumed of the hardware.  \sys does not need a modality for
other address spaces: all of xv6's kernel code runs under one kernel
page table shared by every HART, with the translation tiers of
\autoref{sec:proof} distinguishing the boot-time regimes, and the
switch to a user page table is reasoned about through the page-table-switching
window in the trampoline and the invariant for user-mode execution.

\paragraph{Shared memory and memory models.}

Formal models of the shared-memory behavior of multiprocessors are
well established: x86-TSO~\cite{sewell:x86-tso} is the canonical
store-buffer model, and RISC-V's architected memory model,
RVWMO~\cite{riscv:unpriv}, has formal axiomatic and operational
definitions, including an operational model in
Coq~\cite{pulte:promising-arm} in the style of the promising
semantics~\cite{kang:promising}.
RISC-V also architects TSO itself, as the Ztso
extension~\cite{riscv:unpriv}, a legal strengthening of RVWMO.

Program logics for weak memory have so far been exercised only on small
programs.  Ridge~\cite{ridge:tso} verified a spinlock with
rely-guarantee reasoning over x86-TSO.
iCAP-TSO~\cite{sieczkowski:icap-tso} is the direct precedent for \sys's
memory model: a higher-order separation logic over an operational TSO
semantics, whose ``fiction of sequential consistency'' lets
well-synchronized code be verified with SC-style rules while the racy
code that implements synchronization is verified against the store
buffers directly.  Its programs are written in an idealized language,
and its case studies are locks and small data structures.  GPS and
its Iris-based successors target language-level models such as
C11~\cite{turon:gps,dang:rustbelt-relaxed}, and
AxSL~\cite{hammond:axsl} brings Iris to the relaxed Arm-A memory model
at the ISA level, but for litmus-test-sized programs.
\sys's views are inspired by views from
GPFSL~\cite{dang:rustbelt-relaxed}.  One difference in the way the xv6 kernel
proofs use views is the idea of suspended views, which are used to reason
about context switching and migrating threads between cores.

No prior kernel-scale proof reasons in a weak-memory logic.  Kernel
verification under weak memory has instead gone through reduction
theorems that recover sequential consistency: Cohen and
Schirmer~\cite{cohen:tso-reduction} proved that code obeying a
store-buffer flush discipline exhibits only SC behaviors on TSO,
developed for the (unfinished) Hyper-V verification at the C level, and
SeKVM~\cite{tao:vrm} was verified over Arm relaxed memory by proving
that the kernel follows a synchronization discipline under which its
relaxed executions reduce to SC ones.  Armada~\cite{lorch:armada}
verifies concurrent data structures by TSO-aware refinement, and
VSync~\cite{oberhauser:vsync} model-checks kernel synchronization
primitives under weak-memory models rather than proving them
deductively.  The reduction approach breaks down where a kernel is
intentionally racy: e.g., xv6 uses flag variables in shared memory
for lock-free coordination.  \sys instead verifies xv6 directly in a program logic
over a weak model of shared memory: well-synchronized code is
verified with SC-style reasoning, and xv6's racy code against the
TSO-with-load-load-reordering semantics itself.

\paragraph{Devices and DMA.}

To our knowledge, no prior machine-checked kernel proof reasons about
a DMA-capable device model running concurrently with the CPU.  seL4
assumes DMA away (devices are trusted or confined by an IOMMU);
CertiKOS's device objects~\cite{chen:certikos-drivers} are
interrupt-driven and do not access memory; Verisoft's device
models~\cite{hillebrand:devices} use programmed I/O; and the Bedrock2
lightbulb device~\cite{erbsen:lightbulb} is a stream of MMIO events.
Penninckx et al.~\cite{penninckx:io} express device I/O in separation
logic, but at the C level with abstract I/O actions.
The xv6 proof transfers ownership of DMA buffers between the kernel and
the disk controller's model, so that the proof accounts for every byte
the device reads or writes and for the moment each sector reaches the disk.

\section{Change log}
\label{sec:changelog}

\paragraph{Changes since arXiv:2609.04043v1~\cite{kaashoek:machcsl}.}
Added application-level correctness theorem (\autoref{sec:theorem}).
Added verification of user-level applications (\autoref{sec:user}).
Added non-coherent instruction cache model.
Weakened shared-memory model to allow load-load reordering.
Added subsections \autoref{sec:proof:plic} and \autoref{sec:agent:iter}.
Improved CSL explanation (\autoref{sec:csl-intro}).
Various clarifications.

\paragraph{Changes since arXiv:2609.04043v2.}
Added a more sophisticated top-level theorem that allows redirection to
files, pipes, and sandboxing of unverified binaries using \cc{seccomp}
(\autoref{sec:user-app}).
Ported the development from Rocq to Lean.
Found a bug in \cc{grep} and a bug in Sail's Lean extraction.
Extended conformance testing to the CVA6 implementation in Verilator.

\section{Acknowledgments}

Thanks to Jason Gross and Tej Chajed for inspiring us to work on
agent-assisted verification.
Thanks to Baltasar Dinis, Yun-Sheng Chang, Colin Gordon, Gernot Heiser,
Ismail Kuru, Jay Lorch, Natalie Neamtu, and Deian Stefan
for feedback that helped improve this paper.
This work was supported by NSF award CNS-2225441 and by gifts from Amazon
and Google.

\bibliographystyle{abbrvnat}
\bibliography{n-str,p,n,n-conf}{}

\begin{thebibliography}{109}
\providecommand{\natexlab}[1]{#1}
\providecommand{\url}[1]{\texttt{#1}}
\expandafter\ifx\csname urlstyle\endcsname\relax
  \providecommand{\doi}[1]{doi: #1}\else
  \providecommand{\doi}{doi: \begingroup \urlstyle{rm}\Url}\fi

\bibitem[Alkassar et~al.(2008{\natexlab{a}})Alkassar, Hillebrand, Leinenbach,
  Schirmer, and Starostin]{alkassar:verisoft-approach}
E.~Alkassar, M.~A. Hillebrand, D.~Leinenbach, N.~W. Schirmer, and A.~Starostin.
\newblock The {Verisoft} approach to systems verification.
\newblock In \emph{Proceedings of the 2nd International Conference on Verified
  Software: Theories, Tools, Experiments (VSTTE)}, pages 209--224, Toronto,
  Canada, Oct. 2008{\natexlab{a}}.

\bibitem[Alkassar et~al.(2008{\natexlab{b}})Alkassar, Schirmer, and
  Starostin]{alkassar:paging}
E.~Alkassar, N.~Schirmer, and A.~Starostin.
\newblock Formal pervasive verification of a paging mechanism.
\newblock In \emph{Proceedings of the 14th International Conference on Tools
  and Algorithms for the Construction and Analysis of Systems~(TACAS)}, pages
  109--123, Budapest, Hungary, Mar.--Apr. 2008{\natexlab{b}}.

\bibitem[Alkassar et~al.(2010{\natexlab{a}})Alkassar, Cohen, Hillebrand,
  Kovalev, and Paul]{alkassar:shadow-page-tables}
E.~Alkassar, E.~Cohen, M.~A. Hillebrand, M.~Kovalev, and W.~J. Paul.
\newblock Verifying shadow page table algorithms.
\newblock In \emph{Proceedings of the 10th Formal Methods in Computer-Aided
  Design~(FMCAD)}, pages 267--270, Lugano, Switzerland, Oct.
  2010{\natexlab{a}}.

\bibitem[Alkassar et~al.(2010{\natexlab{b}})Alkassar, Hillebrand, Paul, and
  Petrova]{alkassar:baby-hypervisor}
E.~Alkassar, M.~A. Hillebrand, W.~J. Paul, and E.~Petrova.
\newblock Automated verification of a small hypervisor.
\newblock In \emph{Proceedings of the 3rd Working Conference on Verified
  Software: Theories, Tools, and Experiments~(VSTTE)}, pages 40--54, Edinburgh,
  United Kingdom, Aug. 2010{\natexlab{b}}.

\bibitem[Alkassar et~al.(2010{\natexlab{c}})Alkassar, Paul, Starostin, and
  Tsyban]{alkassar:verisoft}
E.~Alkassar, W.~J. Paul, A.~Starostin, and A.~Tsyban.
\newblock Pervasive verification of an {OS} microkernel: Inline assembly,
  memory consumption, concurrent devices.
\newblock In \emph{Proceedings of the 3rd Working Conference on Verified
  Software: Theories, Tools, and Experiments~(VSTTE)}, pages 71--85, Edinburgh,
  United Kingdom, Aug. 2010{\natexlab{c}}.

\bibitem[Amani et~al.(2016)Amani, Hixon, Chen, Rizkallah, Chubb, O'Connor,
  Beeren, Nagashima, Lim, Sewell, Tuong, Keller, Murray, Klein, and
  Heiser]{amani:cogent}
S.~Amani, A.~Hixon, Z.~Chen, C.~Rizkallah, P.~Chubb, L.~O'Connor, J.~Beeren,
  Y.~Nagashima, J.~Lim, T.~Sewell, J.~Tuong, G.~Keller, T.~Murray, G.~Klein,
  and G.~Heiser.
\newblock \textsc{Cogent}: Verifying high-assurance file system
  implementations.
\newblock In \emph{Proceedings of the 21st International Conference on
  Architectural Support for Programming Languages and Operating
  Systems~(ASPLOS)}, pages 175--188, Atlanta, GA, Apr. 2016.

\bibitem[Armstrong et~al.(2019)Armstrong, Bauereiss, Campbell, Reid, Gray,
  Norton, Mundkur, Wassell, French, Pulte, Flur, Stark, Krishnaswami, and
  Sewell]{armstrong:sail}
A.~Armstrong, T.~Bauereiss, B.~Campbell, A.~Reid, K.~E. Gray, R.~M. Norton,
  P.~Mundkur, M.~Wassell, J.~French, C.~Pulte, S.~Flur, I.~Stark,
  N.~Krishnaswami, and P.~Sewell.
\newblock {ISA} semantics for {ARMv8-A}, {RISC-V}, and {CHERI-MIPS}.
\newblock In \emph{Proceedings of the 46th ACM Symposium on Principles of
  Programming Languages~(POPL)}, Cascais, Portugal, Jan. 2019.

\bibitem[Armstrong et~al.(2021)Armstrong, Campbell, Simner, Pulte, and
  Sewell]{armstrong:isla}
A.~Armstrong, B.~Campbell, B.~Simner, C.~Pulte, and P.~Sewell.
\newblock Isla: Integrating full-scale {ISA} semantics and axiomatic
  concurrency models.
\newblock In \emph{Proceedings of the 33rd International Conference on Computer
  Aided Verification~(CAV)}, pages 303--316, Los Angeles, CA, July 2021.

\bibitem[Bevier(1989)]{bevier:kit}
W.~R. Bevier.
\newblock Kit: A study in operating system verification.
\newblock \emph{IEEE Transactions on Software Engineering}, 15\penalty0
  (11):\penalty0 1382--1396, Nov. 1989.

\bibitem[Bevier et~al.(1989)Bevier, Hunt, Moore, and Young]{bevier:cli-stack}
W.~R. Bevier, W.~A. Hunt, Jr., J.~S. Moore, and W.~D. Young.
\newblock An approach to systems verification.
\newblock \emph{Journal of Automated Reasoning}, 5\penalty0 (4):\penalty0
  411--428, Dec. 1989.

\bibitem[Brun et~al.(2023)Brun, Achermann, Chajed, Howell, Zellweger, and
  Lattuada]{brun:beyond-isolation}
M.~Brun, R.~Achermann, T.~Chajed, J.~Howell, G.~Zellweger, and A.~Lattuada.
\newblock Beyond isolation: {OS} verification as a foundation for correct
  applications.
\newblock In \emph{Proceedings of the 19th Workshop on Hot Topics in Operating
  Systems (HotOS)}, Providence, RI, June 2023.

\bibitem[Chajed et~al.(2017)Chajed, Chen, Chlipala, Kaashoek, Zeldovich, and
  Ziegler]{chajed:fscq-cacm}
T.~Chajed, H.~Chen, A.~Chlipala, M.~F. Kaashoek, N.~Zeldovich, and D.~Ziegler.
\newblock Certifying a file system using {Crash} {Hoare} {Logic}: Correctness
  in the presence of crashes.
\newblock \emph{Communications of the ACM}, 60\penalty0 (4):\penalty0 75--84,
  Apr. 2017.

\bibitem[Chajed et~al.(2019{\natexlab{a}})Chajed, Tassarotti, Kaashoek, and
  Zeldovich]{chajed:argosy}
T.~Chajed, J.~Tassarotti, M.~F. Kaashoek, and N.~Zeldovich.
\newblock Argosy: Verifying layered storage systems with recovery refinement.
\newblock In \emph{Proceedings of the 40th ACM SIGPLAN Conference on
  Programming Language Design and Implementation~(PLDI)}, pages 1037--1051,
  Phoenix, AZ, June 2019{\natexlab{a}}.

\bibitem[Chajed et~al.(2019{\natexlab{b}})Chajed, Tassarotti, Kaashoek, and
  Zeldovich]{chajed:perennial}
T.~Chajed, J.~Tassarotti, M.~F. Kaashoek, and N.~Zeldovich.
\newblock Verifying concurrent, crash-safe systems with {Perennial}.
\newblock In \emph{Proceedings of the 27th ACM Symposium on Operating Systems
  Principles~(SOSP)}, pages 243--258, Huntsville, Ontario, Canada, Oct.
  2019{\natexlab{b}}.

\bibitem[Chajed et~al.(2021)Chajed, Tassarotti, Theng, Jung, Kaashoek, and
  Zeldovich]{chajed:go-journal}
T.~Chajed, J.~Tassarotti, M.~Theng, R.~Jung, M.~F. Kaashoek, and N.~Zeldovich.
\newblock {GoJournal}: a verified, concurrent, crash-safe journaling system.
\newblock In \emph{Proceedings of the 15th USENIX Symposium on Operating
  Systems Design and Implementation~(OSDI)}, pages 423--439, Virtual
  conference, July 2021.

\bibitem[Chajed et~al.(2022)Chajed, Tassarotti, Theng, Kaashoek, and
  Zeldovich]{chajed:daisynfs}
T.~Chajed, J.~Tassarotti, M.~Theng, M.~F. Kaashoek, and N.~Zeldovich.
\newblock Verifying the {DaisyNFS} concurrent and crash-safe file system with
  sequential reasoning.
\newblock In \emph{Proceedings of the 16th USENIX Symposium on Operating
  Systems Design and Implementation~(OSDI)}, pages 447--463, Carlsbad, CA, July
  2022.

\bibitem[Chen et~al.(2015)Chen, Ziegler, Chajed, Chlipala, Kaashoek, and
  Zeldovich]{chen:fscq}
H.~Chen, D.~Ziegler, T.~Chajed, A.~Chlipala, M.~F. Kaashoek, and N.~Zeldovich.
\newblock Using {Crash Hoare Logic} for certifying the {FSCQ} file system.
\newblock In \emph{Proceedings of the 25th ACM Symposium on Operating Systems
  Principles~(SOSP)}, pages 18--37, Monterey, CA, Oct. 2015.

\bibitem[Chen et~al.(2016)Chen, Wu, Shao, Lockerman, and
  Gu]{chen:certikos-drivers}
H.~Chen, X.~Wu, Z.~Shao, J.~Lockerman, and R.~Gu.
\newblock Toward compositional verification of interruptible {OS} kernels and
  device drivers.
\newblock In \emph{Proceedings of the 37th ACM SIGPLAN Conference on
  Programming Language Design and Implementation~(PLDI)}, pages 431--447, Santa
  Barbara, CA, June 2016.

\bibitem[Chen et~al.(2017)Chen, Chajed, Konradi, Wang, \.Ileri, Chlipala,
  Kaashoek, and Zeldovich]{chen:dfscq}
H.~Chen, T.~Chajed, A.~Konradi, S.~Wang, A.~\.Ileri, A.~Chlipala, M.~F.
  Kaashoek, and N.~Zeldovich.
\newblock Verifying a high-performance crash-safe file system using a tree
  specification.
\newblock In \emph{Proceedings of the 26th ACM Symposium on Operating Systems
  Principles~(SOSP)}, pages 270--286, Shanghai, China, Oct. 2017.

\bibitem[Chen et~al.(2025)Chen, Li, Zhang, Narayanan, and
  Burtsev]{chen:atmosphere}
X.~Chen, Z.~Li, J.~Zhang, V.~Narayanan, and A.~Burtsev.
\newblock Atmosphere: Practical verified kernels with {Rust} and {Verus}.
\newblock In \emph{Proceedings of the 31st ACM Symposium on Operating Systems
  Principles~(SOSP)}, Seoul, South Korea, Oct. 2025.

\bibitem[Chlipala(2013)]{chlipala:bedrock2}
A.~Chlipala.
\newblock The {Bedrock} structured programming system: Combining generative
  metaprogramming and {Hoare} logic in an extensible program verifier.
\newblock In \emph{Proceedings of the 18th ACM SIGPLAN International Conference
  on Functional Programming~(ICFP)}, pages 391--402, Boston, MA, Sept. 2013.

\bibitem[Chlipala(2015)]{chlipala:bedrock-web}
A.~Chlipala.
\newblock From network interface to multithreaded web applications: A case
  study in modular program verification.
\newblock In \emph{Proceedings of the 42nd ACM Symposium on Principles of
  Programming Languages~(POPL)}, pages 609--622, Mumbai, India, Jan. 2015.

\bibitem[Choi et~al.(2017)Choi, Vijayaraghavan, Sherman, Chlipala, and
  Arvind]{choi:kami}
J.~Choi, M.~Vijayaraghavan, B.~Sherman, A.~Chlipala, and Arvind.
\newblock Kami: A platform for high-level parametric hardware specification and
  its modular verification.
\newblock In \emph{Proceedings of the 22nd ACM SIGPLAN International Conference
  on Functional Programming~(ICFP)}, Oxford, United Kingdom, Sept. 2017.

\bibitem[Cohen and Schirmer(2010)]{cohen:tso-reduction}
E.~Cohen and N.~Schirmer.
\newblock From total store order to sequential consistency: A practical
  reduction theorem.
\newblock In \emph{Proceedings of the 1st International Conference on
  Interactive Theorem Proving}, Edinburgh, United Kingdom, July 2010.

\bibitem[Cohen et~al.(2013)Cohen, Paul, and
  Schmaltz]{cohen:multicore-hypervisor}
E.~Cohen, W.~Paul, and S.~Schmaltz.
\newblock Theory of multi core hypervisor verification.
\newblock In \emph{Proceedings of the 39th International Conference on Current
  Trends in Theory and Practice of Computer Science (SOFSEM)}, pages 1--27,
  {\v{S}}pindler{\r{u}}v Ml{\'y}n, Czech Republic, Jan. 2013.

\bibitem[Cox et~al.(2026)Cox, Kaashoek, and Morris]{xv6}
R.~Cox, M.~F. Kaashoek, and R.~T. Morris.
\newblock Xv6, a simple {Unix}-like teaching operating system, 2026.
\newblock \url{https://github.com/mit-pdos/xv6-riscv}.

\bibitem[Dang et~al.(2020)Dang, Jourdan, Kaiser, and
  Dreyer]{dang:rustbelt-relaxed}
H.-H. Dang, J.-H. Jourdan, J.-O. Kaiser, and D.~Dreyer.
\newblock {RustBelt} meets relaxed memory.
\newblock In \emph{Proceedings of the 47th ACM Symposium on Principles of
  Programming Languages~(POPL)}, New Orleans, LA, Jan. 2020.

\bibitem[Erbsen et~al.(2021)Erbsen, Gruetter, Choi, Wood, and
  Chlipala]{erbsen:lightbulb}
A.~Erbsen, S.~Gruetter, J.~Choi, C.~Wood, and A.~Chlipala.
\newblock Integration verification across software and hardware for a simple
  embedded system.
\newblock In \emph{Proceedings of the 42nd ACM SIGPLAN Conference on
  Programming Language Design and Implementation~(PLDI)}, Virtual conference,
  June 2021.

\bibitem[Erbsen et~al.(2024)Erbsen, Philipoom, Jamner, Lin, Gruetter,
  Pit-Claudel, and Chlipala]{erbsen:garage-door}
A.~Erbsen, J.~Philipoom, D.~Jamner, A.~Lin, S.~Gruetter, C.~Pit-Claudel, and
  A.~Chlipala.
\newblock Foundational integration verification of a cryptographic server.
\newblock In \emph{Proceedings of the 45th ACM SIGPLAN Conference on
  Programming Language Design and Implementation~(PLDI)}, Copenhagen, Denmark,
  June 2024.

\bibitem[Feiertag and Neumann(1979)]{feiertag:psos}
R.~J. Feiertag and P.~G. Neumann.
\newblock The foundations of a provably secure operating system {(PSOS)}.
\newblock In \emph{Proceedings of the 1979 National Computer Conference}, pages
  329--334, 1979.

\bibitem[Feiertag et~al.(1977)Feiertag, Levitt, and
  Robinson]{feiertag:multilevel}
R.~J. Feiertag, K.~N. Levitt, and L.~Robinson.
\newblock Proving multilevel security of a system design.
\newblock In \emph{Proceedings of the 6th ACM Symposium on Operating Systems
  Principles~(SOSP)}, pages 57--65, West Lafayette, IN, Nov. 1977.

\bibitem[Feng et~al.(2007)Feng, Ferreira, and Shao]{feng:csl-ag}
X.~Feng, R.~Ferreira, and Z.~Shao.
\newblock On the relationship between concurrent separation logic and
  assume-guarantee reasoning.
\newblock In \emph{Proceedings of the 16th European Symposium on
  Programming~(ESOP)}, pages 173--188, Braga, Portugal, Mar. 2007.

\bibitem[Feng et~al.(2008)Feng, Shao, Dong, and Guo]{feng:interrupts}
X.~Feng, Z.~Shao, Y.~Dong, and Y.~Guo.
\newblock Certifying low-level programs with hardware interrupts and preemptive
  threads.
\newblock In \emph{Proceedings of the 29th ACM SIGPLAN Conference on
  Programming Language Design and Implementation~(PLDI)}, pages 170--182,
  Tucson, AZ, June 2008.

\bibitem[Ferraiuolo et~al.(2017)Ferraiuolo, Baumann, Hawblitzel, and
  Parno]{ferraiuolo:komodo}
A.~Ferraiuolo, A.~Baumann, C.~Hawblitzel, and B.~Parno.
\newblock {Komodo}: Using verification to disentangle secure-enclave hardware
  from software.
\newblock In \emph{Proceedings of the 26th ACM Symposium on Operating Systems
  Principles~(SOSP)}, pages 287--305, Shanghai, China, Oct. 2017.

\bibitem[Fox and Myreen(2010)]{fox:armv7}
A.~Fox and M.~O. Myreen.
\newblock A trustworthy monadic formalization of the {ARMv7} instruction set
  architecture.
\newblock In \emph{Proceedings of the 1st International Conference on
  Interactive Theorem Proving}, pages 243--258, Edinburgh, United Kingdom, July
  2010.

\bibitem[Georges et~al.(2024)Georges, Gu{\'e}neau, {Van Strydonck}, Timany,
  Trieu, Devriese, and Birkedal]{georges:cerise}
A.~L. Georges, A.~Gu{\'e}neau, T.~{Van Strydonck}, A.~Timany, A.~Trieu,
  D.~Devriese, and L.~Birkedal.
\newblock Cerise: Program verification on a capability machine in the presence
  of untrusted code.
\newblock \emph{Journal of the ACM}, 71\penalty0 (1):\penalty0 3:1--3:59, Feb.
  2024.

\bibitem[Goel(2016)]{goel:thesis}
S.~Goel.
\newblock \emph{Formal Verification of Application and System Programs Based on
  a Validated x86 {ISA} Model}.
\newblock PhD thesis, The University of Texas at Austin, 2016.

\bibitem[Goel et~al.(2017)Goel, Hunt, and Kaufmann]{goel:x86isa}
S.~Goel, W.~A. Hunt, Jr., and M.~Kaufmann.
\newblock Engineering a formal, executable x86 {ISA} simulator for software
  verification.
\newblock \emph{Provably Correct Systems}, pages 173--209, 2017.

\bibitem[Gu et~al.(2015)Gu, Koenig, Ramananandro, Shao, Wu, Weng, Zhang, and
  Guo]{gu:certikos-layers}
R.~Gu, J.~Koenig, T.~Ramananandro, Z.~Shao, X.~Wu, S.-C. Weng, H.~Zhang, and
  Y.~Guo.
\newblock Deep specifications and certified abstraction layers.
\newblock In \emph{Proceedings of the 42nd ACM Symposium on Principles of
  Programming Languages~(POPL)}, pages 595--608, Mumbai, India, Jan. 2015.

\bibitem[Gu et~al.(2016)Gu, Shao, Chen, Wu, Kim, Sj{\"o}berg, and
  Costanzo]{gu:certikos}
R.~Gu, Z.~Shao, H.~Chen, X.~N. Wu, J.~Kim, V.~Sj{\"o}berg, and D.~Costanzo.
\newblock {CertiKOS}: An extensible architecture for building certified
  concurrent {OS} kernels.
\newblock In \emph{Proceedings of the 12th USENIX Symposium on Operating
  Systems Design and Implementation~(OSDI)}, pages 653--669, Savannah, GA, Nov.
  2016.

\bibitem[Gu et~al.(2018)Gu, Shao, Kim, Wu, Koenig, Sj{\"o}berg, Chen, Costanzo,
  and Ramananandro]{gu:certikos-ccal}
R.~Gu, Z.~Shao, J.~Kim, X.~Wu, J.~Koenig, V.~Sj{\"o}berg, H.~Chen, D.~Costanzo,
  and T.~Ramananandro.
\newblock Certified concurrent abstraction layers.
\newblock In \emph{Proceedings of the 39th ACM SIGPLAN Conference on
  Programming Language Design and Implementation~(PLDI)}, pages 646--661,
  Philadelphia, PA, June 2018.

\bibitem[Guanciale et~al.(2016)Guanciale, Nemati, Dam, and
  Baumann]{guanciale:memory-isolation}
R.~Guanciale, H.~Nemati, M.~Dam, and C.~Baumann.
\newblock Provably secure memory isolation for {Linux} on {ARM}.
\newblock \emph{Journal of Computer Security}, 24\penalty0 (6):\penalty0
  793--837, 2016.

\bibitem[Hammond et~al.(2024)Hammond, Liu, P\'{e}rami, Sewell, Birkedal, and
  Pichon-Pharabod]{hammond:axsl}
A.~Hammond, Z.~Liu, T.~P\'{e}rami, P.~Sewell, L.~Birkedal, and
  J.~Pichon-Pharabod.
\newblock An axiomatic basis for computer programming on the relaxed {Arm-A}
  architecture: The {AxSL} logic.
\newblock In \emph{Proceedings of the 51st ACM Symposium on Principles of
  Programming Languages~(POPL)}, London, United Kingdom, Jan. 2024.

\bibitem[Hance et~al.(2020)Hance, Lattuada, Hawblitzel, Howell, Johnson, and
  Parno]{hance:veribetrkv}
T.~Hance, A.~Lattuada, C.~Hawblitzel, J.~Howell, R.~Johnson, and B.~Parno.
\newblock Storage systems are distributed systems (so verify them that way!).
\newblock In \emph{Proceedings of the 14th USENIX Symposium on Operating
  Systems Design and Implementation~(OSDI)}, pages 99--115, Virtual conference,
  Nov. 2020.

\bibitem[Hawblitzel et~al.(2014)Hawblitzel, Howell, Lorch, Narayan, Parno,
  Zhang, and Zill]{hawblitzel:ironclad}
C.~Hawblitzel, J.~Howell, J.~R. Lorch, A.~Narayan, B.~Parno, D.~Zhang, and
  B.~Zill.
\newblock {Ironclad} {Apps}: End-to-end security via automated full-system
  verification.
\newblock In \emph{Proceedings of the 11th USENIX Symposium on Operating
  Systems Design and Implementation~(OSDI)}, pages 165--181, Broomfield, CO,
  Oct. 2014.

\bibitem[Hillebrand et~al.(2005)Hillebrand, {In der Rieden}, and
  Paul]{hillebrand:devices}
M.~Hillebrand, T.~{In der Rieden}, and W.~Paul.
\newblock Dealing with {I/O} devices in the context of pervasive system
  verification.
\newblock In \emph{Proceedings of the 23rd IEEE International Conference on
  Computer Design (ICCD)}, pages 309--316, San Jose, CA, Oct. 2005.

\bibitem[\.Ileri et~al.(2018)\.Ileri, Chajed, Chlipala, Kaashoek, and
  Zeldovich]{ileri:disksec}
A.~\.Ileri, T.~Chajed, A.~Chlipala, M.~F. Kaashoek, and N.~Zeldovich.
\newblock Proving confidentiality in a file system using {DiskSec}.
\newblock In \emph{Proceedings of the 13th USENIX Symposium on Operating
  Systems Design and Implementation~(OSDI)}, pages 323--338, Carlsbad, CA, Oct.
  2018.

\bibitem[Jensen et~al.(2013)Jensen, Benton, and Kennedy]{jensen:high-level-sep}
J.~B. Jensen, N.~Benton, and A.~Kennedy.
\newblock High-level separation logic for low-level code.
\newblock In \emph{Proceedings of the 40th ACM Symposium on Principles of
  Programming Languages~(POPL)}, pages 301--314, Rome, Italy, Jan. 2013.

\bibitem[Jung et~al.(2015)Jung, Swasey, Sieczkowski, Svendsen, Turon, Birkedal,
  and Dreyer]{jung:iris-1}
R.~Jung, D.~Swasey, F.~Sieczkowski, K.~Svendsen, A.~Turon, L.~Birkedal, and
  D.~Dreyer.
\newblock Iris: Monoids and invariants as an orthogonal basis for concurrent
  reasoning.
\newblock In \emph{Proceedings of the 42nd ACM Symposium on Principles of
  Programming Languages~(POPL)}, Mumbai, India, Jan. 2015.

\bibitem[Jung et~al.(2018)Jung, Krebbers, Jourdan, Bizjak, Birkedal, and
  Dreyer]{jung:iris-jfp}
R.~Jung, R.~Krebbers, J.-H. Jourdan, A.~Bizjak, L.~Birkedal, and D.~Dreyer.
\newblock Iris from the ground up: a modular foundation for higher-order
  concurrent separation logic.
\newblock \emph{Journal of Functional Programming}, 28:\penalty0 e20, 2018.

\bibitem[Kaashoek and Zeldovich(2026)]{kaashoek:machcsl}
M.~F. Kaashoek and N.~Zeldovich.
\newblock Extending concurrent separation logic to the hardware level to verify
  the {xv6} {OS} kernel on {RISC-V} with {AI} agents.
\newblock {arXiv:2609.04043 [cs.LO]}, Sept. 2026.
\newblock Available at \url{https://arxiv.org/abs/2609.04043}.

\bibitem[Kang et~al.(2017)Kang, Hur, Lahav, Vafeiadis, and
  Dreyer]{kang:promising}
J.~Kang, C.-K. Hur, O.~Lahav, V.~Vafeiadis, and D.~Dreyer.
\newblock A promising semantics for relaxed-memory concurrency.
\newblock In \emph{Proceedings of the 44th ACM Symposium on Principles of
  Programming Languages~(POPL)}, pages 175--189, Paris, France, Jan. 2017.

\bibitem[Karger et~al.(1990)Karger, Zurko, Bonin, Mason, and Kahn]{karger:vmm}
P.~A. Karger, M.~E. Zurko, D.~W. Bonin, A.~H. Mason, and C.~E. Kahn.
\newblock A {VMM} security kernel for the {VAX} architecture.
\newblock In \emph{Proceedings of the 11th IEEE Symposium on Security and
  Privacy}, pages 2--19, Oakland, CA, May 1990.

\bibitem[Klein et~al.(2009)Klein, Elphinstone, Heiser, Andronick, Cock, Derrin,
  Elkaduwe, Engelhardt, Norrish, Kolanski, Sewell, Tuch, and
  Winwood]{klein:sel4}
G.~Klein, K.~Elphinstone, G.~Heiser, J.~Andronick, D.~Cock, P.~Derrin,
  D.~Elkaduwe, K.~Engelhardt, M.~Norrish, R.~Kolanski, T.~Sewell, H.~Tuch, and
  S.~Winwood.
\newblock {seL4}: Formal verification of an {OS} kernel.
\newblock In \emph{Proceedings of the 22nd ACM Symposium on Operating Systems
  Principles~(SOSP)}, pages 207--220, Big Sky, MT, Oct. 2009.

\bibitem[Klein et~al.(2014)Klein, Andronick, Elphinstone, Murray, Sewell,
  Kolanski, and Heiser]{klein:sel4-tocs}
G.~Klein, J.~Andronick, K.~Elphinstone, T.~Murray, T.~Sewell, R.~Kolanski, and
  G.~Heiser.
\newblock Comprehensive formal verification of an {OS} microkernel.
\newblock \emph{ACM Transactions on Computer Systems}, 32\penalty0
  (1):\penalty0 2:1--70, Feb. 2014.

\bibitem[Klein et~al.(2018)Klein, Andronick, Fernandez, Kuz, Murray, and
  Heiser]{klein:sel4-cacm}
G.~Klein, J.~Andronick, M.~Fernandez, I.~Kuz, T.~Murray, and G.~Heiser.
\newblock Formally verified software in the real world.
\newblock \emph{Communications of the ACM}, 61\penalty0 (10):\penalty0 68--77,
  Oct. 2018.

\bibitem[Kolanski(2011)]{kolanski:phd}
R.~Kolanski.
\newblock \emph{Verification of Programs in Virtual Memory Using Separation
  Logic}.
\newblock PhD thesis, University of New South Wales, July 2011.

\bibitem[Kolanski and Klein(2008)]{kolanski:mapped-sep-logic}
R.~Kolanski and G.~Klein.
\newblock Mapped separation logic.
\newblock In \emph{Proceedings of the 2nd International Conference on Verified
  Software: Theories, Tools, and Experiments~(VSTTE)}, pages 15--29, Toronto,
  Canada, Oct. 2008.

\bibitem[Kolanski and Klein(2009)]{kolanski:mapping-sep}
R.~Kolanski and G.~Klein.
\newblock Types, maps and separation logic.
\newblock In \emph{Proceedings of the 22nd International Conference on Theorem
  Proving in Higher Order Logics}, pages 276--292, Munich, Germany, Aug. 2009.

\bibitem[Krebbers et~al.(2017)Krebbers, Jung, Bizjak, Jourdan, Dreyer, and
  Birkedal]{krebbers:iris3.0}
R.~Krebbers, R.~Jung, A.~Bizjak, J.-H. Jourdan, D.~Dreyer, and L.~Birkedal.
\newblock The essence of higher-order concurrent separation logic.
\newblock In \emph{Proceedings of the 26th European Symposium on
  Programming~(ESOP)}, pages 696--723, Uppsala, Sweden, Apr. 2017.

\bibitem[Kumar et~al.(2014)Kumar, Myreen, Norrish, and Owens]{kumar:cakeml}
R.~Kumar, M.~O. Myreen, M.~Norrish, and S.~Owens.
\newblock {CakeML}: a verified implementation of {ML}.
\newblock In \emph{Proceedings of the 41st ACM Symposium on Principles of
  Programming Languages~(POPL)}, pages 179--192, San Diego, CA, Jan. 2014.

\bibitem[Kuru(2025)]{kuru:thesis}
I.~Kuru.
\newblock \emph{Modal Abstractions for Operating System Kernels}.
\newblock PhD thesis, Drexel University, Mar. 2025.

\bibitem[Kuru and Gordon(2025)]{kuru:modal}
I.~Kuru and C.~S. Gordon.
\newblock Modal abstractions for virtualizing memory addresses.
\newblock \emph{Proceedings of the {ACM} on Programming Languages}, 9\penalty0
  (OOPSLA2):\penalty0 2338--2366, Oct. 2025.

\bibitem[LeBlanc et~al.(2025)LeBlanc, Lorch, Hawblitzel, Huang, Tao, Zeldovich,
  and Chidambaram]{leblanc:power}
H.~LeBlanc, J.~R. Lorch, C.~Hawblitzel, C.~Huang, Y.~Tao, N.~Zeldovich, and
  V.~Chidambaram.
\newblock {PoWER} never corrupts: Tool-agnostic verification of crash
  consistency and corruption detection.
\newblock In \emph{Proceedings of the 19th USENIX Symposium on Operating
  Systems Design and Implementation~(OSDI)}, Boston, MA, July 2025.

\bibitem[Leinenbach and Santen(2009)]{leinenbach:hyperv}
D.~Leinenbach and T.~Santen.
\newblock {Verifying} the {Microsoft} {Hyper-V} hypervisor with {VCC}.
\newblock In \emph{Proceedings of the 16th International Symposium on Formal
  Methods~(FM)}, pages 806--809, Eindhoven, The Netherlands, Nov. 2009.

\bibitem[Li et~al.(2021)Li, Li, Gu, Nieh, and Hui]{li:sekvm}
S.~Li, X.~Li, R.~Gu, J.~Nieh, and J.~Z. Hui.
\newblock Formally verified memory protection for a commodity multiprocessor
  hypervisor.
\newblock In \emph{Proceedings of the 30th USENIX Security Symposium},
  Vancouver, Canada, Aug. 2021.

\bibitem[Li et~al.(2022)Li, Li, Dall, Gu, Nieh, Sait, and
  Stockwell]{li:armv9cca}
X.~Li, X.~Li, C.~Dall, R.~Gu, J.~Nieh, Y.~Sait, and G.~Stockwell.
\newblock Design and verification of the {Arm} confidential compute
  architecture.
\newblock In \emph{Proceedings of the 16th USENIX Symposium on Operating
  Systems Design and Implementation~(OSDI)}, Carlsbad, CA, July 2022.

\bibitem[Lorch et~al.(2020)Lorch, Chen, Kapritsos, Parno, Qadeer, Sharma,
  Wilcox, and Zhao]{lorch:armada}
J.~R. Lorch, Y.~Chen, M.~Kapritsos, B.~Parno, S.~Qadeer, U.~Sharma, J.~R.
  Wilcox, and X.~Zhao.
\newblock Armada: Low-effort verification of high-performance concurrent
  programs.
\newblock In \emph{Proceedings of the 41st ACM SIGPLAN Conference on
  Programming Language Design and Implementation~(PLDI)}, pages 197--210,
  London, United Kingdom, June 2020.

\bibitem[Lööw et~al.(2019)Lööw, Kumar, Tan, Myreen, Norrish, Abrahamsson,
  and Fox]{loow:cakeml}
A.~Lööw, R.~Kumar, Y.~K. Tan, M.~O. Myreen, M.~Norrish, O.~Abrahamsson, and
  A.~Fox.
\newblock Verified compilation on a verified processor.
\newblock In \emph{Proceedings of the 40th ACM SIGPLAN Conference on
  Programming Language Design and Implementation~(PLDI)}, pages 1041--1053,
  Phoenix, AZ, June 2019.

\bibitem[Mansky et~al.(2020)Mansky, Honor{\'e}, and Appel]{mansky:vst-certikos}
W.~Mansky, W.~Honor{\'e}, and A.~W. Appel.
\newblock Connecting higher-order separation logic to a first-order outside
  world.
\newblock In \emph{Proceedings of the 29th European Symposium on
  Programming~(ESOP)}, Dublin, Ireland, Apr. 2020.

\bibitem[McCauley and Drongowski(1979)]{mccauley:ksos}
E.~J. McCauley and P.~J. Drongowski.
\newblock {KSOS}---the design of a secure operating system.
\newblock In \emph{Proceedings of the 1979 National Computer Conference}, pages
  345--354, 1979.

\bibitem[Millen(1976)]{millen:kernel-validation}
J.~K. Millen.
\newblock Security kernel validation in practice.
\newblock \emph{Communications of the ACM}, 19\penalty0 (5):\penalty0 243--250,
  May 1976.

\bibitem[Myreen and Gordon(2007)]{myreen:machine-code}
M.~O. Myreen and M.~J.~C. Gordon.
\newblock {Hoare} logic for realistically modelled machine code.
\newblock In \emph{Proceedings of the 13th International Conference on Tools
  and Algorithms for the Construction and Analysis of Systems~(TACAS)}, pages
  568--582, Braga, Portugal, Mar.--Apr. 2007.

\bibitem[Nelson et~al.(2017)Nelson, Sigurbjarnarson, Zhang, Johnson, Bornholt,
  Torlak, and Wang]{nelson:hyperkernel}
L.~Nelson, H.~Sigurbjarnarson, K.~Zhang, D.~Johnson, J.~Bornholt, E.~Torlak,
  and X.~Wang.
\newblock Hyperkernel: Push-button verification of an {OS} kernel.
\newblock In \emph{Proceedings of the 26th ACM Symposium on Operating Systems
  Principles~(SOSP)}, pages 252--269, Shanghai, China, Oct. 2017.

\bibitem[Nelson et~al.(2019)Nelson, Bornholt, Gu, Baumann, Torlak, and
  Wang]{nelson:serval}
L.~Nelson, J.~Bornholt, R.~Gu, A.~Baumann, E.~Torlak, and X.~Wang.
\newblock Scaling symbolic evaluation for automated verification of systems
  code with {Serval}.
\newblock In \emph{Proceedings of the 27th ACM Symposium on Operating Systems
  Principles~(SOSP)}, pages 225--242, Huntsville, Ontario, Canada, Oct. 2019.

\bibitem[Nemati et~al.(2015)Nemati, Guanciale, and Dam]{nemati:armv7-mmu}
H.~Nemati, R.~Guanciale, and M.~Dam.
\newblock Trustworthy virtualization of the {ARMv7} memory subsystem.
\newblock In \emph{Proceedings of the 41st International Conference on Current
  Trends in Theory and Practice of Computer Science (SOFSEM)}, pages 578--589,
  Pec pod Sn{\v{e}}{\v{z}}kou, Czech Republic, Jan. 2015.

\bibitem[Nienhuis et~al.(2020)Nienhuis, Joannou, Bauereiss, Fox, Roe, Campbell,
  Naylor, Norton, Moore, Neumann, Stark, Watson, and Sewell]{nienhuis:cheri}
K.~Nienhuis, A.~Joannou, T.~Bauereiss, A.~Fox, M.~Roe, B.~Campbell, M.~Naylor,
  R.~M. Norton, S.~W. Moore, P.~G. Neumann, I.~Stark, R.~N.~M. Watson, and
  P.~Sewell.
\newblock Rigorous engineering for hardware security: Formal modelling and
  proof in the {CHERI} design and implementation process.
\newblock In \emph{Proceedings of the 41st IEEE Symposium on Security and
  Privacy}, pages 1003--1020, San Francisco, CA, May 2020.

\bibitem[Ntzik et~al.(2015)Ntzik, da~Rocha~Pinto, and Gardner]{ntzik:faults}
G.~Ntzik, P.~da~Rocha~Pinto, and P.~Gardner.
\newblock Fault-tolerant resource reasoning.
\newblock In \emph{Proceedings of the 13th Asian Symposium on Programming
  Languages and Systems (APLAS)}, pages 169--188, Pohang, South Korea,
  Nov.--Dec. 2015.

\bibitem[Oberhauser et~al.(2021)Oberhauser, de~Lima~Chehab, Behrens, Fu,
  Paolillo, Oberhauser, Bhat, Wen, Chen, Kim, and Vafeiadis]{oberhauser:vsync}
J.~Oberhauser, R.~L. de~Lima~Chehab, D.~Behrens, M.~Fu, A.~Paolillo,
  L.~Oberhauser, K.~Bhat, Y.~Wen, H.~Chen, J.~Kim, and V.~Vafeiadis.
\newblock {VSync}: Push-button verification and optimization for
  synchronization primitives on weak memory models.
\newblock In \emph{Proceedings of the 26th International Conference on
  Architectural Support for Programming Languages and Operating
  Systems~(ASPLOS)}, Virtual conference, Apr. 2021.

\bibitem[Penninckx et~al.(2015)Penninckx, Jacobs, and Piessens]{penninckx:io}
W.~Penninckx, B.~Jacobs, and F.~Piessens.
\newblock Sound, modular and compositional verification of the input/output
  behavior of programs.
\newblock In \emph{Proceedings of the 24th European Symposium on
  Programming~(ESOP)}, pages 158--182, London, United Kingdom, Apr. 2015.

\bibitem[Pulte et~al.(2019)Pulte, Pichon-Pharabod, Kang, Lee, and
  Hur]{pulte:promising-arm}
C.~Pulte, J.~Pichon-Pharabod, J.~Kang, S.-H. Lee, and C.-K. Hur.
\newblock {Promising-ARM/RISC-V}: A simpler and faster operational concurrency
  model.
\newblock In \emph{Proceedings of the 40th ACM SIGPLAN Conference on
  Programming Language Design and Implementation~(PLDI)}, pages 1--15, Phoenix,
  AZ, June 2019.

\bibitem[Ridge(2010)]{ridge:tso}
T.~Ridge.
\newblock A rely-guarantee proof system for {x86-TSO}.
\newblock In \emph{Proceedings of the 3rd Working Conference on Verified
  Software: Theories, Tools, and Experiments~(VSTTE)}, Edinburgh, United
  Kingdom, Aug. 2010.

\bibitem[Ridge et~al.(2015)Ridge, Sheets, Tuerk, Giugliano, Madhavapeddy, and
  Sewell]{ridge:sibylfs}
T.~Ridge, D.~Sheets, T.~Tuerk, A.~Giugliano, A.~Madhavapeddy, and P.~Sewell.
\newblock {SibylFS}: formal specification and oracle-based testing for {POSIX}
  and real-world file systems.
\newblock In \emph{Proceedings of the 25th ACM Symposium on Operating Systems
  Principles~(SOSP)}, pages 38--53, Monterey, CA, Oct. 2015.

\bibitem[{RISC-V International}(2024)]{riscv:unpriv}
{RISC-V International}.
\newblock The {RISC-V} instruction set manual, volume {I}: Unprivileged
  architecture.
\newblock \url{https://riscv.org/specifications/ratified/}, 2024.

\bibitem[{RISC-V International}(2026)]{sail-riscv}
{RISC-V International}.
\newblock Formal specification of the {RISC-V} {ISA}.
\newblock \url{https://github.com/riscv/sail-riscv}, 2026.

\bibitem[Sammler et~al.(2022)Sammler, Hammond, Lepigre, Campbell,
  Pichon-Pharabod, Dreyer, Garg, and Sewell]{sammler:islaris}
M.~Sammler, A.~Hammond, R.~Lepigre, B.~Campbell, J.~Pichon-Pharabod, D.~Dreyer,
  D.~Garg, and P.~Sewell.
\newblock Islaris: Verification of machine code against authoritative {ISA}
  semantics.
\newblock In \emph{Proceedings of the 43rd ACM SIGPLAN Conference on
  Programming Language Design and Implementation~(PLDI)}, pages 825--840, San
  Diego, CA, June 2022.

\bibitem[Schellhorn et~al.(2014)Schellhorn, Ernst, Pf\"{a}hler, Haneberg, and
  Reif]{schellhorn:flashix}
G.~Schellhorn, G.~Ernst, J.~Pf\"{a}hler, D.~Haneberg, and W.~Reif.
\newblock Development of a verified flash file system.
\newblock In \emph{Proceedings of the ABZ Conference}, pages 9--24, Toulouse,
  France, June 2014.

\bibitem[Schiller(1975)]{schiller:pdp11-kernel}
W.~L. Schiller.
\newblock The design and specification of a security kernel for the
  {PDP-11/45}.
\newblock Technical Report MTR-2934, MITRE Corp., Bedford, MA, Mar. 1975.

\bibitem[Sewell et~al.(2010)Sewell, Sarkar, Owens, {Zappa Nardelli}, and
  Myreen]{sewell:x86-tso}
P.~Sewell, S.~Sarkar, S.~Owens, F.~{Zappa Nardelli}, and M.~O. Myreen.
\newblock {x86-TSO}: A rigorous and usable programmer's model for x86
  multiprocessors.
\newblock \emph{Communications of the ACM}, 53\penalty0 (7):\penalty0 89--97,
  July 2010.

\bibitem[Sewell et~al.(2013)Sewell, Myreen, and Klein]{sewell:sel4-tv}
T.~Sewell, M.~Myreen, and G.~Klein.
\newblock Translation validation for a verified {OS} kernel.
\newblock In \emph{Proceedings of the 34th ACM SIGPLAN Conference on
  Programming Language Design and Implementation~(PLDI)}, pages 471--482,
  Seattle, WA, June 2013.

\bibitem[Sieczkowski et~al.(2015)Sieczkowski, Svendsen, Birkedal, and
  Pichon-Pharabod]{sieczkowski:icap-tso}
F.~Sieczkowski, K.~Svendsen, L.~Birkedal, and J.~Pichon-Pharabod.
\newblock A separation logic for fictional sequential consistency.
\newblock In \emph{Proceedings of the 24th European Symposium on
  Programming~(ESOP)}, London, United Kingdom, Apr. 2015.

\bibitem[Sigurbjarnarson et~al.(2016)Sigurbjarnarson, Bornholt, Torlak, and
  Wang]{sigurbjarnarson:yggdrasil}
H.~Sigurbjarnarson, J.~Bornholt, E.~Torlak, and X.~Wang.
\newblock Push-button verification of file systems via crash refinement.
\newblock In \emph{Proceedings of the 12th USENIX Symposium on Operating
  Systems Design and Implementation~(OSDI)}, pages 1--16, Savannah, GA, Nov.
  2016.

\bibitem[Sigurbjarnarson et~al.(2018)Sigurbjarnarson, Nelson, Castro-Karney,
  Bornholt, Torlak, and Wang]{sigurbjarnarson:nickel}
H.~Sigurbjarnarson, L.~Nelson, B.~Castro-Karney, J.~Bornholt, E.~Torlak, and
  X.~Wang.
\newblock Nickel: A framework for design and verification of information flow
  control systems.
\newblock In \emph{Proceedings of the 13th USENIX Symposium on Operating
  Systems Design and Implementation~(OSDI)}, pages 287--306, Carlsbad, CA, Oct.
  2018.

\bibitem[Simner et~al.(2022)Simner, Armstrong, Pichon-Pharabod, Pulte,
  Grisenthwaite, and Sewell]{simner:relaxed-vm}
B.~Simner, A.~Armstrong, J.~Pichon-Pharabod, C.~Pulte, R.~Grisenthwaite, and
  P.~Sewell.
\newblock Relaxed virtual memory in {Armv8-A}.
\newblock In \emph{Proceedings of the 31st European Symposium on
  Programming~(ESOP)}, pages 143--173, Munich, Germany, Apr. 2022.

\bibitem[Skorstengaard et~al.(2019)Skorstengaard, Devriese, and
  Birkedal]{skorstengaard:stktokens}
L.~Skorstengaard, D.~Devriese, and L.~Birkedal.
\newblock {StkTokens}: Enforcing well-bracketed control flow and stack
  encapsulation using linear capabilities.
\newblock In \emph{Proceedings of the 46th ACM Symposium on Principles of
  Programming Languages~(POPL)}, Cascais, Portugal, Jan. 2019.

\bibitem[Syeda and Klein(2020)]{syeda:tlb}
H.~T. Syeda and G.~Klein.
\newblock Formal reasoning under cached address translation.
\newblock \emph{Journal of Automated Reasoning}, 64:\penalty0 911--945, 2020.

\bibitem[Tan et~al.(2016)Tan, Myreen, Kumar, Fox, Owens, and
  Norrish]{tan:cakeml}
Y.~K. Tan, M.~O. Myreen, R.~Kumar, A.~Fox, S.~Owens, and M.~Norrish.
\newblock A new verified compiler backend for {CakeML}.
\newblock In \emph{Proceedings of the 21st ACM SIGPLAN International Conference
  on Functional Programming~(ICFP)}, pages 60--73, Nara, Japan, Sept. 2016.

\bibitem[Tao et~al.(2021)Tao, Yao, Li, Li, Nieh, and Gu]{tao:vrm}
R.~Tao, J.~Yao, X.~Li, S.-W. Li, J.~Nieh, and R.~Gu.
\newblock Formal verification of a multiprocessor hypervisor on {Arm} relaxed
  memory hardware.
\newblock In \emph{Proceedings of the 28th ACM Symposium on Operating Systems
  Principles~(SOSP)}, pages 866--881, Virtual conference, Oct. 2021.

\bibitem[{The Coq Development Team}(2023)]{coq}
{The Coq Development Team}.
\newblock \emph{The {Coq} Proof Assistant, version 8.17.1}, June 2023.
\newblock URL \url{https://doi.org/10.5281/zenodo.8161141}.

\bibitem[Turon et~al.(2014)Turon, Vafeiadis, and Dreyer]{turon:gps}
A.~Turon, V.~Vafeiadis, and D.~Dreyer.
\newblock {GPS}: Navigating weak memory with ghosts, protocols, and separation.
\newblock In \emph{Proceedings of the 29th Annual ACM Conference on
  Object-Oriented Programming, Systems, Languages, and Applications~(OOPSLA)},
  pages 691--707, Portland, OR, Oct. 2014.

\bibitem[{Van Strydonck} et~al.(2022){Van Strydonck}, Georges, Gu{\'e}neau,
  Trieu, Timany, Piessens, Birkedal, and Devriese]{vanstrydonck:full-system}
T.~{Van Strydonck}, A.~L. Georges, A.~Gu{\'e}neau, A.~Trieu, A.~Timany,
  F.~Piessens, L.~Birkedal, and D.~Devriese.
\newblock Proving full-system security properties under multiple attacker
  models on capability machines.
\newblock In \emph{Proceedings of the 35th IEEE Computer Security Foundations
  Symposium~(CSF)}, pages 80--95, Haifa, Israel, Aug. 2022.

\bibitem[Vasudevan et~al.(2016)Vasudevan, Chaki, Maniatis, Jia, and
  Datta]{vasudevan:uberspark}
A.~Vasudevan, S.~Chaki, P.~Maniatis, L.~Jia, and A.~Datta.
\newblock {{\"u}berSpark}: Enforcing verifiable object abstractions for
  automated compositional security analysis of a hypervisor.
\newblock In \emph{Proceedings of the 25th USENIX Security Symposium}, pages
  87--104, Austin, TX, Aug. 2016.

\bibitem[Walker et~al.(1980)Walker, Kemmerer, and Popek]{walker:uclaunix}
B.~J. Walker, R.~A. Kemmerer, and G.~J. Popek.
\newblock Specification and verification of the {UCLA} {Unix} security kernel.
\newblock \emph{Communications of the ACM}, 23\penalty0 (2):\penalty0 118--131,
  Feb. 1980.

\bibitem[Xia et~al.(2020)Xia, Zakowski, He, Hur, Malecha, Pierce, and
  Zdancewic]{xia:itrees}
L.~Xia, Y.~Zakowski, P.~He, C.-K. Hur, G.~Malecha, B.~C. Pierce, and
  S.~Zdancewic.
\newblock Interaction trees: Representing recursive and impure programs in
  {Coq}.
\newblock In \emph{Proceedings of the 47th ACM Symposium on Principles of
  Programming Languages~(POPL)}, pages 51:1--51:32, New Orleans, LA, Jan. 2020.

\bibitem[Xu et~al.(2016)Xu, Fu, Feng, Zhang, Zhang, and Li]{xu:certiucos}
F.~Xu, M.~Fu, X.~Feng, X.~Zhang, H.~Zhang, and Z.~Li.
\newblock A practical verification framework for preemptive {OS} kernels.
\newblock In \emph{Proceedings of the 28th International Conference on Computer
  Aided Verification~(CAV)}, pages 59--79, Toronto, Canada, July 2016.

\bibitem[Yang and Hawblitzel(2010)]{yang:verve}
J.~Yang and C.~Hawblitzel.
\newblock Safe to the last instruction: Automated verification of a type-safe
  operating system.
\newblock In \emph{Proceedings of the 31st ACM SIGPLAN Conference on
  Programming Language Design and Implementation~(PLDI)}, pages 99--110,
  Toronto, Canada, June 2010.

\bibitem[Zeldovich(2026)]{rocq-warm}
N.~Zeldovich.
\newblock {rocq-warm}.
\newblock \url{https://github.com/zeldovich/rocq-warm}, Sept. 2026.

\bibitem[Zhou et~al.(2024)Zhou, Anjali, Chen, Gong, Hawblitzel, and
  Cui]{zhou:verismo}
Z.~Zhou, Anjali, W.~Chen, S.~Gong, C.~Hawblitzel, and W.~Cui.
\newblock {VeriSMo}: A verified security module for confidential {VMs}.
\newblock In \emph{Proceedings of the 18th USENIX Symposium on Operating
  Systems Design and Implementation~(OSDI)}, Santa Clara, CA, July 2024.

\bibitem[Zou et~al.(2019)Zou, Ding, Du, Fu, Gu, and Chen]{zou:atomfs}
M.~Zou, H.~Ding, D.~Du, M.~Fu, R.~Gu, and H.~Chen.
\newblock Using concurrent relational logic with helper for verifying the
  {AtomFS} file system.
\newblock In \emph{Proceedings of the 27th ACM Symposium on Operating Systems
  Principles~(SOSP)}, Huntsville, Ontario, Canada, Oct. 2019.

\end{thebibliography}

\end{document}